\documentclass[leqno]{amsart}

\usepackage[foot]{amsaddr}

\usepackage{color}
\usepackage{geometry}
\usepackage{subfiles}
\usepackage{subcaption}
\usepackage{graphicx}
\usepackage[section]{placeins}
\usepackage[linesnumbered,ruled,vlined]{algorithm2e}
\usepackage{hyperref}
\usepackage{stmaryrd}
\usepackage{amsmath,amssymb,amsfonts,amsthm,textcomp}
\usepackage{mathtools}
\usepackage{tensor}
\usepackage{multicol}
\usepackage{empheq}
\usepackage{tikz,pgfplots}
\usetikzlibrary{calc}
\usepackage{enumitem}
\newtheorem{remark}{Remark}
\usepackage{longtable}
\usepackage{mathrsfs}

\usepackage{xparse}
\usepackage{fancyhdr}
\usepackage{amssymb}
\usepackage{pifont}
\newcommand{\cmark}{\ding{51}}%
\newcommand{\xmark}{\ding{55}}%

\usepackage{bigints}

\usepackage[numbers]{natbib}

\usepackage[most]{tcolorbox}

\newtcbtheorem{Summary}{\bfseries Box}{enhanced,drop shadow={black!50!white},
  coltitle=black,
  top=0.3in,
  attach boxed title to top left=
  {xshift=1.5em,yshift=-\tcboxedtitleheight/2},
  boxed title style={size=small,colback=black!50!white}
}{summary}

\NewDocumentCommand{\dgal}{sO{}m}{%
  \IfBooleanTF{#1}
    {\dgalext{#3}}
    {\dgalx[#2]{#3}}%
}

\NewDocumentCommand{\dgalext}{m}{%
  \sbox0{%
    \mathsurround=0pt 
    $\left\{\vphantom{#1}\right.\kern-\nulldelimiterspace$ %
  }%
  \sbox2{\{}%
  \ifdim\ht0=\ht2
    \{\kern-.625\wd2 \{#1\}\kern-.625\wd2 \}%
  \else
    \left\{\kern-.7\wd0\left\{#1\right\}\kern-.7\wd0\right\}%
  \fi
}

\NewDocumentCommand{\dgalx}{om}{%
  \sbox0{\mathsurround=0pt$#1\{$}%
  \sbox2{\{}%
  \ifdim\ht0=\ht2
    \{\kern-.625\wd2 \{#2\}\kern-.625\wd2 \}%
  \else
    \mathopen{#1\{\kern-.7\wd0 #1\{}
    #2
    \mathclose{#1\}\kern-.7\wd0 #1\}}
  \fi
}

\graphicspath{
			{./Figures/intro/}
			{./Figures/3D_meshsize/}
			{./Figures/3D_Experimental/}
			{./Figures/3D_SingleSeed_aniso/}
             }

\usepackage{mathtools}
\emergencystretch=\maxdimen
\newfont{\tenbfsl}{cmbxti9 scaled 1200}
\newfont{\tenbbb}{msbm10}
\newfont{\svnbbb}{msbm8}

\theoremstyle{remark}

\theoremstyle{definition}

\newcounter{syn}[section] 
\renewcommand{\thesyn}{\arabic{section}.\arabic{syn}}

\makeatletter
\def\threevdots{\mskip+4mu\vbox{\baselineskip2.25\p@ \lineskiplimit\z@
  \kern4.9\p@\hbox{.}\hbox{.}\hbox{.}}\mskip+3.8mu}
\makeatother

\begin{document}

\title[Three-dimensional experimental-scale electrodeposition ]{Three-dimensional experimental-scale phase-field modelling of dendrite formation in rechargeable lithium-metal batteries}

\author{Marcos E. Arguello$^{\square,\ddagger,\star}$}
\email{m.arguello@postgrad.curtin.edu.au, m.arguello.19@abdn.ac.uk }
\author{Nicol\'as A. Labanda$^{\diamondsuit,\spadesuit}$}
\email{nlabanda@srk.com.au}
\author{Victor M. Calo$^{\diamondsuit}$}
\email{victor.calo@curtin.edu.au}
\author{Monica Gumulya$^{\blacktriangledown}$}
\email{m.gumulya@curtin.edu.au}
\author{Ranjeet Utikar$^{\square}$}
\email{r.utikar@curtin.edu.au}
\author{Jos Derksen$^{\ddagger}$}
\email{jderksen@abdn.ac.uk}
\address{$^{\square}$ WA School of Mines, Mineral, Energy and Chemical Engineering, Curtin University, P.O. BOX U1987, Perth, WA 6845, Australia.}
\address{$^{\ddagger}$ School of Engineering, University of Aberdeen, Elphinstone Road, AB24 3UE Aberdeen, United Kingdom.}
\address{$^{\diamondsuit}$ School of Electrical Engineering, Computing and Mathematical Sciences, Curtin University, P.O. Box U1987, Perth, WA 6845, Australia}
\address{$^{\spadesuit}$ SRK Consulting, West Perth, Western Australia, Australia.}
\address{$^{\blacktriangledown}$ Occupation, Environment and Safety, School of Population Health, Curtin University, P.O. Box U1987, Perth, WA 6845, Australia}
\address[$^{\star}$]{Corresponding author: Marcos E. Arguello, m.arguello@postgrad.curtin.edu.au}

\date{\today}

\begin{abstract}
\noindent
We perform phase-field simulations of the electrodeposition process that forms dendrites within metal-anode batteries including anisotropic representation.  We describe the evolution of a phase field, the lithium-ion concentration, and an electric potential, during a battery charge cycle,  solving equations using time-marching algorithms with automatic time-step adjustment and implemented on an open-source finite element library.  A modified lithium crystal surface anisotropy representation for phase-field electrodeposition model is proposed and evaluated through different numerical tests, exhibiting low sensitivity to the numerical parameters.  Change of dendritic morphological behaviour is captured by a variation of the simulated inter-electrode distance.  A set of simulations are presented to validate the proposed formulation,  showing their agreement with experimentally-observed lithium dendrite growth rates, and morphologies reported in the literature.

\end{abstract}

\maketitle





\section*{Symbols List}
\label{section:symbols}

 \begin{longtable}[c]{ c  c c }
 \caption*{}\\

 \multicolumn{3}{c}{}\\
 \hline
Symbol & Description & Units\\
 \hline
 \endfirsthead

 \multicolumn{3}{c}{}\\
 \hline
Symbol & Description & Units\\
 \hline
 \endhead

 \hline
 \endfoot

 \hline
 \multicolumn{3}{c}{}\\
 \endlastfoot
 
$\text{A}^-$ &  Anion species & $\left[-\right]$ \\

$C_m^l$ & Site density electrolyte & $\left[mol/m^3\right]$ \\

$C_m^s$ & Site density electrode & $\left[mol/m^3\right]$ \\

$C_0$ & Bulk Li-ion concentration & $\left[mol/m^3\right]$ \\

$D^{eff}$ & Effective diffusivity & $\left[m^2/s\right]$ \\

$\vec{E}$ & Electric field vector & $\left[V/m\right]$ \\

$E_0$ & Energy density normalization constant & $\left[J/m^3\right]$ \\

$F$ & Faraday constant & $\left[sA/mol\right]$ \\

$f_{ch}$ & Helmholtz free energy density & $\left[J/m^3\right]$ \\

$f_{grad}$ & Surface energy density & $\left[J/m^3\right]$ \\

$f_{elec}$ & Electrostatic energy density & $\left[J/m^3\right]$ \\

$g\left(\xi\right)$ & Double-well function & $\left[J/m^3\right]$ \\

$\text{h}$ & Mesh size & $\left[m\right]$ \\

$H$ & Dendrite height &  $\left[m\right]$ \\

$h\left(\xi\right)$ & Interpolation function & $\left[-\right]$ \\

$h_0$ & Length normalization constant & $\left[m\right]$ \\

$i$ & Current density & $\left[A/m^2\right]$ \\

$i_0$ & Exchange current density & $\left[A/m^2\right]$ \\

$L_\eta$ & Kinetic coefficient & $\left[1/s\right]$ \\

$L_\sigma$ & Interfacial mobility & $\left[m^3/\left(J s\right)\right]$ \\

$l_x$ & Longitudinal battery cell size (x direction) & $\left[m\right]$ \\

$l_{x_{u}}$ & Region of interest & $\left[m\right]$ \\

$l_y$ & Lateral battery cell size (y direction) & $\left[m\right]$ \\

$l_z$ & Lateral battery cell size (z direction) & $\left[m\right]$ \\

$\text{M}$ &  Metal atom species & $\left[-\right]$ \\

$\text{M}^+$ &  Cation species & $\left[-\right]$ \\

$n$ & Valence & $\left[-\right]$ \\

$R$ & Gas constant & $\left[J/\left(mol \ K\right)\right]$ \\

$\mathscr{R}$ & Phase-field interface thickness to mesh resolution ratio & $\left[-\right]$ \\

$T$ & Temperature & $\left[K\right]$ \\

$t$ & Time & $\left[s\right]$ \\

$t_0$ & Time normalization constant & $\left[s\right]$ \\

$W$ & Barrier height & $\left[J/m^3\right]$ \\

$\alpha$ & Charge transfer coefficient & $\left[-\right]$ \\

$\gamma$ & Surface Energy & $\left[J/m^2\right]$ \\

$\delta_{aniso}$ & Anisotropy strength & $\left[-\right]$ \\

$\delta_{PF}$ & Phase-field diffuse interface thickness & $\left[m\right]$ \\

${\widetilde{\zeta}}_+$ & Normalized Li-ion concentration & $\left[-\right]$ \\

$\eta$ & Total overpotential & $\left[V\right]$ \\

$\kappa$ & Gradient energy coefficient variable & $\left[J/m\right]$ \\

$\kappa_{0}$ & Gradient energy coefficient constant & $\left[J/m\right]$ \\

$\xi$ & Phase-field order parameter & $\left[-\right]$ \\

$\sigma^{eff}$ & Effective conductivity & $\left[S/m\right]$ \\

$\phi$ & Electric potential & $\left[V\right]$ \\

$\phi_b$ & Charging voltage & $\left[V\right]$ \\

$\Psi$ & Gibbs free energy & $\left[J\right]$ \\

$\omega$ & Anisotropy mode & $\left[-\right]$ \\

 \end{longtable}

\section{Introduction}
\label{section:Introduction}

Global energy demand continues to rise due to industrial activity and the world's population expansion, with an average growth rate of about 1\% to 2\% per year since 2010 (pre-Covid19 pandemic levels)~\cite{ iea2021global}. The increasing consumption of non-renewable energy reserves, such as coal, gas, and oil~\cite{ smil2016energy}, and awareness of climate change~\cite{ iea2022global, Kumar20}, have triggered a steep growth in renewable energy sources (6\% average annual growth worldwide over the past decade)~\cite{ owidenergy}, along with an urgent need for the development of improved energy storage systems~\cite{ pires2014power}. Globally, around one-quarter of our electricity comes from renewables, which include hydropower, wind, solar, biomass, ocean energy, biofuel, and geothermal~\cite{ iea2021global}.

\begin{figure} [h!]
\centering%
\begin{subfigure}[b]{0.56\linewidth}
    \centering%
{\includegraphics[height = 5.7cm]{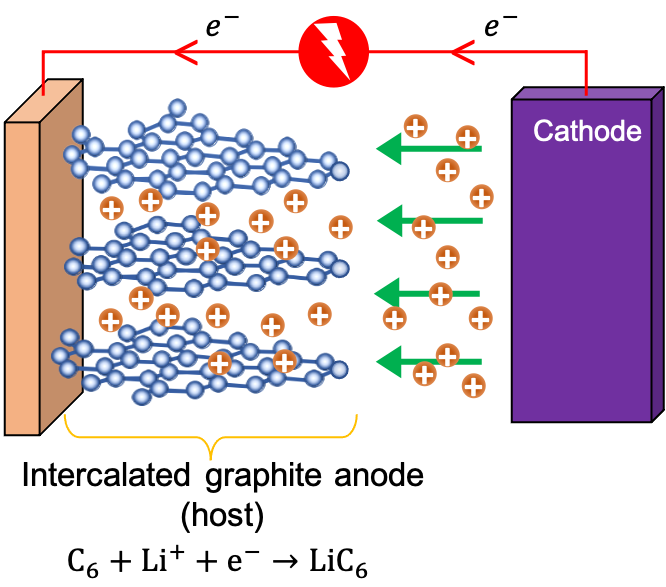}}
\caption{Li-ion Battery.}
\label{fig:Li-ion_Scheme}
\end{subfigure}
\begin{subfigure}[b]{0.4\linewidth}
    \centering%
{\includegraphics[height = 5.9cm]{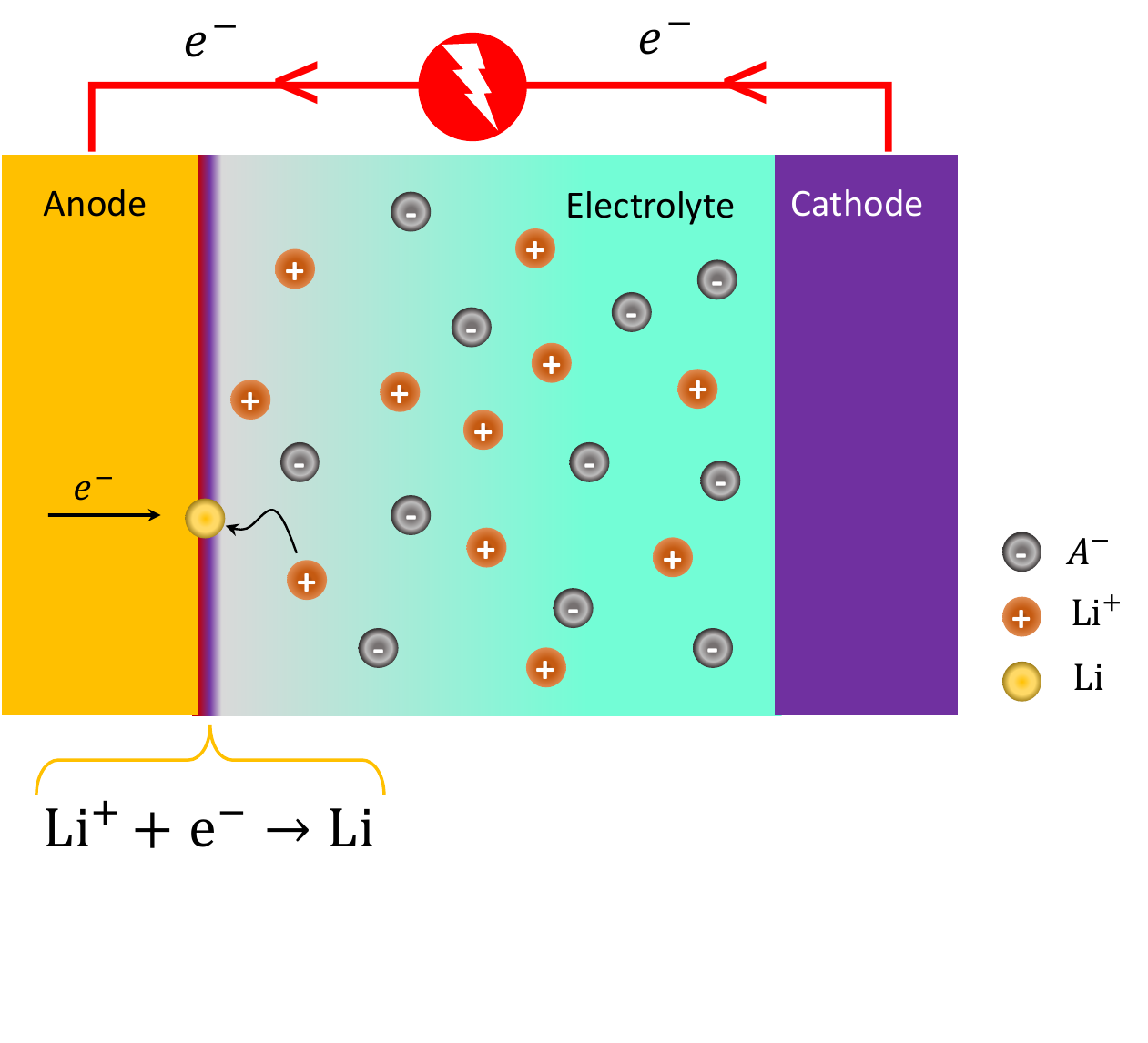}}
\caption{Li-metal Battery.}
\label{fig:Li-metal_Scheme}
\end{subfigure}
\caption{Schematic comparing charging mechanism and anode's structures between conventional Li-ion (\subref{fig:Li-ion_Scheme}), and Li-metal (\subref{fig:Li-metal_Scheme}) batteries. Grey, orange, and yellow spheres represent $\text{A}^-$ anions, $\text{M}^+$ cations, and $\text{M}$ atom, respectively.}
\label{fig:Li-ionVsLMB_Schematic}
\end{figure}

New chemistry and designs, such as metal anode batteries, are under active research to achieve an energy density of 500 Wh/kg and manufacturing costs lower than \$100/kWh~\cite{ Yanguang2017}. Despite enormous efforts, today's highest energy density remains below 400 Wh/kg, with an average growth rate of about 5\% per year since 1970~\cite{ winter2018before}. As the energy density limitation (299 Wh/kg) of conventional lithium-ion batteries based on intercalated graphite anode cannot meet the current market demand, researchers are refocusing on lithium metal batteries (LMBs)~\cite{ zhang2020towards}. LMBs can achieve ultra-high energy densities by avoiding the use of a graphite lattice to host $\text{Li}^+$ (intercalation process), as illustrated by the comparative schematic of Figure~\ref{fig:Li-ionVsLMB_Schematic}. The graphite material (host) drastically reduces the energy density of conventional Li-ion batteries by adding extra weight to the battery pack that does not participate in the electrochemical reaction~\citep{ winter2018before, lin2017ovzivitev}. 

Although lithium metal has been an attractive anode alternative in rechargeable batteries since the early 1970s, its commercialization has been hindered due to several shortcomings. The greatest challenge to achieving the commercial realization of lithium-metal batteries is related to their stability and safety~\citep{ lin2017ovzivitev}. These issues are closely linked to lithium anode: dendrite formation due to the uneven deposition of lithium, dead lithium formed after dendrites breakage, formation of unstable solid electrolyte interphase (SEI), and volume expansion of the metal anode. Additionally, these mechanisms interact, causing synergistic detrimental effects~\citep{ wang2021confronting}.  

Dendrite formation in LMBs is the consequence of lithium's uneven deposition, associated with thermodynamic and kinetic factors, such as the inhomogeneous distribution of Li-ion concentration and electric potential on the electrode surface. Furthermore, the morphology of the electrodeposited lithium is influenced by different factors such as the magnitude and frequency of the applied current density, electrolyte concentration, temperature, pressure, ion transport, and mechanical properties in the electrolyte~\cite{ frenck2019factors, JANA2017552}. Understanding dendrite formation in LMBs combines theory, experiment, and computation~\citep{ franco2019boosting}. Within computational research, based on classical chemical reaction kinetics, the phase-field method is a suitable modeling technique for studying mesoscale ($\mu m$) electro-kinetic phenomena, such as dendritic electrodeposition of lithium, with a reasonable computational cost. Phase-field (diffuse-interface) models can simulate the morphology evolution of Li electrodeposit due to reaction-driven phase transformation within metal anode batteries and rationalize morphology patterns of dendrites observed experimentally, under both low and high applied current densities~\cite{ PhysRevE.69.021603, PhysRevE.69.021604, SHIBUTA2007511, OKAJIMA2010118, PhysRevE.86.051609, Bazant2013, doi:10.1063/1.4905341, ELY2014581, Zhang_2014, CHEN2015376, PhysRevE.92.011301, doi:10.1021/acsenergylett.8b01009, YAN2018193, YURKIV2018609, MU2019100921, jana2019electrochemomechanics, wang2019phase, zhang2019dendrite, guan2015simulation, guan2018phase, liu2019phase, mu2020simulation, CHEN2021229203, Zhang_2021, https://doi.org/10.1002/advs.202003301, qiao2022quantitative, arguello2022phase, ARGUELLO2022104892, ZHANG2022285, li2022understanding}.

While progress has occurred in phase-field modeling of lithium dendrites in recent years, there are still several issues related to the evolution of dendritic patterns in lithium metal electrodes that remain unresolved~\cite{ CHEN2021229203}. The fundamental failure mechanism of lithium anode remains unclear and controversial~\citep{ wang2021confronting}. A significant effort has gone into using 2D models to rationalize 3D dendritic patterns observed experimentally~\cite{ ARGUELLO2022104892}. Furthermore, various strategies exist to suppress Li dendrites' growth and weaken side reactions. Some of these strategies address the battery operating conditions, including pulse charging lithium dendrite suppression~\cite{ qiao2022quantitative} and control of internal temperature~\cite{ YAN2018193}. Other alternatives focus on the electrode (anode), such as modeling of 3D conductive structured lithium metal anode~\cite{ zhang2019dendrite, ZHANG2022285}, and low porosity and stable SEI structure~\cite{ MU2019100921}. Besides,  other approaches center on the electrolyte management and separator design, proposing a compositionally graded electrolyte~\cite{ doi:10.1021/acsenergylett.8b01009}, dendrite suppression using flow field (forced advection) \citep{wang2019phase}, the study of separator pore size inhibition effect on lithium dendrite~\cite{ li2022understanding}.

Given the inherent 3D nature of lithium dendrite morphologies \citep{ jana2019electrochemomechanics, ding2016situ, TATSUMA20011201}, it is critical to develop phase-field models to understand the impact of 3D effects on triggering the formation of these patterns. Nevertheless, few papers attempt to simulate the full 3D lithium dendrite growth process using phase-field models. For instance, Mu at al.~\cite{  mu2020simulation} performed parallel three-dimensional phase-field simulations of dendritic lithium evolution under different electrochemical states, including charging, suspending, and discharging states. Later, Liu et al.~\cite{ https://doi.org/10.1002/advs.202003301} presented a phase-field model to study the three-dimensional effect of exchange current density on electrodeposition behavior of lithium metal; however, without focusing on dendritic morphologies. Recently, Arguello et al.~\cite{  ARGUELLO2022104892} presented 3D phase-field simulations using an open-source finite element library, to describe hazardous three-dimensional dendritic patterns in LMBs. The authors used time step adaptivity, mesh rationalization, parallel computation, and balanced phase-field interface thickness to mesh resolution ratio. The high computational cost of simulating the detailed lithium electrodeposition is a well-known challenge that has limited the domain size of phase-field simulations~\cite{ PhysRevE.92.011301, YURKIV2018609}. Thus, higher-than-normal dendrite growth rates were reported in the literature for 3D phase-field modeling of dendrite growth, due to the short separation between electrodes used~\cite{ mu2020simulation, ARGUELLO2022104892}. Recently, experimental observations by Chae et al.~\cite{ chae2022modification} have revealed a change in the lithium deposition behavior and morphology from "hazardous" needle- and moss-like dendritic structures to "safer" morphologies (smooth and round shaped surface) as interelectrode spacing increases. Therefore, simulating the dendrite formation at the experimental scale has significant practical relevance.

In this paper, we present a 3D phase-field model that describes the anisotropic dendritic electrodeposition of lithium, building on~\cite{ arguello2022phase, ARGUELLO2022104892}. This work constitutes a step toward the physics-based, quantitative models to rationalize hazardous three-dimensional dendritic patterns needed to achieve the commercial realization of Li-metal batteries~\cite{  ARGUELLO2022104892}. We organize the paper as follows: Section~\ref{section:formu_imple} presents the basic equations describing the lithium-battery dendrite growth process and details its implementation where we introduce a modified representation of the surface anisotropy of lithium metal. Section~\ref{section:3D_Single_nuclei} describes the system layout and properties, together with the implementation of symmetric boundary conditions for a detailed study of symmetric dendritic patterns~\cite{ TATSUMA20011201, jana2019electrochemomechanics}. We discuss numerical simulations of spike-like lithium-battery dendrites growth in Section~\ref{section:Senst_3D},  where we analyze the sensitivity of the simulation results for a series of spatial resolutions and phase-field interface thickness. Section~\ref{section:Aniso_Improved} evaluates the behavior of the surface anisotropy representation model for metal anode battery simulations through different numerical tests. We compare thedendritic patterns with the results obtained in preceding simulation work~\cite{ ARGUELLO2022104892}. Towards the goal of experimental-scale simulations, we test the modified surface anisotropy model under a larger interelectrode distance in Section~\ref{section:3D_LargeScale}. The lithium electrodeposition behavior changes as we increase the interelectrode distance. We study and describe the lithium dendrite propagation rates and morphologies obtained under different charging voltages. Finally, we draw conclusions in Section~\ref{section:concl}.

\section{Formulation \& implementation}
\label{section:formu_imple}

\subsection{Surface anisotropy representation for phase-field electrodeposition models}
\label{subsection:Aniso_Improved_GovEq}

In this section, we present a modified representation of the 3D surface anisotropy of lithium crystal. We start by considering the surface energy expression, following~\cite{ arguello2022phase, ARGUELLO2022104892}: $\text{f}_{\text{grad}}=\frac{1}{2}\kappa\left(\xi\right)\left(\mathrm{\nabla}\xi\right)^2$, where its variational derivative (surface anisotropy of lithium crystal) is: $\frac{\delta \text{f}_{\text{grad}}}{\delta\xi}=\kappa\left(\xi\right)\nabla^2\xi$, consistent with most recent phase-field models of dendritic electrodeposition~\cite{ Zhang_2014, CHEN2015376, PhysRevE.92.011301, doi:10.1021/acsenergylett.8b01009, MU2019100921, CHEN2021229203}.  However, a more accurate representation of $\frac{\delta \text{f}_{\text{grad}}}{\delta\xi}$ may include an additional term, as originally derived by Kobayashi~\cite{ KOBAYASHI1993410} for 2D crystal growth. In 3D, we use the variational derivative version derived by George \& Warren~\cite{ GEORGE2002264} to simulate the surface anisotropy of crystal growth
\begin{equation}
\frac{\delta \text{f}_{\text{grad}}}{\delta\xi}= \frac{\delta}{\delta\xi} \left[ \frac{1}{2}a^2\left(\mathrm{\nabla}\xi\right)^2 \right]=\mathrm{\nabla}\cdot\left(a^2\nabla\xi\right)+\sum_{i=1}^{3}\frac{\partial}{\partial x_i} \left[ a \frac{\partial a}{\partial \left( \frac{\partial\xi}{\partial x_i}\right)} \left(\mathrm{\nabla}\xi\right)^2 \right] ,
\label{eq:SurfEnerg_VarDeriv3D}
\end{equation}
where $a^2=\kappa\left(\xi\right)$ is the three-dimensional surface anisotropy or gradient coefficient. The first term after the last equality remains the same as in the previous surface anisotropy expression; however, we add a second term (derivation due to $\kappa$ variable with $\xi$). We calculate the partial derivative $\frac{\partial a}{\partial \left( \frac{\partial\xi}{\partial x_i}\right)}$ in~\eqref{eq:SurfEnerg_VarDeriv3D}. We express the 3D surface anisotropy coefficient (four-fold anisotropy)~\cite{ GEORGE2002264, ARGUELLO2022104892}, as:
\begin{equation}
a\left(\xi\right)=\sqrt{\kappa_0}\left(1-3\delta_{\text{aniso}}\right)\left[1+\frac{4\delta_{\text{aniso}}}{1-3\delta_{\text{aniso}}}\left(\frac{\sum_{i=1}^{3}\left( \frac{\partial\xi}{\partial x_i}\right)^4}{||\nabla\xi||^4}\right)\right],
\label{eq:KAPPA3D_A}
\end{equation}
where $x_1=x$, $x_2=y$, and $x_3=z$; $\kappa_0$ relates to the Lithium surface tension $\gamma$; and $\delta_{\text{aniso}}$ is the strength of anisotropy~\cite{TRAN201948, ZHENG202040}. Thus, we apply the quotient derivative rule to~\eqref{eq:KAPPA3D_A} and arrive at the partial derivative expression we use in~\eqref{eq:SurfEnerg_VarDeriv3D}; subsequently:
\begin{equation}
\begin{aligned}
\frac{\partial a}{\partial \left( \frac{\partial\xi}{\partial x_i}\right)}
=& \ 4\sqrt{\kappa_0}\delta_{\text{aniso}} \left[\frac{4 \left(\frac{\partial\xi}{\partial x_i}\right)^3 ||\nabla\xi||^4- \left(\frac{\partial\xi}{\partial x_i}\right)^4 4 ||\nabla\xi||^3 \frac{\partial ||\nabla\xi||}{\partial\left(\frac{\partial\xi}{\partial x_i}\right)}}{||\nabla\xi||^8}\right] \\
=& \ 4\sqrt{\kappa_0}\delta_{\text{aniso}} \left[\frac{4 \left(\frac{\partial\xi}{\partial x_i}\right)^3 ||\nabla\xi||^4-4 \left(\frac{\partial\xi}{\partial x_i}\right)^4 ||\nabla\xi||^3 \frac{\frac{\partial\xi}{\partial x_i}}{||\nabla\xi||}}{||\nabla\xi||^8}\right] \\
=& \ 4\sqrt{\kappa_0}\delta_{\text{aniso}} \left[\frac{4 \left(\frac{\partial\xi}{\partial x_i}\right)^3||\nabla\xi||^2- 4\left(\frac{\partial\xi}{\partial x_i}\right)^5}{||\nabla\xi||^6}\right] \\
=& \ 4\sqrt{\kappa_0}\delta_{\text{aniso}}\left[ \frac{4\left(n_i^3-n_i^5\right)}{{||\nabla\xi||}} \right] \ \text{for} \ i=1,2,3 \ ,
\end{aligned}
\label{eq:KAPPA3D_A_DERIVED}
\end{equation}
where $n_i=\frac{\frac{\partial\xi}{\partial x_i}}{||\nabla\xi||}$ for $i=1,2,3$.

Although its extensive use in phase-field models of crystal growth (solidification)~\cite{ GEORGE2002264, TAKAKI201321}, only Wang et al.~\cite{ wang2015dendrite} apply these models to a 2D phase-field simulation of dendrite growth in the recharging process of zinc–air batteries. This limited use of this known model is because it induces minor morphological changes in 2D electrodeposition process; compare the similarity of the 2D dendritic morphologies reported by Wang et al.~\cite{ wang2015dendrite} including the additional surface anisotropy term, and Zhang et al.~\cite{Zhang_2014} not using it. However, as we show later, its effect is crucial when modeling 3D dendritic growth.

We modify the phase-field Butler-Volmer equation (reactive Allen-Cahn)~\cite{ CHEN2015376, arguello2022phase, ARGUELLO2022104892} by including the additional surface anisotropy term:
\begin{equation}
\begin{aligned}
\frac{\partial\xi}{\partial t}&=-L_\sigma\left\{\frac{\partial g\left(\xi\right)}{\partial\xi}-\mathrm{\nabla}\cdot\left(a^2\nabla\xi\right)-\sum_{i=1}^{3}\frac{\partial}{\partial x_i} \left[ a \frac{\partial a}{\partial \left( \frac{\partial\xi}{\partial x_i}\right)} \left(\mathrm{\nabla}\xi\right)^2 \right]\right\}\\\
&\qquad\qquad\qquad\qquad-L_\eta\frac{\partial h\left(\xi\right)}{\partial\xi}\left[e^{\left(\frac{\left(1-\alpha\right)nF\phi}{RT}\right)}-{\widetilde{\zeta}}_+\ e^{\left(\frac{-\alpha nF\phi}{RT}\right)}\right] \ .
\end{aligned}
\label{eq:linealPF_Aniso}
\end{equation}

\subsection{Governing equations}
\label{subsection:Gov_eq}

Based on phase-field theory, the governing equations for dendrite growth in lithium-metal batteries were discussed in~\cite{ arguello2022phase, ARGUELLO2022104892}. Thus, herein, we only summarize this problem's basic equations, modifying just the phase-field equation for inclusion of the surface anisotropy model (see~\cite{ arguello2022phase, ARGUELLO2022104892} for further details and references). Additionally, we include a symbols list with descriptions and associated units for reference.

\begin{figure}[h]
    \centering%
{\includegraphics[height = 6.5cm]{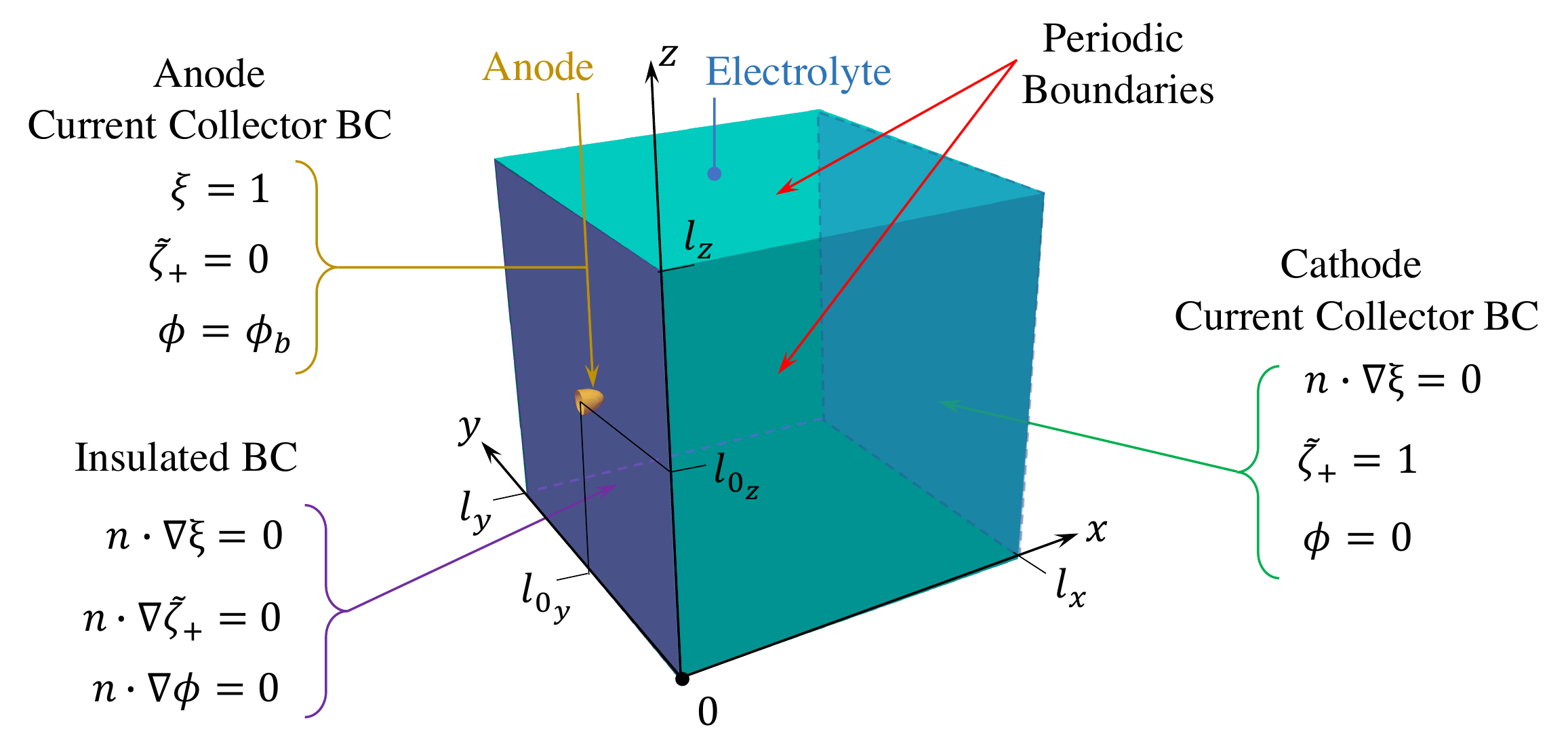}}
\caption{System layout \& boundary conditions for artificial nucleation simulations as defined in~\cite{ARGUELLO2022104892}. Reproduced with Journal's permission}
\label{fig:BC_Seed3D}
\end{figure}

The lithium-metal batteries dendrite problem using phase field formulation is simulated by the following set of equations:\textit{Find $\Xi=\left(\xi, {\widetilde{\zeta}}_+,\phi\right)$ fulfilling}
\begin{equation}
  \left\{
    \begin{aligned}
      \displaystyle \frac{\partial\xi}{\partial t} &= \displaystyle  -L_\sigma\left\{\frac{\partial g\left(\xi\right)}{\partial\xi}-\mathrm{\nabla}\cdot\left(a^2\nabla\xi\right)-\sum_{i=1}^{3}\frac{\partial}{\partial x_i} \left[ a \frac{\partial a}{\partial \left( \frac{\partial\xi}{\partial x_i}\right)} \left(\mathrm{\nabla}\xi\right)^2 \right]\right\} \\
      &\qquad\qquad-L_\eta\frac{\partial h\left(\xi\right)}{\partial\xi}\left[e^{\left(\frac{\left(1-\alpha\right)nF\phi}{RT}\right)}-{\widetilde{\zeta}}_+\ e^{\left(\frac{-\alpha nF\phi}{RT}\right)}\right] ,
      && \text{ in }V \times I  \\
      \displaystyle \frac{\partial{\widetilde{\zeta}}_+}{\partial t}   &= \displaystyle  \nabla\cdot\left[ D^{\text{eff}}\left(\xi\right)\ \mathrm{\nabla}{\widetilde{\zeta}}_+ +D^{\text{eff}}\left(\xi\right)\frac{n F }{R T}{\widetilde{\zeta}}_+\mathrm{\nabla\phi}\right] \\
      &\qquad\qquad-\frac{\text{C}_m^s}{C_0}\frac{\partial\xi}{\partial t} ,
      &&  \text{ in }V \times I   \\
      \displaystyle n F \text{C}_m^s\frac{\partial\xi}{\partial t}  &= \displaystyle \mathrm{\nabla}\cdot\left[\sigma^{\text{eff}}\left(\xi\right)\ \mathrm{\nabla\phi}\right]   ,
      &&  \text{ in }V \times I   \\ 
      \xi &= \xi_{D}  ,
      && \text{ on } \partial V_{D} \times I   \\ 
      {\widetilde{\zeta}}_{+} &= \widetilde{\zeta}_{+D}  ,
      &&  \text{ on } \partial V_{D} \times I  \\
      \phi &= \phi_{D}    ,
      &&\text{ on } \partial V_{D} \times I   \\ 
      \nabla \xi   \cdot \boldsymbol{n} &= 0   ,
      &&  \text{ on }  \partial V_{N}  \times I  \\
      \nabla {\widetilde{\zeta}}_+ \cdot \boldsymbol{n}  &= 0 ,
      &&  \text{ on }  \partial V_{N} \times I \\
      \nabla \phi   \cdot \boldsymbol{n} &= 0  ,
      &&  \text{ on }  \partial V_{N} \times I  \\ 
      \xi \left( \boldsymbol{x} , t_0 \right) &= \xi_{0}   ,
      &&   \text{ in }V  \\
      {\widetilde{\zeta}}_{+} \left( \boldsymbol{x} , t_0 \right) &= {\widetilde{\zeta}}_{+0}  ,
      &&   \text{ in }V  \\
      \phi \left( \boldsymbol{x} , t_0 \right) &= \phi _{0}   ,
      &&  \text{ in }V 
\end{aligned}\right. 
\end{equation}
where $\xi$ is the phase-field order parameter, ${\widetilde{\zeta}}_+$ is the lithium-ion concentration, and $\phi$ is the electric potential; $V$ is the problem domain with boundary $\partial V = \partial V_{N} \cup \partial V_{D}$, the subscript $N$ and $D$ related to the Neumann and Dirichlet parts (see \ref{fig:BC_Seed3D}), with outward unit normal $\boldsymbol{n}$, and $I$ is the time interval.

\subsection{Implementation details}
\label{subsection:implement}

Following~\cite{ ARGUELLO2022104892}, we discretize and solve the set of partial differential equations describing the coupled electrochemical interactions during a battery charge cycle using an open-source finite element library (FEniCS environment)~\cite{ alnaes2015fenics}. We use eight-node (tri-linear) hexahedral elements. We use a message passing interface package MPI4py~\cite{ 9439927, DALCIN20111124, DALCIN2008655, DALCIN20051108} for parallelization and solve nonlinear using SNES from PETSc~\cite{petsc2021}. We perform the simulations using a laptop with a 2.4 GHz processor with 8-core Intel Core i9 and 16 GB 2667 MHz DDR4 RAM (see~\cite{ARGUELLO2022104892} for further details).




\section{System layout \& properties}
\label{section:3D_Single_nuclei}

Generally, the computational domain for a battery simulation comprises the anode and cathode regions and the space between the electrodes filled with an electrolyte~\cite{ Trembacki_2019}. However, in most phase-field simulations of metal electrodeposition, including those we perform, the cathode region reduces to a current collector boundary condition on the electrolyte side of the domain (see Figure~\ref{fig:BC_Seed3D}). We model a battery cell with a traditional sandwich architecture, undergoing a recharging process under fixed applied electro potential. We represent this cell as an $l_x\times l_y\times l_z$ hexagonal domain. Figure~\ref{fig:BC_Seed3D} summarizes the boundary conditions we apply~\cite{ ARGUELLO2022104892}. The initial structure consists of artificial nucleation regions as part of the problem's initial condition. Ellipsoidal protrusions (seeds) with semi-axes $r_x$, $r_y$, $r_z$, and center $\left(0,l_{0_y},l_{0_z}\right)$, exist at the surface of the anode undergoing electrodeposition (see Figure~\ref{fig:BC_Seed3D}); seeding is a widely-used strategy in phase-field simulations of electrodeposition~\cite{ Zhang_2014, CHEN2015376, YURKIV2018609, MU2019100921}, which reduces the computational cost as the lithium metal is only able to electrodeposit on the same nuclei, enhancing dendrite growth and allowing for detailed study of its morphology~\cite{ ARGUELLO2022104892}.

\begin{table}[h]
\caption{Simulation Parameters~\cite{ARGUELLO2022104892}. Reproduced with Journal's permission.} 
\centering 
\begin{tabular}{l c c c c} 
\hline\hline 
Description & Symbol & Real Value & Normalized & Source  \\ [0.5ex] 
\hline 
Exc. current density & $\text{i}_{0}$ & $30\left[A/m^2\right]$ & $30$ &~\cite{ Monroe_2003}  \\ 
Surface tension & $\gamma$ & $0.556\left[J/m^2\right]$ & $0.22$ &~\cite{ VITOS1998186,tran2016surface}\\
Barrier height & $W$ & $W=\frac{12\gamma}{\delta_{PF}}=4.45\times10^{6}\left[J/m^3\right]$ & $1.78$ & computed  \\ 
Gradient energy coefficient & $\kappa_{0}$ & $\kappa_{0}=\frac{3\gamma\delta_{PF}}{2}=1.25\times10^{-6}\left[J/m\right]$ & $0.5$ & computed  \\
Anisotropy strength & $\delta_{aniso}$ & $0.044$ & $0.044$ &~\cite{ tran2016surface,TRAN201948}  \\
Anisotropy mode & $\omega$ & $4$ & $4$ &~\cite{ TRAN201948,ZHENG202040}  \\
Kinetic coefficient  & $L_{\eta}$ & $L_{\eta}=\text{i}_{0}\frac{\gamma}{\text{nF}\text{C}_m^s}=1.81\times10^{-3}\left[1/s\right]$ & $1.81\times10^{-3}$ & computed    \\
Site density electrode & $\text{C}_m^s$ & $7.64\times10^{4}\left[mol/m^3\right]$ & $76.4$ &~\cite{ doi:10.1021/acsenergylett.8b01009} \\
Bulk Li-ion concentration & $\text{C}_{0}$ & $10^{3}\left[mol/m^3\right]$ & $1$ & computed  \\
Conductivity electrode & $\sigma^s$ & $10^{7}\left[S/m\right]$ & $10^{7}$ &~\cite{ CHEN2015376} \\
Conductivity electrolyte & $\sigma^l$ & $1.19\left[S/m\right]$ & $1.19$ &~\cite{ Valo_en_2005}   \\
Diffusivity electrode & $D^s$ & $7.5\times10^{-13}\left[m^2/s\right]$ & $0.75$ &~\cite{ CHEN2015376}   \\
Diffusivity electrolyte & $D^l$ & $3.197\times10^{-10}\left[m^2/s\right]$ & $319.7$ &~\cite{ Valo_en_2005}  \\ [1ex] 
\hline 
\end{tabular}
\label{table:material_parameters} 
\end{table}

The initial condition for a transition zone between solid electrode ($\xi=1$) and liquid electrolyte ($\xi=0$), vary in the $x$ spatial direction according to: $\frac{1}{2}\left[1-\text{tanh}\left(x\sqrt{\frac{W}{2\kappa_{0}}}\right)\right]$~\citep{ doi:10.1146/annurev.matsci.32.101901.155803}. For the artificial nucleation case, we modify the initial condition formula, replacing “$x$” by $h_0\left[\left(\frac{x}{r_x}\right)^2+\left(\frac{y-l_{0_y}}{r_y}\right)^2+\left(\frac{z-l_{0_z}}{r_z}\right)^2-1\right]$ within the hyperbolic tangent argument, to account for a smooth transition between the solid seed (lithium metal anode) and the surrounding liquid electrolyte region~\cite{ ARGUELLO2022104892}. We assume the cell's electrolyte to be 1M $\text{LiPF}_\text{6}$ EC/DMC 1:1 volume ratio solution~\cite{ doi:10.1021/acsenergylett.8b01009}. The electrode phase is a pure solid (neglecting any solid phase nanoporosity). Furthermore, Table~\ref{table:material_parameters} presents the phase-field model parameters~\cite{ ARGUELLO2022104892}. The normalization constants for length, time, energy and concentration scales are $h_0=1\left[\mu m\right]$, $t_0=1\left[s\right]$, $E_0=2.5\times10^6 \left[J/m^3\right]$, and $C_0=1\times10^3\left[mol/m^3\right]$, respectively~\cite{ doi:10.1021/acsenergylett.8b01009}.

Table~\ref{table:steup_parameters} provides a list of the simulation settings and numerical parameters used in each numerical test presented in this work. For reference, the numerical tests have been enumerated in the order in which they appear in this paper.

\begin{table}[h]
\caption{Summary of simulations settings \& numerical parameters.} 
\centering 
\begin{tabular}{c c c c c c c c c} 
\hline\hline 
Test & Symmetric & Modified & Inter- & Charging & Interfacial & Mesh & Mesh & Phase-field  \\ [0.5ex] 
 & BCs & Surface & electrode & Voltage & Mobility & Size & Rotation &  Interface  \\ [0.5ex] 
 & & Anisotropy & Distance & &  &  &  & Thickness  \\ [0.5ex] 
\hline 
\# & $\left[-\right]$ & $\left[-\right]$ & $l_x\left[\mu m\right]$ & $\phi_b\left[V\right]$ & $L_{\sigma}\left[m^3/\left(Js\right)\right]$ & h$\left[\mu m\right]$ & $\left[-\right]$ & $\delta_{PF}\left[\mu m\right]$ \\ [0.5ex] 
\hline 
1 & \cmark & \xmark & 80 & -0.7 & $2.5\times 10^{-3}$ & 0.5 & \xmark & 2  \\ 
2 & \cmark & \xmark & 80 & -0.7 & $2.5\times 10^{-3}$ & 0.33 & \xmark & 2  \\ 
3 & \cmark & \xmark & 80 & -0.7 & $2.5\times 10^{-3}$ & 0.25 & \xmark & 2  \\ 
4 & \cmark & \xmark & 80 & -0.7 & $2.5\times 10^{-3}$ & 0.5 & \xmark & 1.5  \\
5 & \cmark & \xmark & 80 & -0.7 & $2.5\times 10^{-3}$ & 0.25 & \xmark & 1.5  \\
6 & \cmark & \xmark & 80 & -0.7 & $2.5\times 10^{-3}$ & 0.25 & \xmark & 1  \\
7  & \xmark & \cmark & 80 & -0.7 & $2.5\times 10^{-3}$ & 0.5 & \xmark & 1.5 \\
8 &  \xmark & \xmark & 80 & -0.7 & $2.5\times 10^{-3}$ & 0.5 & \xmark & 1.5 \\
9 &  \xmark & \xmark & 80 & -0.7 & $2.5\times 10^{-3}$ & 0.5 & \cmark & 1.5 \\
10 & \xmark & \cmark & 80 & -0.7 & $2.5\times 10^{-3}$ & 0.5 & \cmark & 1.5 \\
11 & \xmark & \cmark & 80 & -0.7 & $2.5\times 10^{-3}$ & 0.5 & \xmark & 1.5 \\
12 & \cmark & \cmark & 80 & -0.7 & $2.5\times 10^{-3}$ & 0.25 & \xmark & 1 \\
13 & \xmark & \cmark & 5000 & -0.7 & $2.5\times 10^{-4}$ & 0.4 & \xmark & 1.5 \\
14 & \xmark & \cmark & 5000 & -1.4 & $2.5\times 10^{2}$ & 0.4 & \xmark & 1.5 \\ [1ex] 
\hline 
\end{tabular}
\label{table:steup_parameters} 
\end{table}

\begin{figure} [h!]
    \centering%
{\includegraphics[height = 6.5cm]{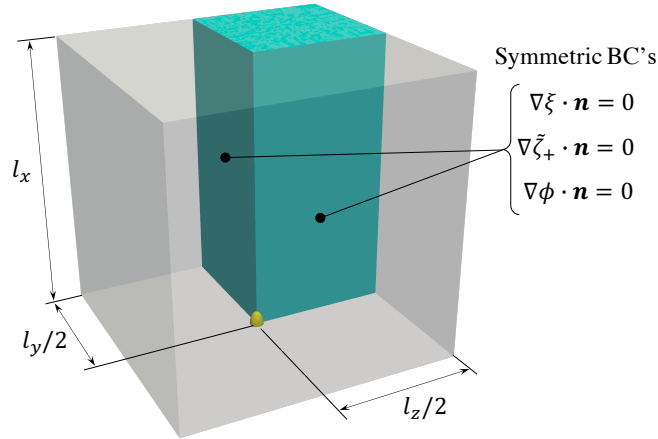}}
\caption{Symmetric boundary conditions for 3D spike-like simulations.}
\label{fig:SymmetricBCs_SingleSeed3D}
\end{figure}

\subsection{Symmetric boundary conditions}
\label{subsubsection:Symmetry_3D}

The symmetric nature of spike-like lithium morphology~\cite{ TATSUMA20011201, jana2019electrochemomechanics, ARGUELLO2022104892} allows us to reduce the computational cost by using symmetry boundary conditions to model only one-quarter of the domain. Thus, we split the domain in four, and apply Neumann boundary conditions ($\nabla \xi   \cdot \boldsymbol{n} = 0; \ \nabla {\widetilde{\zeta}}_+   \cdot \boldsymbol{n} = 0; \ \nabla \phi   \cdot \boldsymbol{n} = 0$) to those boundaries facing the center of the domain as depicted in Figure~\ref{fig:SymmetricBCs_SingleSeed3D}. Therefore, we reduce the size of our computational domain to 25\% ($l_x,l_y/2,l_z/2$). We verify our strategy by comparing the previous 3D simulation result using the whole domain (see~\cite{ ARGUELLO2022104892}) and those obtained using symmetric boundary conditions (see Figure~\ref{fig:Delta15_R3_SingleSeed3D}). Thus, the symmetric boundary conditions reduce the computational cost, which allows us to use finer meshes in the sensitivity analysis.

\section{Phase-field interface thickness to mesh resolution ratio: A sensitivity analysis}
\label{section:Senst_3D}

The phase-field interface thickness significantly affects the simulated reaction rates~\cite{ arguello2022phase}. Wider interfaces (larger $\delta_{PF}$) increase the reactive area in the simulation, which induces faster electrodeposition rates. Numerical evidence shows that 1D interface-thickness-independent growth (convergent results) are possible well before reaching the physical nanometer interface width~\cite{MORIGAKI200213, arguello2022phase}.

In this section, we perform a sensitivity analysis to study possible mesh-induced effects on the simulated 3D dendrite morphology, propagation rates (dendrite's height vs time), electrodeposition rates (dendrite's volume vs time), and energy levels. We compare 3D simulation results for different spatial resolutions and phase-field interface thicknesses. We use the anisotropy model from~\cite{ARGUELLO2022104892} to enable the comparison with the results reported therein.

\begin{figure}[h!]
\begin{subfigure}[b]{0.32\linewidth}
    \centering%
	{\includegraphics[height = 4cm]{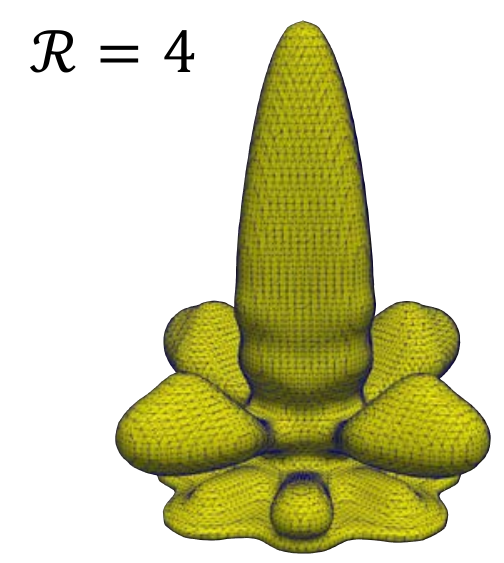}}
	\caption{$\delta_{PF} = 2\left[\mu m\right]; \ t=0.76\left[s\right]$.}
	\label{fig:Delta20_R4_SingleSeed3D}
\end{subfigure}
\begin{subfigure}[b]{0.32\linewidth}
    \centering%
    {\includegraphics[height = 4cm]{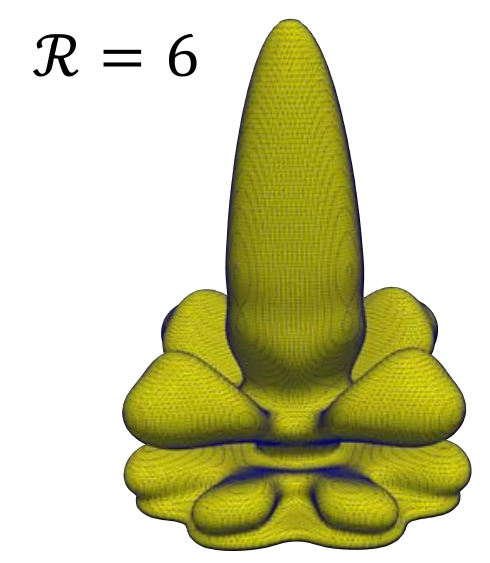}}
    \caption{$\delta_{PF} = 2\left[\mu m\right]; \ t=0.89\left[s\right]$.}
    \label{fig:Delta20_R6_SingleSeed3D}
\end{subfigure}
\begin{subfigure}[b]{0.32\linewidth}
    \centering%
    {\includegraphics[height = 4cm]{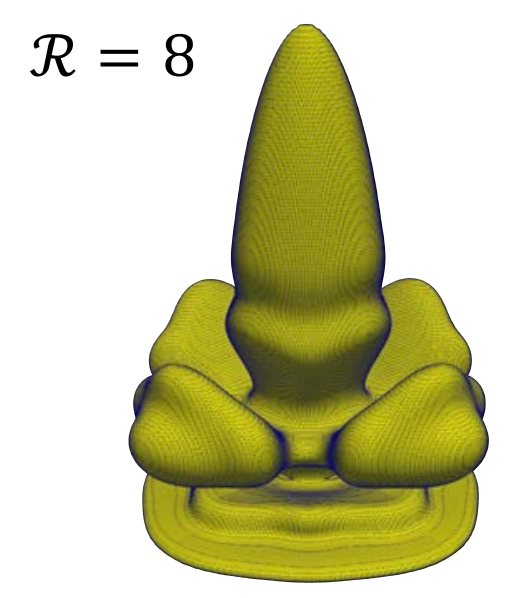}}
    \caption{$\delta_{PF} = 2\left[\mu m\right]; \ t=1.06\left[s\right]$.}
    \label{fig:Delta20_R8_SingleSeed3D}
\end{subfigure} \\
\centering%
\begin{subfigure}[b]{0.32\linewidth}
    \centering%
    {\includegraphics[height = 4cm]{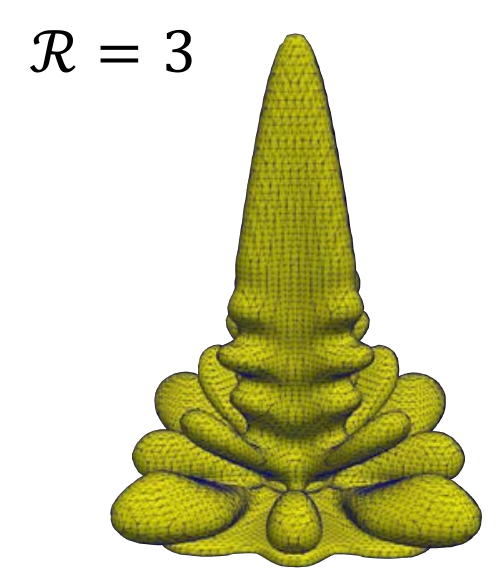}}
    \caption{$\delta_{PF} = 1.5\left[\mu m\right]; \ t=0.58\left[s\right]$.}
    \label{fig:Delta15_R3_SingleSeed3D}
\end{subfigure}
\centering%
\begin{subfigure}[b]{0.32\linewidth}
    \centering%
    {\includegraphics[height = 4cm]{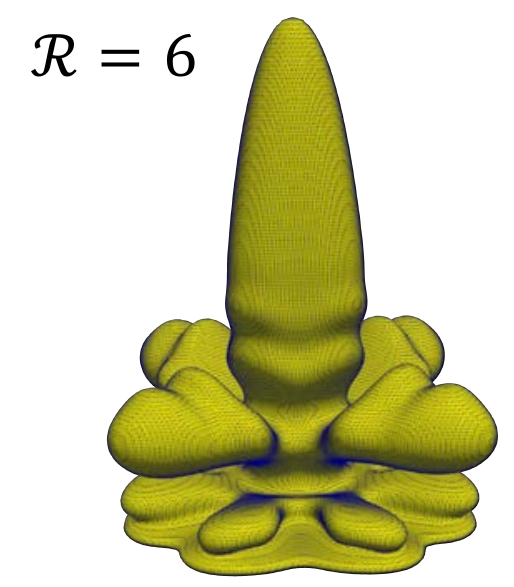}}
    \caption{$\delta_{PF} = 1.5\left[\mu m\right]; \ t=0.76\left[s\right]$.}
     \label{fig:Delta15_R6_SingleSeed3D}
\end{subfigure} 
\begin{subfigure}[b]{0.32\linewidth}
    \centering%
    {\includegraphics[height = 4cm]{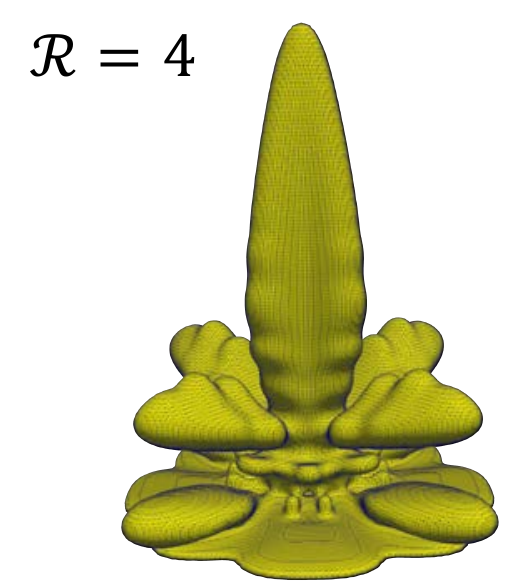}}
    \caption{$\delta_{PF} = 1.0\left[\mu m\right]; \ t=0.52\left[s\right]$.}
    \label{fig:Delta10_R4_SingleSeed3D}
\end{subfigure}
\caption{Sensitivity analysis of 3D spike-like lithium dendrite morphology, for different phase-field interface thickness to mesh resolution ratios ($\mathscr{R}=\delta_{PF}/\text{h}$), under $\phi_b=-0.7\left[V\right]$ charging potential. Simulated morphologies for $\delta_{PF}=2, \ 1.5 \ \text{and} \ 1  \left[\mu m\right]$ phase-field interface thickness; mesh grid overlaid with dendrite's morphology. We use dendrite's common height ($H=45\left[\mu m\right]$) as the basis of our comparison. Tests 1 to 6.}
\label{fig:MeshvsDeltaPF__SingleSeed3D}
\end{figure}

We select a geometric unit that characterizes a real cell structure~\cite{ YURKIV2018609, Trembacki_2019}. We choose a computational domain of $80 \times 80 \times 80 \left[\mu m^3\right]$. Consequently, given the domain size, we expect growth rates up to two orders of magnitude faster than those that occur in physical scale cells under the same applied voltage, due to the short separation between electrodes $l_x = 80 \left[\mu m\right]$~\cite{ARGUELLO2022104892}. Figure~\ref{fig:MeshvsDeltaPF__SingleSeed3D} presents a collection of 3D spike-like lithium dendrite morphologies (isosurface plot of the phase-field variable $\xi=0.5$), obtained by varying the phase-field interface thickness ($\delta_{PF}=$ 1, 1.5 and 2  $\left[\mu m\right]$), and mesh sizes (h = 0.5, 0.375 and 0.25 $\left[\mu m\right]$) (Tests 1 to 6). Thus, we test different combinations of phase-field interface thickness to mesh resolution ratios ($\mathscr{R}=\delta_{PF}/\text{h}=3 \ \text{to} \ 8$). We compare dendrite morphologies at the moment they reach a height of $H=45\left[\mu m\right]$. 

Figure~\ref{fig:MeshvsDeltaPF__SingleSeed3D} shows spike-like patterns that exhibit morphological similarity. Each consists of the main trunk and four side branches growing in each horizontal direction (as a result of the body-centered cubic (bcc) crystallographic arrangement of lithium metal~\cite{ YURKIV2018609}). Smaller phase-field interface thicknesses produce more slender dendritic morphologies (cf. Figures~\ref{fig:Delta10_R4_SingleSeed3D} ($\delta_{PF}=1\left[\mu m\right]$) and~\ref{fig:Delta20_R8_SingleSeed3D} ($\delta_{PF}=2\left[\mu m\right]$)). We use Paraview's mean curvature measurement~\cite{ayachit2015paraview} to analyse the dendrite's tip radius (isosurface plot of the phase-field variable $\xi=0.5$). Thus, the measured dendrite's tip radius in Figure~\ref{fig:Delta10_R4_SingleSeed3D} ($\delta_{PF}=1\left[\mu m\right]$) is about $r_{tip_{1}}=2.7\left[\mu m\right]$, while the computed dendrites' tip radius in Figure~\ref{fig:Delta20_R8_SingleSeed3D} ($\delta_{PF}=2\left[\mu m\right]$) is about $r_{tip_{2}}=5.3\left[\mu m\right]$ (49\% larger). The dendrite's maximum cross sectional area, main trunk, in Figure~\ref{fig:Delta10_R4_SingleSeed3D} ($\delta_{PF}=1\left[\mu m\right]$) is approximately $A_{max_{1}}=154\left[\mu m^2\right]$, while the computed cross sectional area in Figure~\ref{fig:Delta10_R4_SingleSeed3D} ($\delta_{PF}=2\left[\mu m\right]$) is about $A_{max_{2}}=254\left[\mu m^2\right]$ (40\% larger). In addition, Figure~\ref{fig:MeshvsDeltaPF__SingleSeed3D} shows that increasing the resolution ratio delivers thicker dendritic morphologies ($\mathscr{R}=\delta_{PF}/\text{h}$), keeping the phase-field interface thickness constant (more elements at the interface). We compare the morphologies in the first row of Figure~\ref{fig:MeshvsDeltaPF__SingleSeed3D} ($\delta_{PF}=2\left[\mu m\right]$), against those on the second row of Figure~\ref{fig:MeshvsDeltaPF__SingleSeed3D} ($\delta_{PF}=1.5\left[\mu m\right]$); in both cases finer mesh resolutions (higher $\mathscr{R}$) lead to less slender and less branched dendritic morphologies.

\begin{figure} [h!]
    \centering%
{\includegraphics[height = 8cm]{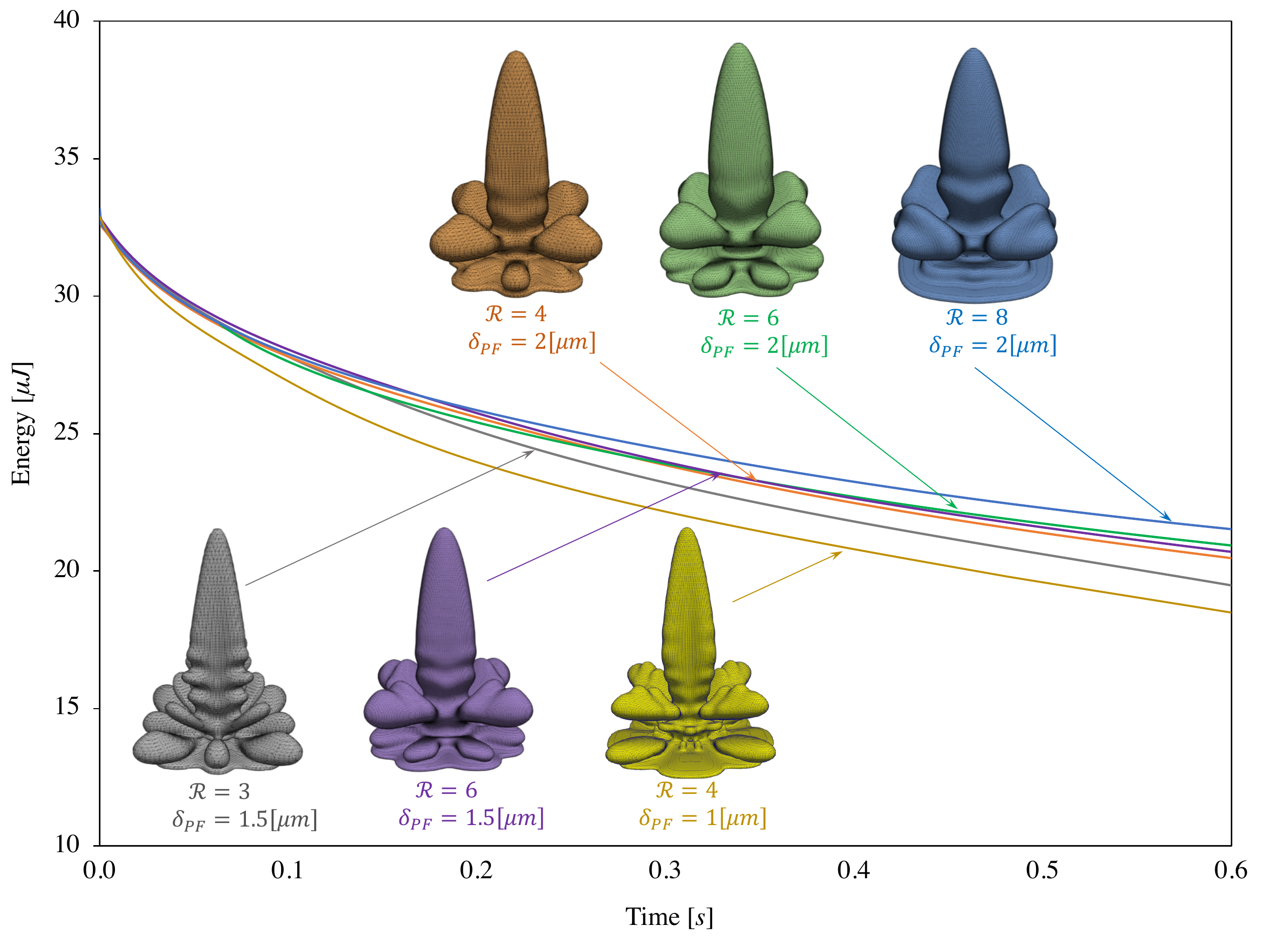}}
\caption{Comparison of energy time series for 3D spike-like dendrite growth simulations, for different phase-field interface thickness to mesh resolution ratios ($\mathscr{R}=\delta_{PF}/\text{h}$), under $\phi_b=-0.7\left[V\right]$ charging potential. Dendrite morphologies at height $H=45\left[\mu m\right]$ for reference (colours by phase-field interface thickness and mesh size). Tests 1 to 6.}
\label{fig:EnegryComparison_SingleSeed3D}
\end{figure}

Figure~\ref{fig:EnegryComparison_SingleSeed3D} shows the evolution of the Gibbs free energy of the system $\Psi=\int_{V}\left[\text{f}_{\text{ch}}\left(\xi,\zeta_i\right)+\frac{1}{2}\kappa\left(\xi\right)\left(\mathrm{\nabla}\xi\right)^2+\text{f}_{\text{elec}}\left(\xi,\zeta_i,\phi\right)\right]dV$~\cite{ ARGUELLO2022104892}. We plot the total energy curve for different simulation set-ups (phase-field interface thickness $\delta_{PF}$ and mesh resolution ratio $\mathscr{R}$), as the figure indicates. Figure~\ref{fig:EnegryComparison_SingleSeed3D} shows that in all cases, the systems' discrete free energy does not increase with time (adaptivity delivers discrete energy stable results). We obtain a maximum energy difference of about 9\% ($t=0.6\left[s\right]$) between the simulations with the maximum ($\delta_{PF}=2\left[\mu m\right]$ \& $\mathscr{R}=8$) and minimum ($\delta_{PF}=1\left[\mu m\right]$ \& $\mathscr{R}=4$) total energy levels (see Figure~\ref{fig:EnegryComparison_SingleSeed3D}). In addition, those dendrites sharing a similar level of total energy (represented in green, orange, and purple) exhibit closer morphological resemblance, as the figure shows. 

\begin{figure} [h!]
    \centering%
{\includegraphics[height = 8cm]{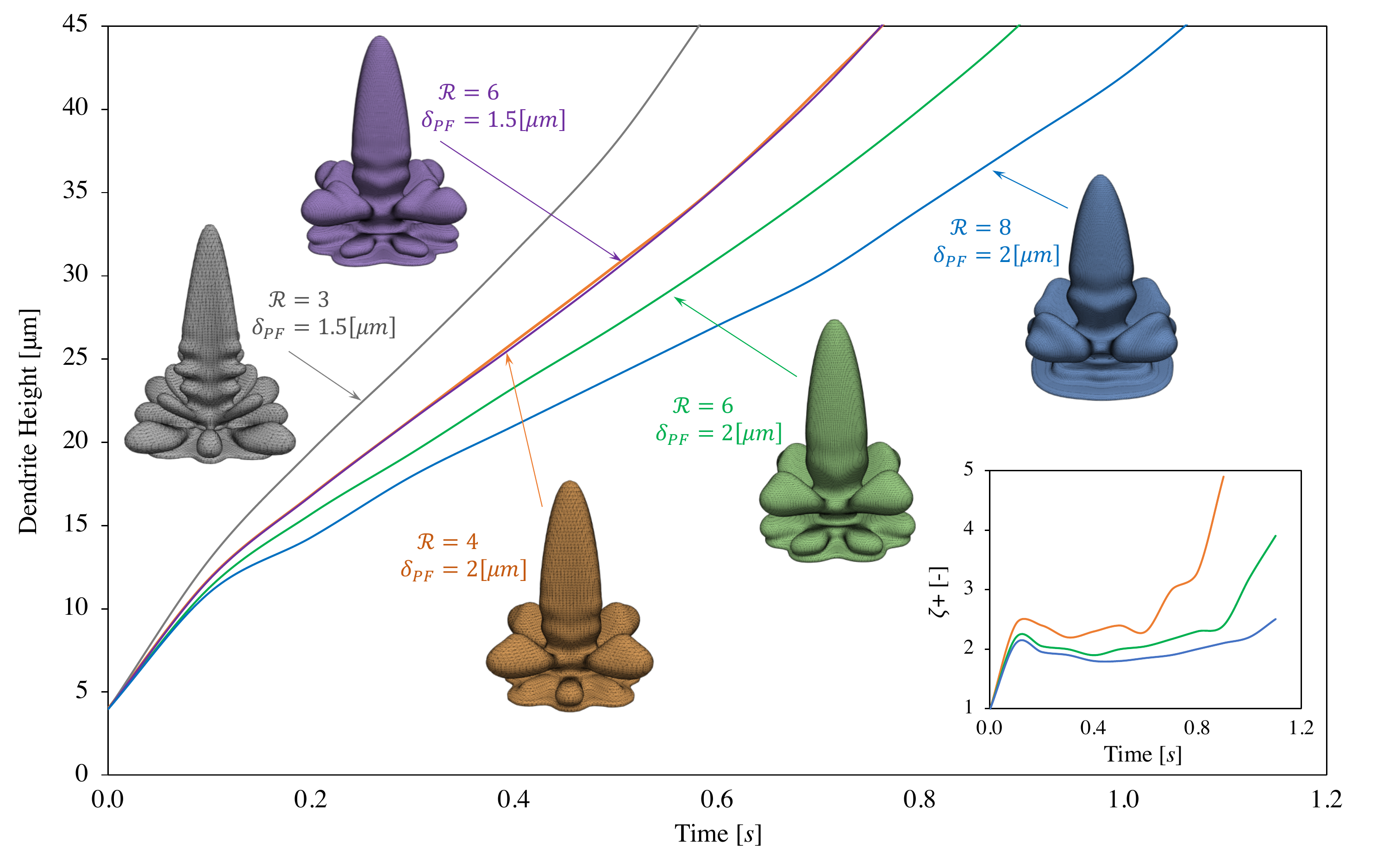}}
\caption{Comparison of 3D spike-like dendrite propagation rate, for different phase-field interface thickness to mesh resolution ratios ($\mathscr{R}=\delta_{PF}/\text{h}$), under $\phi_b=-0.7\left[V\right]$ charging potential. Dendrite morphologies at common height $H=45\left[\mu m\right]$ for reference (colours by phase-field interface thickness and mesh size). The inset shows maximum Li-ion concentration as a function of time for different $\mathscr{R}=\delta_{PF}/\text{h}$ ratios, using the same phase-field interface thickness ($\delta_{PF}=2 \left[\mu m\right]$). Tests 1 to 6.}
\label{fig:GrowthRate_SingleSeed3D}
\end{figure}

Figure~\ref{fig:GrowthRate_SingleSeed3D} shows the effect of the phase-field interface thickness and mesh resolution ratio on the dendrite's propagation rate ($H \ \text{vs} \ t$). Smaller phase-field interface thickness produces significantly faster propagation rates. For example, simulation using smaller interface thickness ($\delta_{PF}=1\left[\mu m\right]$) exhibits up-to 100\% faster growth rates than those obtained under larger interface thickness ($\delta_{PF}=2\left[\mu m\right]$). Figure~\ref{fig:GrowthRate_SingleSeed3D} shows that slower dendrite propagation rates occur as we increase the mesh resolution ratio ($\mathscr{R}=\delta_{PF}/\text{h}$), keeping the phase-field interface thickness constant. The inset in Figure~\ref{fig:GrowthRate_SingleSeed3D} plots the maximum Li-ion concentration surrounding the dendrite's tips for $\delta_{PF}=2\left[\mu m\right]$ where the enriched Li-ion concentration decreases as we increase the mesh resolution $\mathscr{R}=\delta_{PF}/\text{h}$ (more accurate solution), leading to slower propagation rates ($H \ \text{vs} \ t$). 

\begin{figure} [h!]
    \centering%
{\includegraphics[height = 8cm]{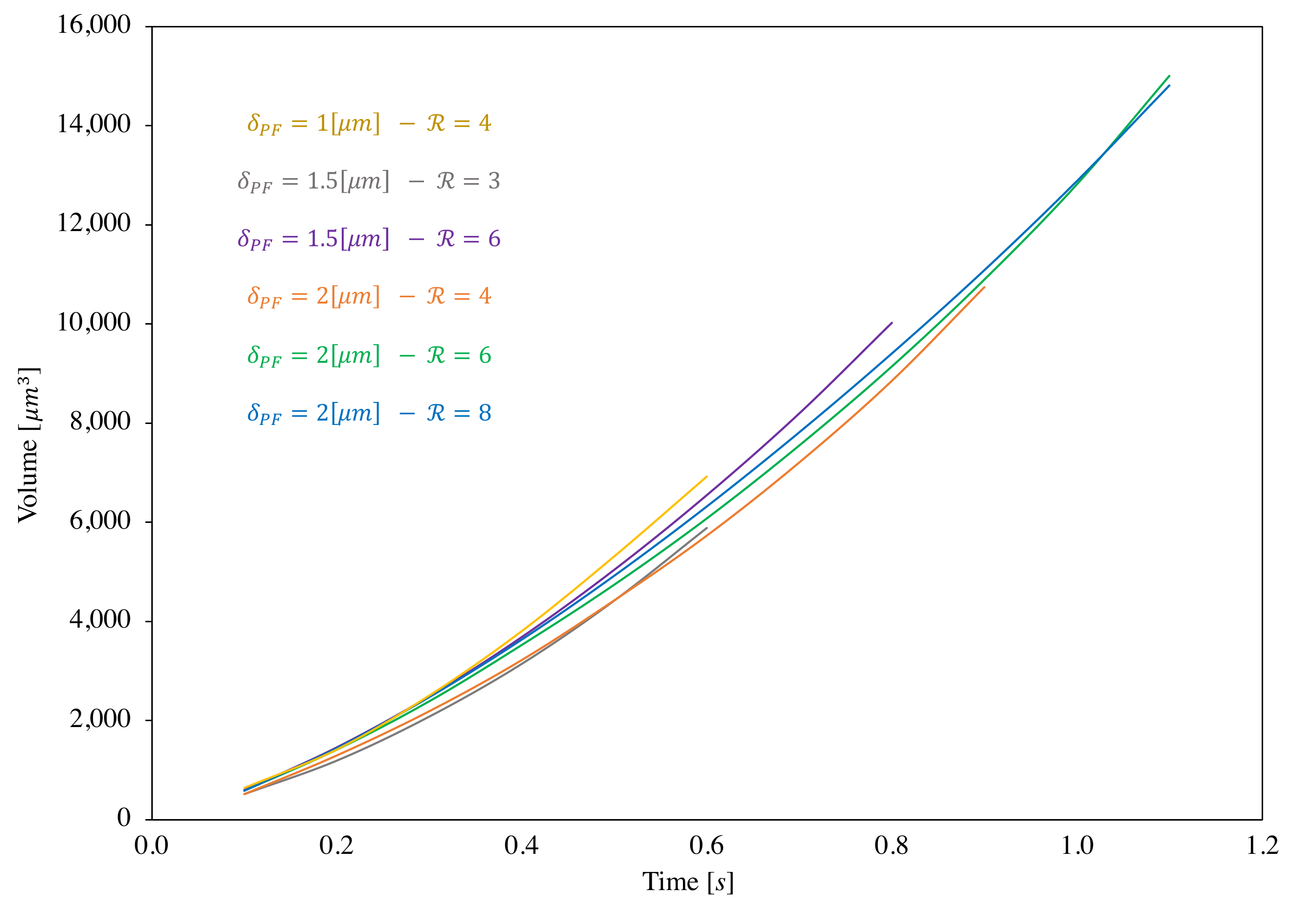}}
\caption{Comparison electrodeposited volume for 3D spike-like dendrite growth simulation, for different phase-field interface thickness to mesh resolution ratios ($\mathscr{R}=\delta_{PF}/\text{h}$), under $\phi_b=-0.7\left[V\right]$ charging potential. Tests 1 to 6.}
\label{fig:VolumeRate_SingleSeed3D}
\end{figure}

Figure~\ref{fig:VolumeRate_SingleSeed3D} shows the dendrite's volume evolution as a proxy of the overall electrodeposition rate (volume of lithium metal deposited over time). The effect of the phase-field interface thickness and mesh resolution ratio on the overall electrodeposition rate is less significant (percentage-wise) than it is for the dendrite's propagation rate (dendrite's height over time). For example, Figure~\ref{fig:VolumeRate_SingleSeed3D} shows a maximum electrodeposition rate difference of less than 20\% (volume vs time) between the fastest and slowest simulation results. Thus,  faster dendrite's propagation rates occur for smaller phase-field interface thicknesses due to the lithium metal being deposited / spread over a smaller surface area (more slender dendritic morphologies), rather than differences in the overall electrodeposition rate (minor effect). 

This analysis shows (see Figure~\ref{fig:VolumeRate_SingleSeed3D}) that for phase field interface thickness $2\left[\mu m\right]$ or smaller, the simulated electrodepostion rate (volume of lithium metal deposited over time) is relatively insensitive to the numerical parameters ($\delta_{PF}$ and $\mathscr{R}$). On the other hand, the simulated dendrite propagation rate shows stronger numerical dependencies (see Figure~\ref{fig:GrowthRate_SingleSeed3D}), affecting the level of realism of our results. Thus, propagation predictions presented here should only be taken as a comparison indicator between numerical tests, as we work towards smaller phase-field interface thickness to increase the accuracy of our simulations. 

\section{3D simulations using modified surface anisotropy}
\label{section:Aniso_Improved}

We evaluate the performance of the surface anisotropy representation model for metal anode battery simulations (see Section~\ref{subsection:Aniso_Improved_GovEq}). We perform numerical tests to gain insight into the benefits of this modification compared with the results previously obtained in Section~\ref{section:Senst_3D} and the preceding 3D simulation work~\cite{ ARGUELLO2022104892}. These studies consist of 3D phase-field simulations of lithium dendrite formation during battery charge state to explore three-dimensional highly branched "spike-like" dendritic patterns, commonly observed experimentally. These patterns form under high current density loads, which correspond to fast battery charge~\cite{ jana2019electrochemomechanics, ding2016situ, TATSUMA20011201}. 

\begin{figure}[h!]
\begin{subfigure}[b]{0.32\linewidth}
    \centering%
{\includegraphics[height = 4.5cm]{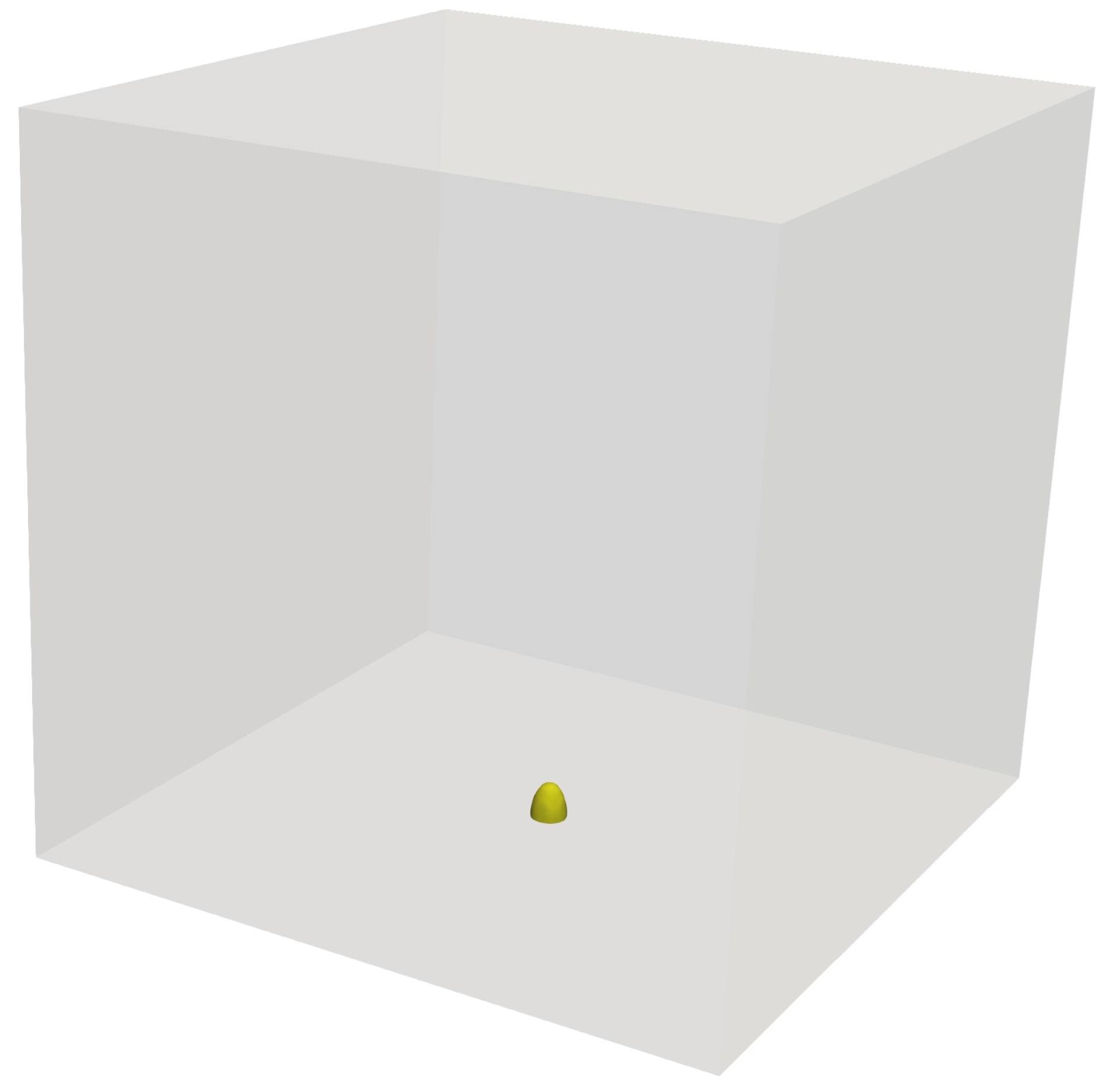}}
\caption{$t = 0.0\left[s\right]$.}
\end{subfigure}
\begin{subfigure}[b]{0.32\linewidth}
    \centering%
    {\includegraphics[height = 4.5cm]{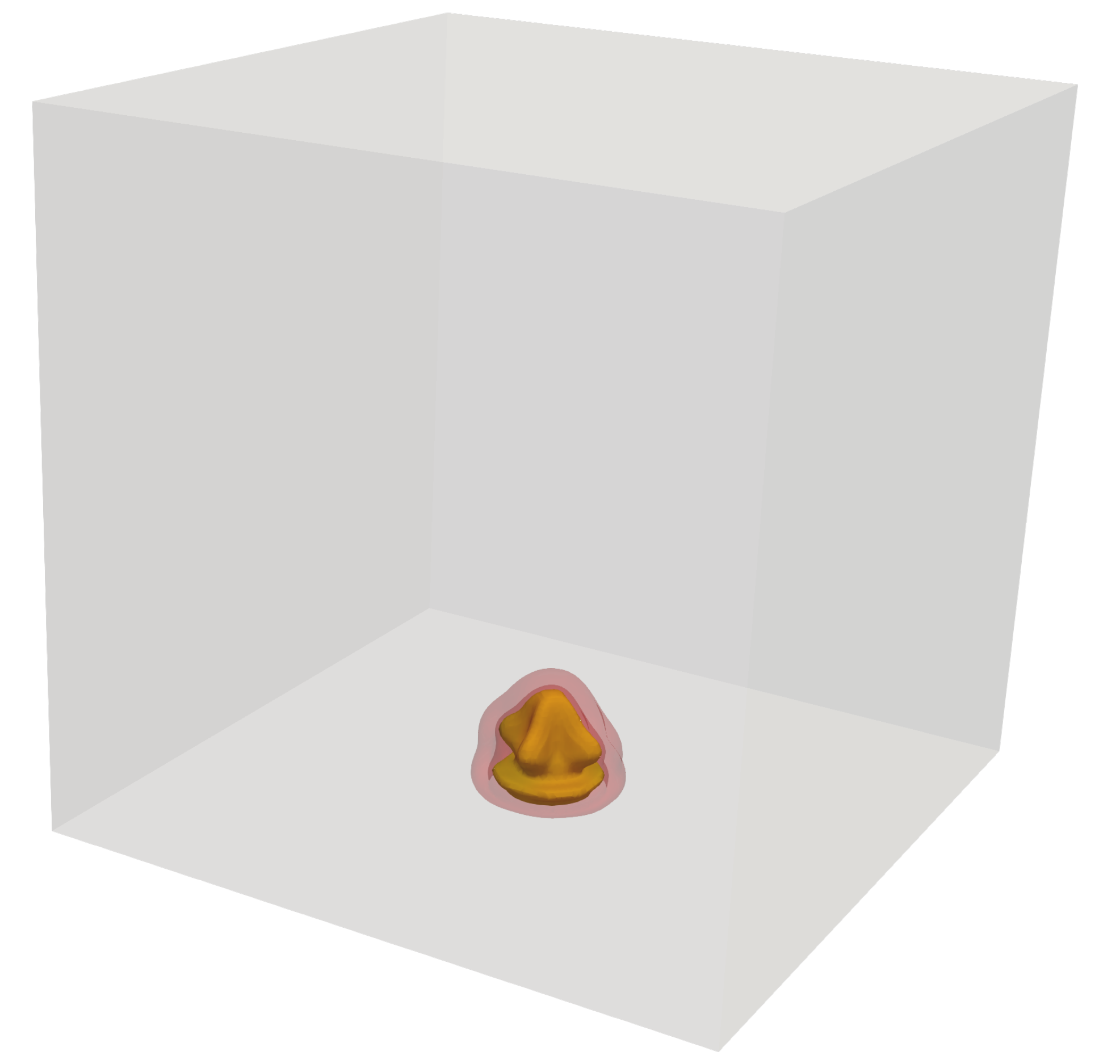}}
    \caption{$t = 0.1\left[s\right]$.}
\end{subfigure}
\begin{subfigure}[b]{0.32\linewidth}
    \centering%
    {\includegraphics[height = 4.5cm]{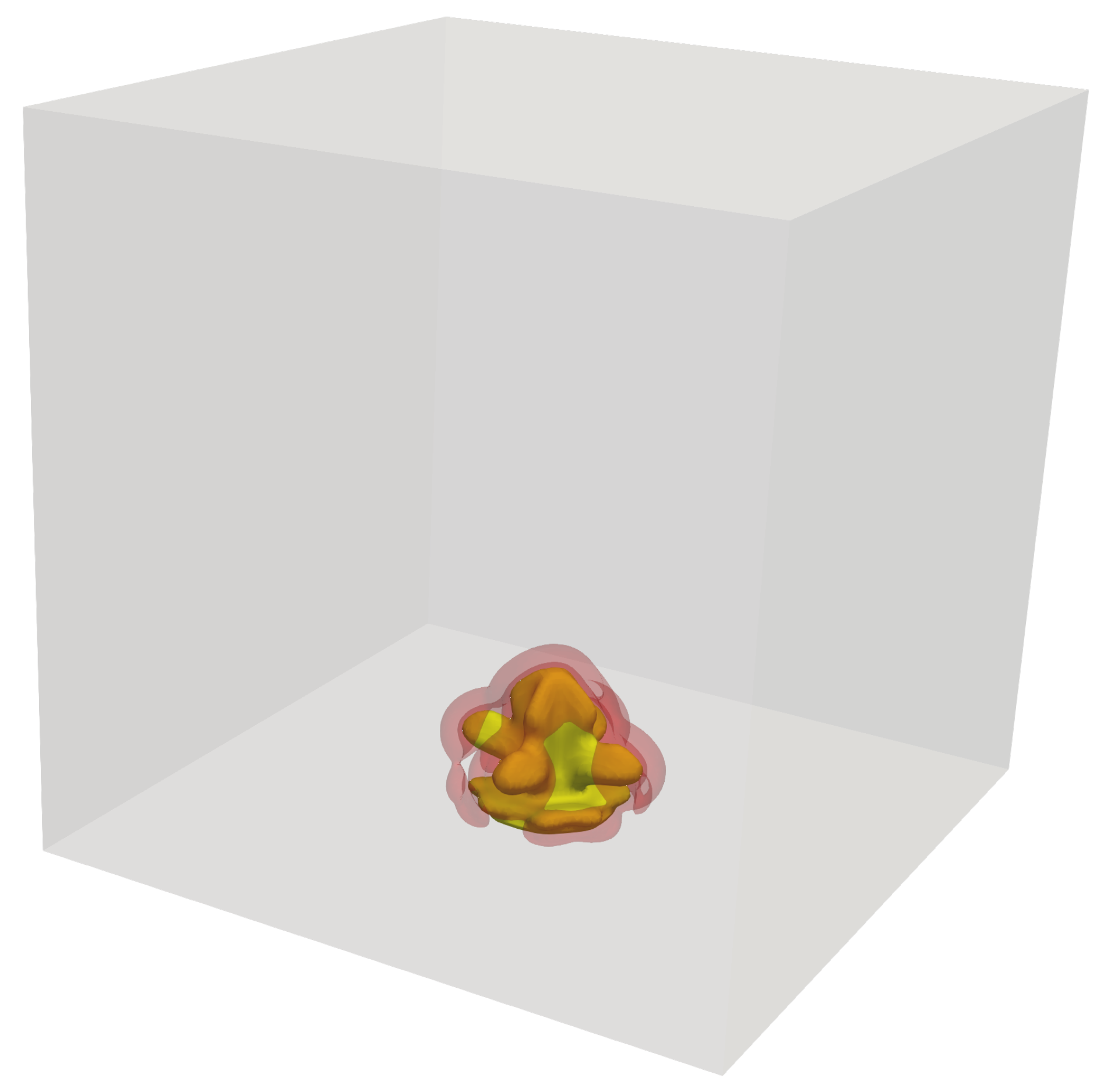}}
    \caption{$t = 0.2\left[s\right]$.}
\end{subfigure}
\begin{subfigure}[b]{0.32\linewidth}

\vspace{0.3cm}
    \centering%
    {\includegraphics[height = 4.5cm]{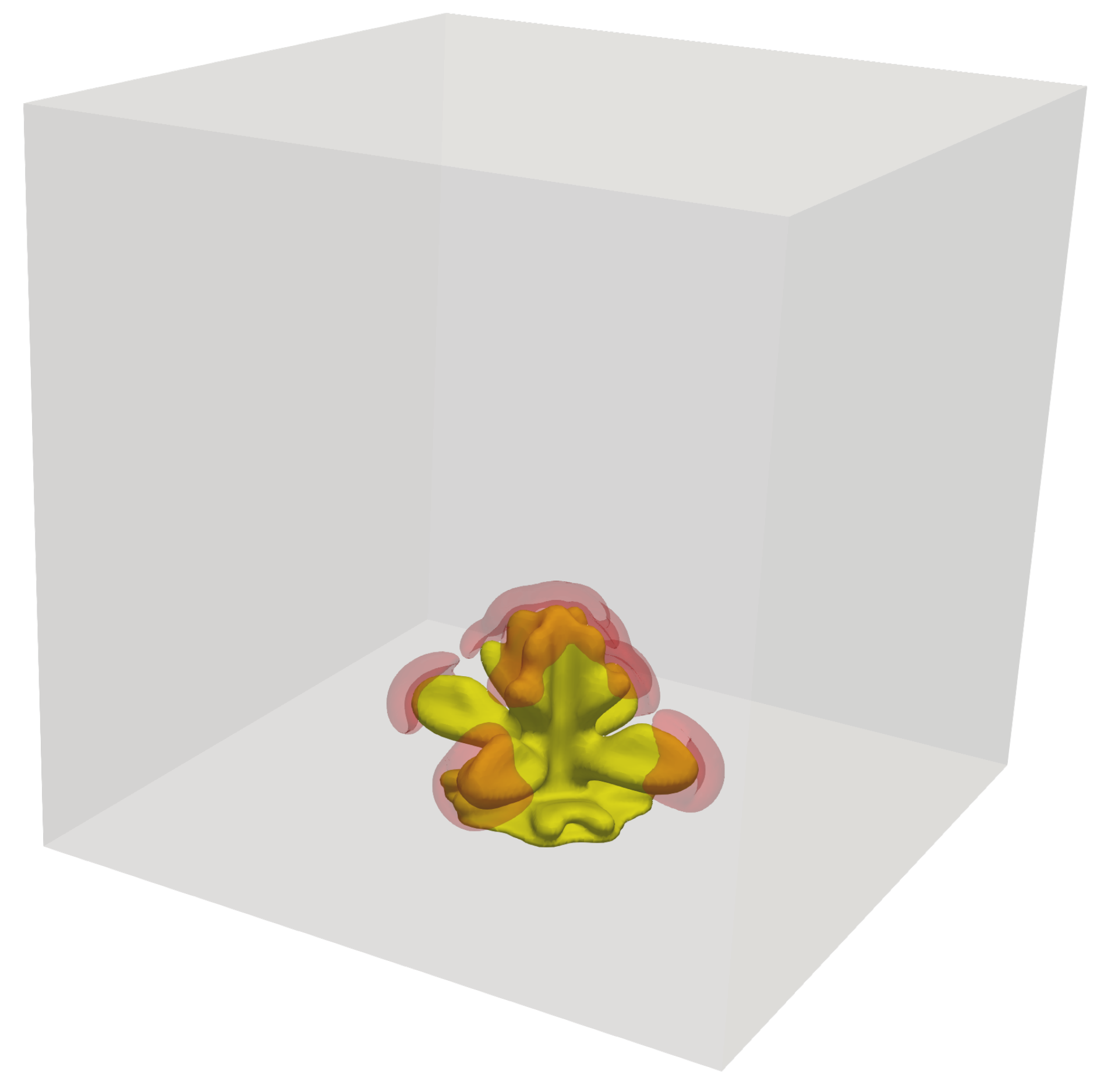}}
    \caption{$t = 0.4\left[s\right]$.}
\end{subfigure}
\begin{subfigure}[b]{0.32\linewidth}
    \centering%
    {\includegraphics[height = 4.5cm]{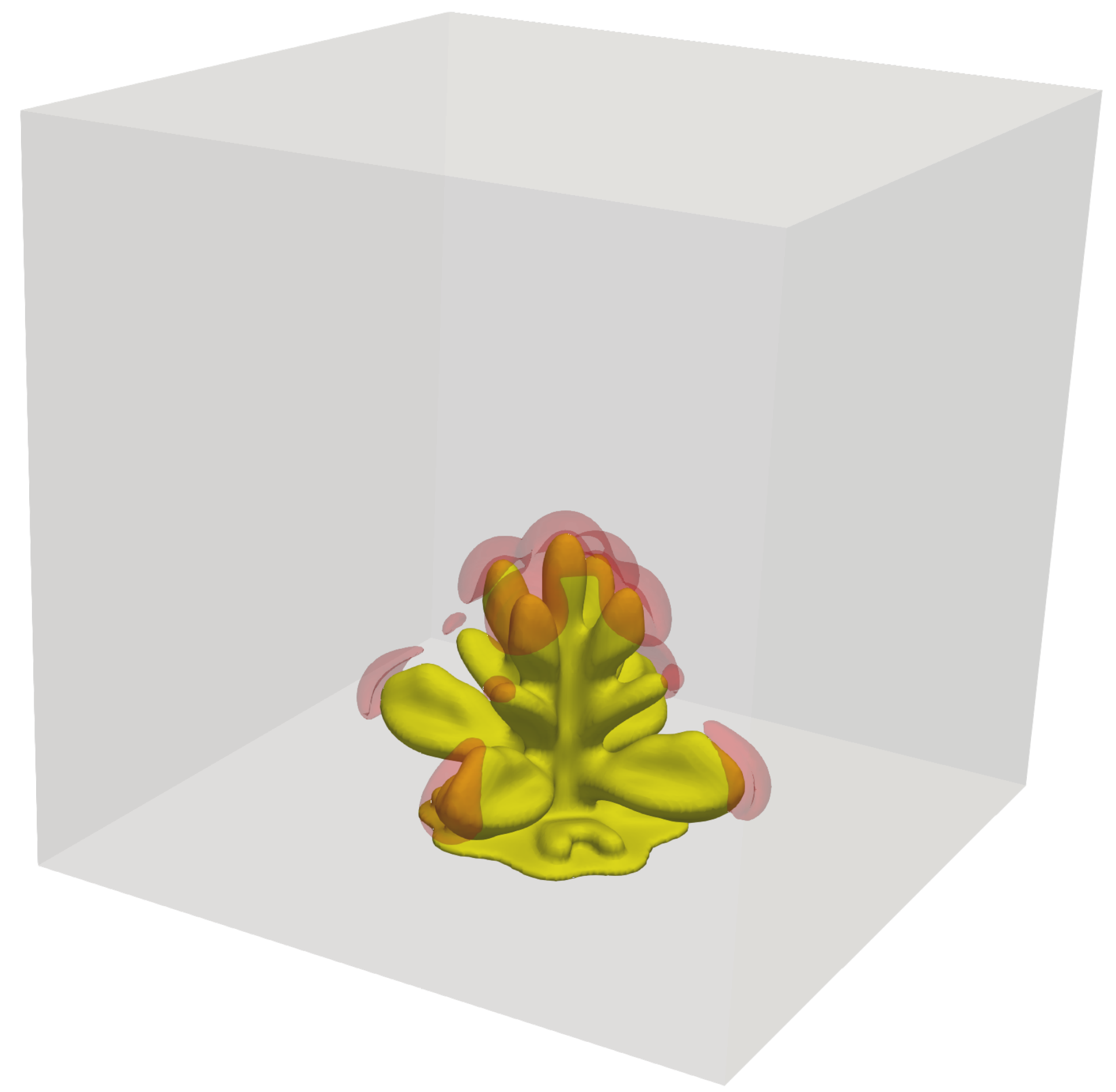}}
    \caption{$t = 0.6\left[s\right]$.}
\end{subfigure}
\begin{subfigure}[b]{0.32\linewidth}
    \centering%
    {\includegraphics[height = 4.5cm]{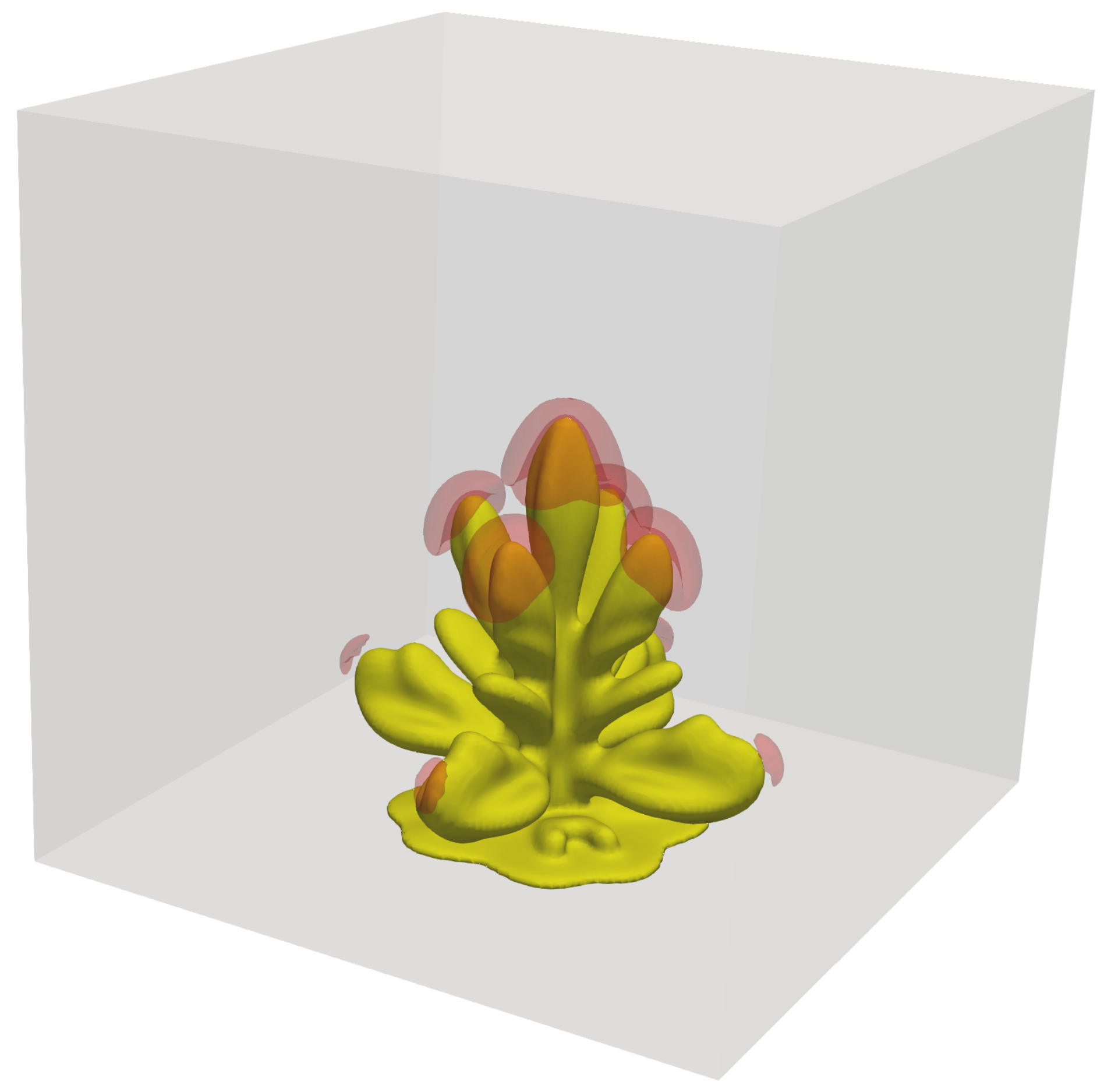}}
    \caption{$t = 0.8\left[s\right]$.}
\end{subfigure}
\caption{Spike-like lithium dendrite formation with modified surface anisotropy representation, under $\phi_b=-0.7\left[V\right]$ charging potential. The electrodeposited lithium is represented with a yellow isosurface plot of the phase-field variable $\xi=0.5$. Electrolyte regions with enriched concentration of lithium-ion ($\widetilde{\zeta}_{+}>1$) represented with orange volumes. Cube domain set as $80 \times 80 \times 80 \left[\mu m^3\right]$. Phase-field interface thickness $\delta_{PF}=1.5 \left[\mu m\right]$ \& mesh size $\text{h}=0.5 \left[\mu m\right]$. Test 7.}
\label{fig:SingleSeedAniso_evolut}
\end{figure}

\subsection{Comparison of simulated patterns: Surface anisotropy model}
\label{subsection:Aniso3D_Compare}

 We study the performance of the modified surface anisotropy representation~\eqref{eq:linealPF_Aniso} using a 3D numerical experiment (Test 7) and comparing the resulting morphologies with those obtained for single nucleus simulations using the standard anisotropy representation (see Figure~\ref{fig:Delta15_R3_SingleSeed3D}).  We use the simulation setup of Section~\ref{section:3D_Single_nuclei}.

Figure~\ref{fig:SingleSeedAniso_evolut} shows the morphological evolution of the simulated lithium dendrite and the enriched lithium-ion concentration ($\widetilde{\zeta}_{+}>1$) in the vicinity of the dendrite tips, with peak values of $\widetilde{\zeta}_{+}=2.3$. In agreement with previous numerical experiments, the simulation forms a spike-like, symmetric, and branched pattern. The dendrite morphology consists of the main trunk and sets of four equal orthogonal branches developing to the sides. The side branches grow up to 18$\left[\mu m\right]$ long (60\% longer than in previous simulations), and 5 to 10$\left[\mu m\right]$ width. Furthermore, the side branches growth is not perpendicular to the main truck but at an angle of about $25^\circ$ to $50^\circ$, with a separation of about 4 to 8$\left[\mu m\right]$ between branches. These results show improved morphological similarity with dendritic patterns observed in lithium experiments performed by Tatsuma et al.~\cite{ TATSUMA20011201} (within 10\% to 20\% difference compared to dendrite's measured features) . 

\begin{figure} [h!]
\centering%
\begin{subfigure}[b]{0.45\linewidth}
    \centering%
{\includegraphics[height = 5cm]{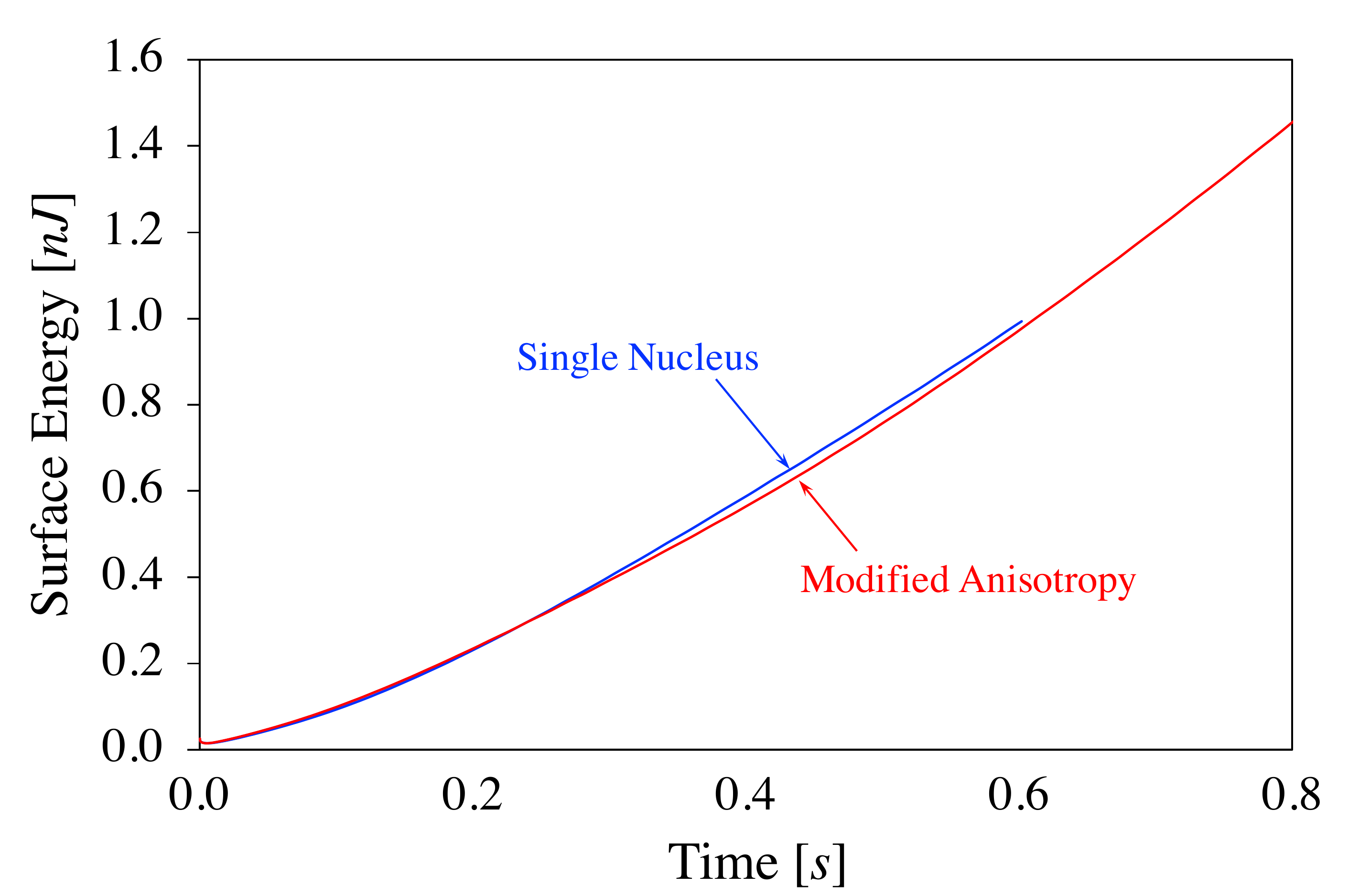}}
\caption{Surface Energy vs t.}
\label{fig:SurfEnergy_3DSeedAniso}
\end{subfigure}
\begin{subfigure}[b]{0.45\linewidth}
    \centering%
{\includegraphics[height = 5cm]{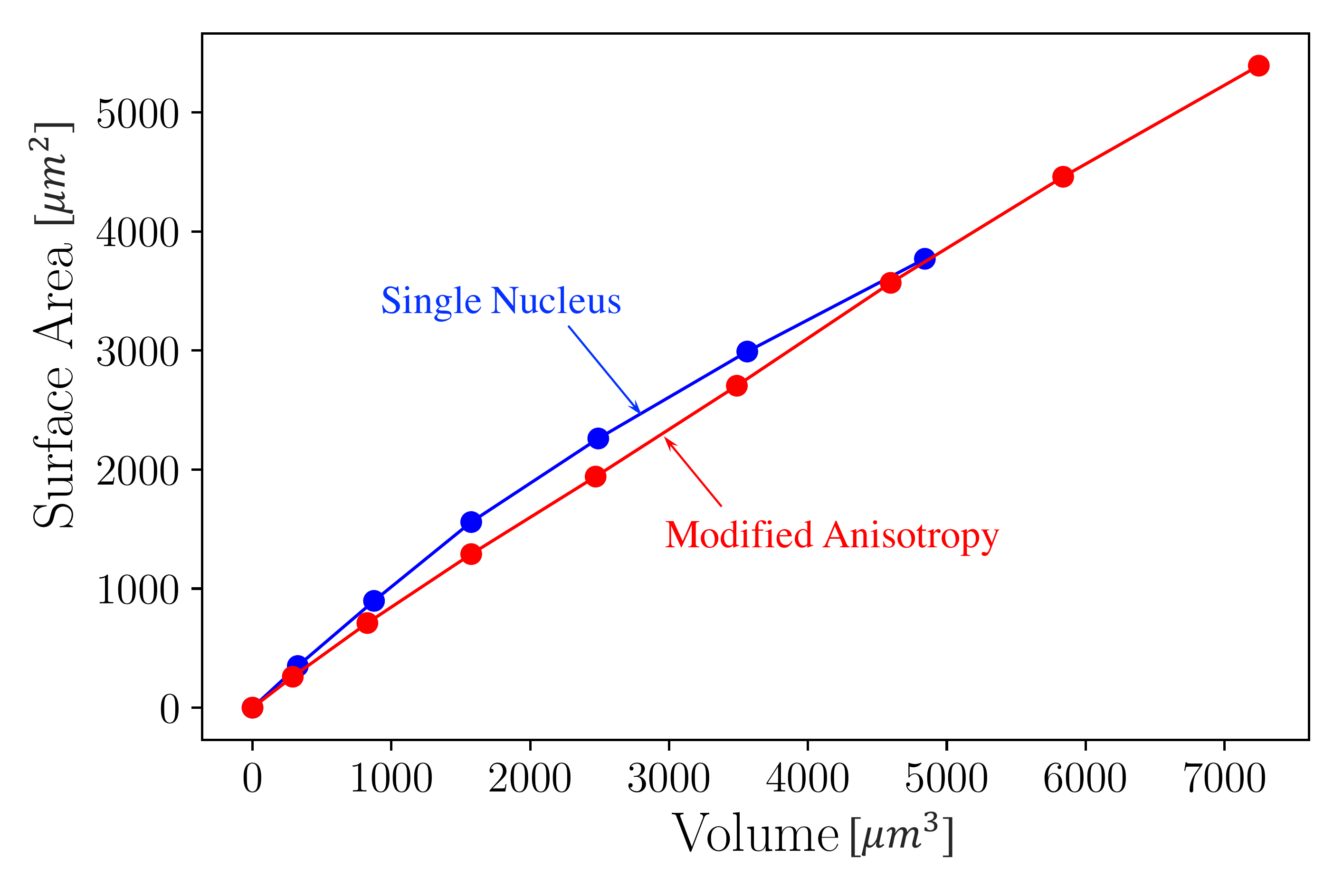}}
\caption{Volume vs Surface Area.}
\label{fig:VolvsSurf_3DSeedAniso}
\end{subfigure}
\caption{Comparison between 3D simulations of lithium dendrite growth (single nucleus initial vs modified surface anisotropy), in terms of the evolution of the surface energy \subref{fig:SurfEnergy_3DSeedAniso}, and volume vs surface area ratio over time \subref{fig:VolvsSurf_3DSeedAniso}. Tests 4 \& 7.}
\label{fig:3D_Aniso_CompareVsSingleSeed}
\end{figure}

Figure~\ref{fig:SurfEnergy_3DSeedAniso} plots the evolution of the surface energy for the 3D lithium patterns we simulate, revealing equivalent energy levels (less than 4\% difference) when compared against the results previously obtained using the initial, non-modified, anisotropy representation. Thus, the numerical experiment demonstrates that the modified anisotropy representation did not significantly affect the surface energy. Additionally, Figure~\ref{fig:VolvsSurf_3DSeedAniso} characterizes the morphology by tracking the dendrites' volume-specific area ($\mu m^2 / \mu m^3$). We compute the volume-specific area average ratios of 0.83 and 0.78 $\left[\mu m^2 / \mu m^3\right]$  for the single nucleus and modified anisotropy simulations, respectively. The slightly lower $\text{surface area/volume}$ ratio of the modified anisotropy representation (-6\%) indicates the dendrite growth has fewer branches (fewer but larger branches).

\subsection{Mesh orientation effect for different surface anisotropy representations}
\label{subsection:Aniso3D_MeshOrientation}

\begin{figure} [h!]
\centering%
\begin{subfigure}[b]{0.48\linewidth}
    \centering%
{\includegraphics[height = 5.5cm]{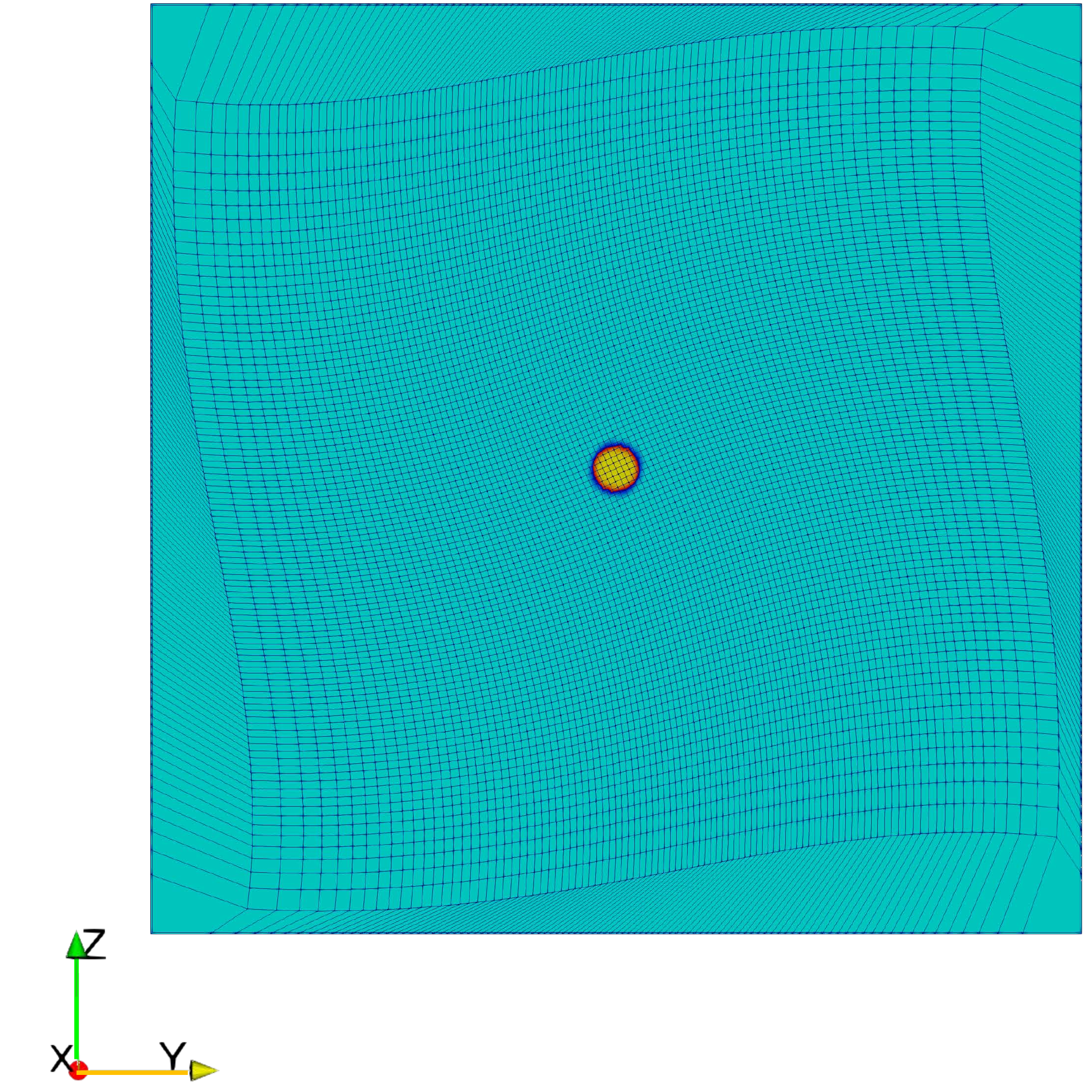}}
\caption{}
\label{fig:Mesh_Rot3D_Top}
\end{subfigure}
\begin{subfigure}[b]{0.48\linewidth}
    \centering%
{\includegraphics[height = 6.2cm]{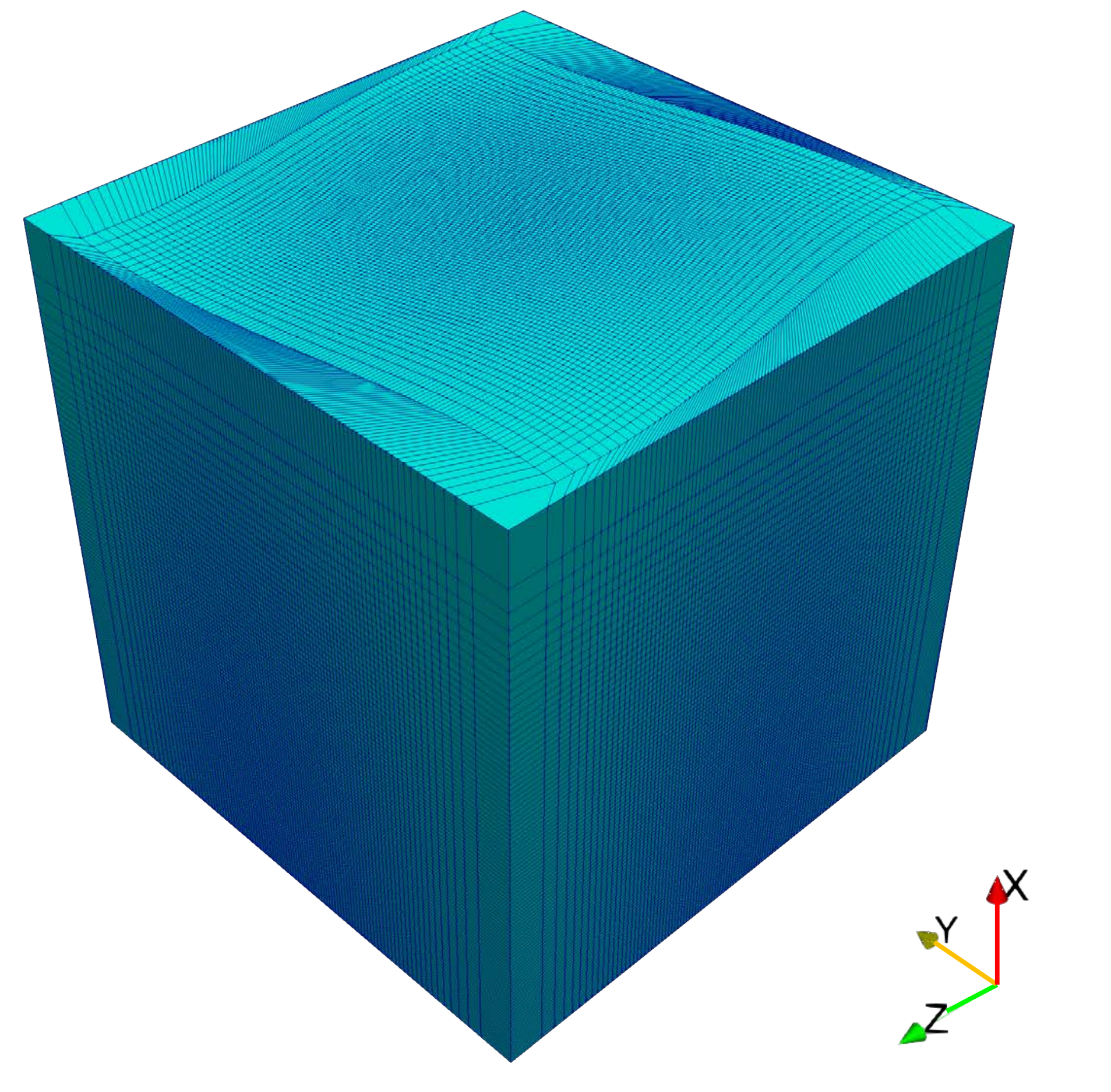}}
\caption{}
\label{fig:Mesh_Rot3D_View}
\end{subfigure}
\caption{Bottom \subref{fig:Mesh_Rot3D_Top} \& perspective \subref{fig:Mesh_Rot3D_View} views of the 3D mesh with 25$^\circ$ rotation around the $x$-axis (node's mapping). Cube domain set as $80 \times 80 \times 80 \left[\mu m^3\right]$.}
\label{fig:Mesh_Rot3D}
\end{figure}

\begin{figure} [h!]
    \centering%
{\includegraphics[height = 6.2cm]{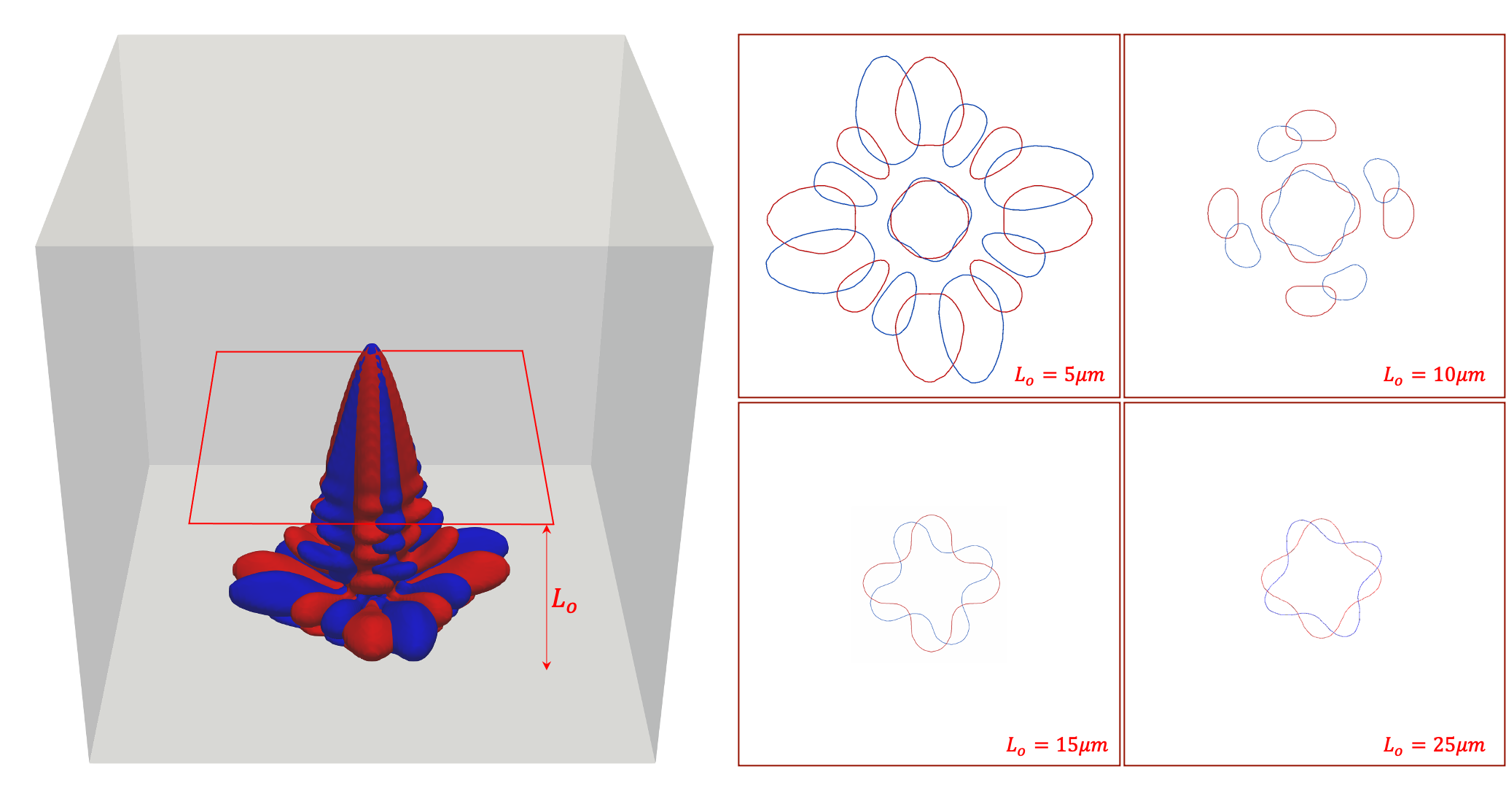}}
\caption{Overlay of 3D simulated dendrite morphologies (non-modified anisotropy representation), obtained under Cartesian mesh (red - Test 8), and 25$^\circ$ rotated mesh around the x-axis (blue - Test 9). Horizontal slices of the dendrite's contour plots at positions $L_O=5, \ 10, \ 15\ \& \ 25[\mu m]$ depict the angular offset between the morphologies.} 
\label{fig:SlicedRot_3D_NoAniso}
\end{figure}

We further compare the behavior of the standard (Test 9) and modified anisotropy representation behavior (Test 10) by studying the mesh orientation's effect on each simulated pattern. So far, the simulations results use structured meshes aligned with the Cartesian axes. Unlike previous 3D simulations, we now proceed to redistribute the mesh (node's mapping) by performing a 25$^\circ$ rotation around the $x$-axis, as depicted in Figure~\ref{fig:Mesh_Rot3D}. Thus, here we test the dendrite's sensitivity to the mesh orientation. Figure~\ref{fig:SlicedRot_3D_NoAniso} compares the dendrite morphologies using the standard anisotropy representation, using Cartesian (Test 8), as well as 25$^\circ$ rotated mesh distribution (Test 9) (see Figure~\ref{fig:Mesh_Rot3D}). We compare dendrite's cross sections (horizontal slices) at positions $L_O=5, \ 10, \ 15\ \& \ 25[\mu m]$. The analysis reveals an alignment of dendrites' side branches to the mesh orientation (angular offset), with no major differences in terms of the simulated dendrite's shapes.

\begin{figure} [h!]
\centering%
\begin{subfigure}[b]{0.9\linewidth}
    \centering%
{\includegraphics[height = 6.2cm]{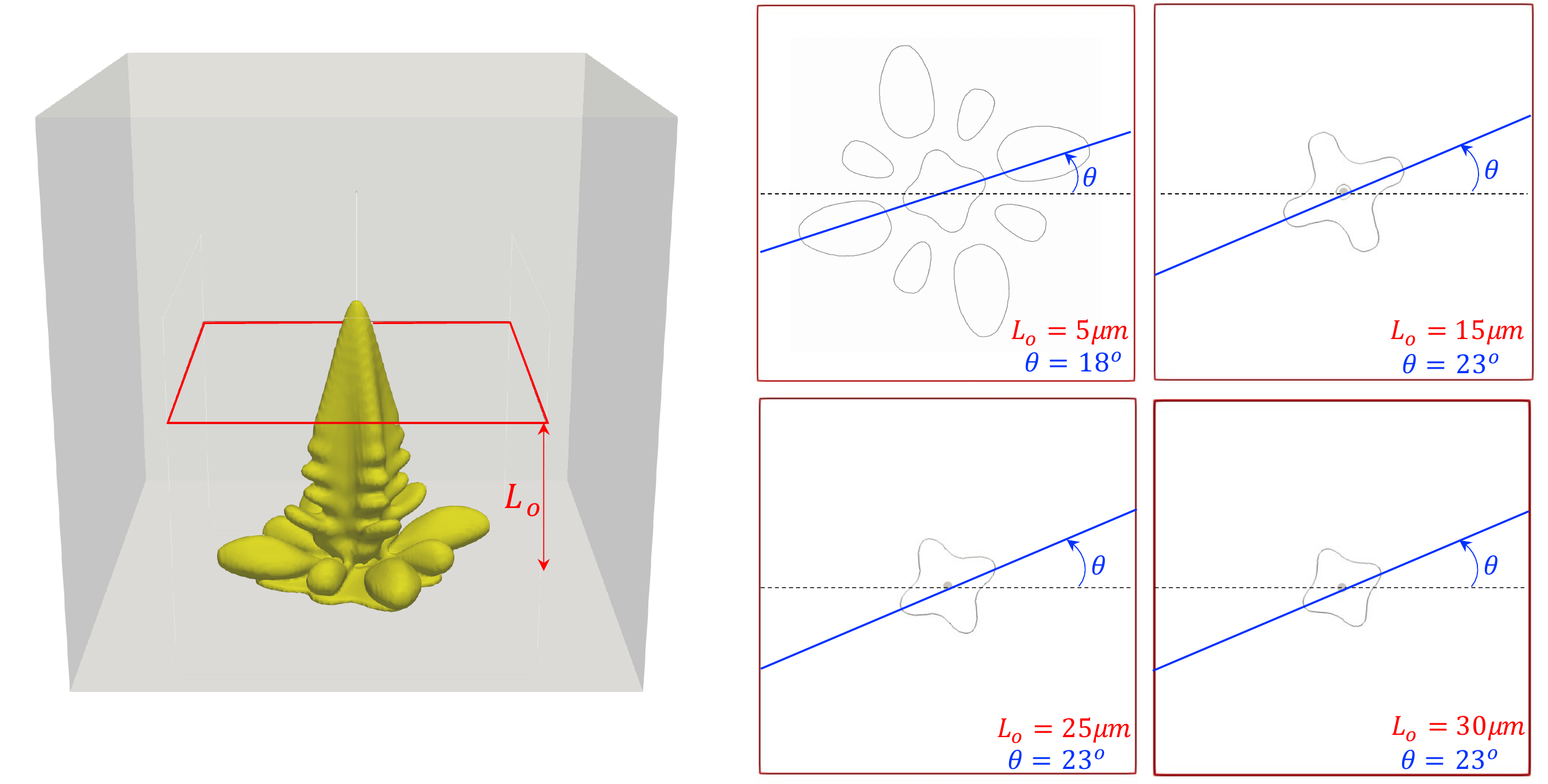}}
\caption{Test 9; $t=0.58\left[s\right]$.} 
\label{fig:SlicedRot_SingleSeed3D}
\end{subfigure} 

\vspace{0.5cm}
\centering%
\begin{subfigure}[b]{0.88\linewidth}
    \centering%
{\includegraphics[height = 6.2cm]{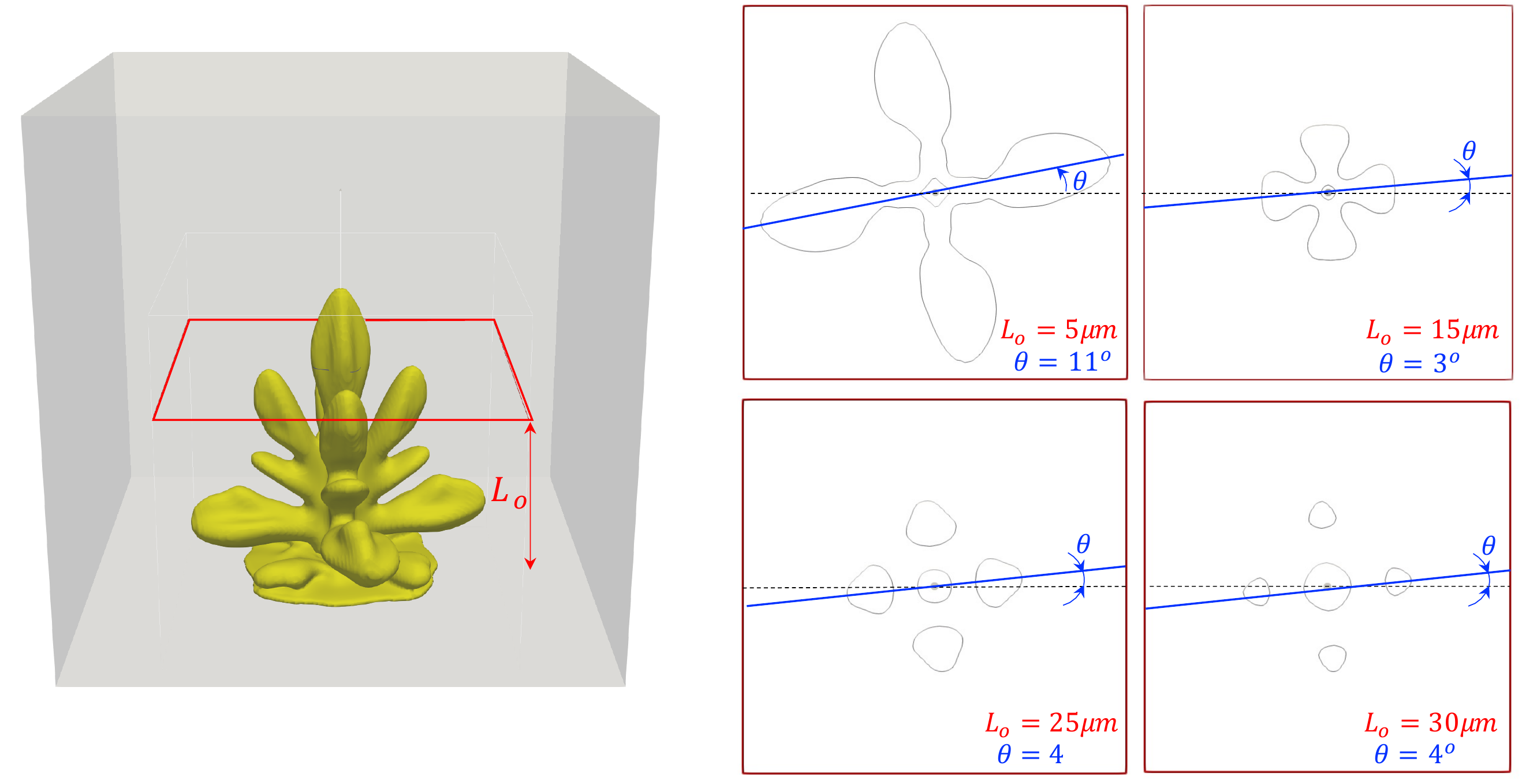}}
\caption{Test 10; $t=0.80\left[s\right]$.}
\label{fig:SlicedRot_Aniso3D}
\end{subfigure}
\caption{3D simulation results using a single artificial protrusion with the initial \subref{fig:SlicedRot_SingleSeed3D} (Test 9) and the modified surface anisotropy representation \subref{fig:SlicedRot_Aniso3D} (Test 10), under a 25$^\circ$ rotated mesh around the x-axis (longitudinal). Horizontal slices of the dendrite's contour plot at positions $L_O=5, \ 15, \ 25 \ \& \ 30 [\mu m]$ depict the orientation $\theta$ of the side branches. We use dendrite's common height ($H=45\left[\mu m\right]$) as the basis of our comparison.} 
\label{fig:SlicedRot_3D}
\end{figure}

Figure~\ref{fig:SlicedRot_3D} shows simulated dendrite morphologies under the rotated mesh distribution, using the standard (Test 9) and modified (Test 10) surface anisotropy representations. Now, we analyze the dendrite's side branches orientation $\theta$ at fixed positions $L_O=5, \ 15, \ 25 \ \& \ 30 [\mu m]$ (horizontal slices of dendrite's contour plot). We define the orientation $\theta$ as the inclination of the line that crosses the geometry by passing through its center and connecting the two farthest points of the contour (see Figure~\ref{fig:SlicedRot_3D}). We compare dendrite morphologies at the moment they reach a height of $H=45\left[\mu m\right]$. The analysis of the dendrite's side branches (horizontal slices) in Figure~\ref{fig:SlicedRot_3D}  reveals that the standard anisotropy representation is more sensitive to the orientation of the mesh. For example, orientation analysis in Figure~\ref{fig:SlicedRot_SingleSeed3D} depicts dendrite's rotation angles of around $\theta=23^\circ$, evidently aligned with the 25$^\circ$ of rotation imposed to the mesh. The side branches, due to the modified surface anisotropy representation~\eqref{eq:linealPF_Aniso} exhibit significantly smaller rotations of about $\theta=4^\circ$, using the same simulation conditions (see Figure~\ref{fig:SlicedRot_Aniso3D}). Thus, the modified anisotropy model shows reduced sensitivity with respect to the mesh. Additionally, mesh refinement reduces the simulations' sensitivity to the mesh orientation. 

\subsection{3D Orientation of lithium crystal: A surface anisotropy-based strategy}
\label{subsection:Aniso3D_CrystalOrientation}

Given the random nature of the nucleation process, we need to deal with some degree of randomness and uncertainty when determining the preferred growth direction of the dendrite's crystal in the battery. The orientation of the crystal, determined by the orientation of the surface anisotropy, will direct the preferred direction of growth of the lithium dendrite. Thus, we adapt this well-known crystal growth model for solidification in~\cite{TAKAKI201321} to electrodeposition dendrite growth. We define a material system of coordinates $\left(\widetilde{x},\widetilde{y},\widetilde{z}\right)$, in which each axis corresponds to the $\langle100\rangle$ direction of a cubic lattice. The following coordinate transformation $\mathbb{T}$ is used between the coordinate systems of $\left(x,y,z\right)$ and $\left(\widetilde{x},\widetilde{y},\widetilde{z}\right)$:
\begin{equation} \label{eq:3Daniso_orientation}
  \begin{aligned}
    \underbrace{
      \begin{Bmatrix}
        \frac{\delta\xi}{\delta\widetilde{x}} \\
        \frac{\delta\xi}{\delta\widetilde{y}} \\
        \frac{\delta\xi}{\delta\widetilde{z}}
      \end{Bmatrix}}_{\frac{\partial\xi}{\partial \widetilde{x_i}}}
    &=  
    \underbrace{
        \begin{aligned} 
          \begin{bmatrix}
            1 & 0 & 0 \\
            0 & \cos{\theta_x} & \sin{\theta_x} \\
            0 & -\sin{\theta_x} & \cos{\theta_x}
          \end{bmatrix}  
          \begin{bmatrix}
            \cos{\theta_y} & 0 & \sin{\theta_y} \\
            0 & 1 & 0 \\
            -\sin{\theta_y} & 0 & \cos{\theta_y}
          \end{bmatrix}  
          \begin{bmatrix}
            \cos{\theta_z} & \sin{\theta_z}& 0 \\
            -\sin{\theta_z} & \cos{\theta_z} & 0 \\
            0 & 0 & 1
          \end{bmatrix}   
        \end{aligned}
    }_{\mathbb{T}}
    \underbrace{
      \begin{Bmatrix}
       \frac{\delta\xi}{\delta x} \\
        \frac{\delta\xi}{\delta y} \\
        \frac{\delta\xi}{\delta z}
      \end{Bmatrix}}_{\frac{\partial\xi}{\partial x_i}}.
  \end{aligned}
\end{equation}
where $\theta_x$, $\theta_y$, and $\theta_z$ are the rotation angles around the x, y, and z axes, respectively. 

\begin{figure} [h!]
\centering%
\begin{subfigure}[b]{0.48\linewidth}
    \centering%
{\includegraphics[height = 6cm]{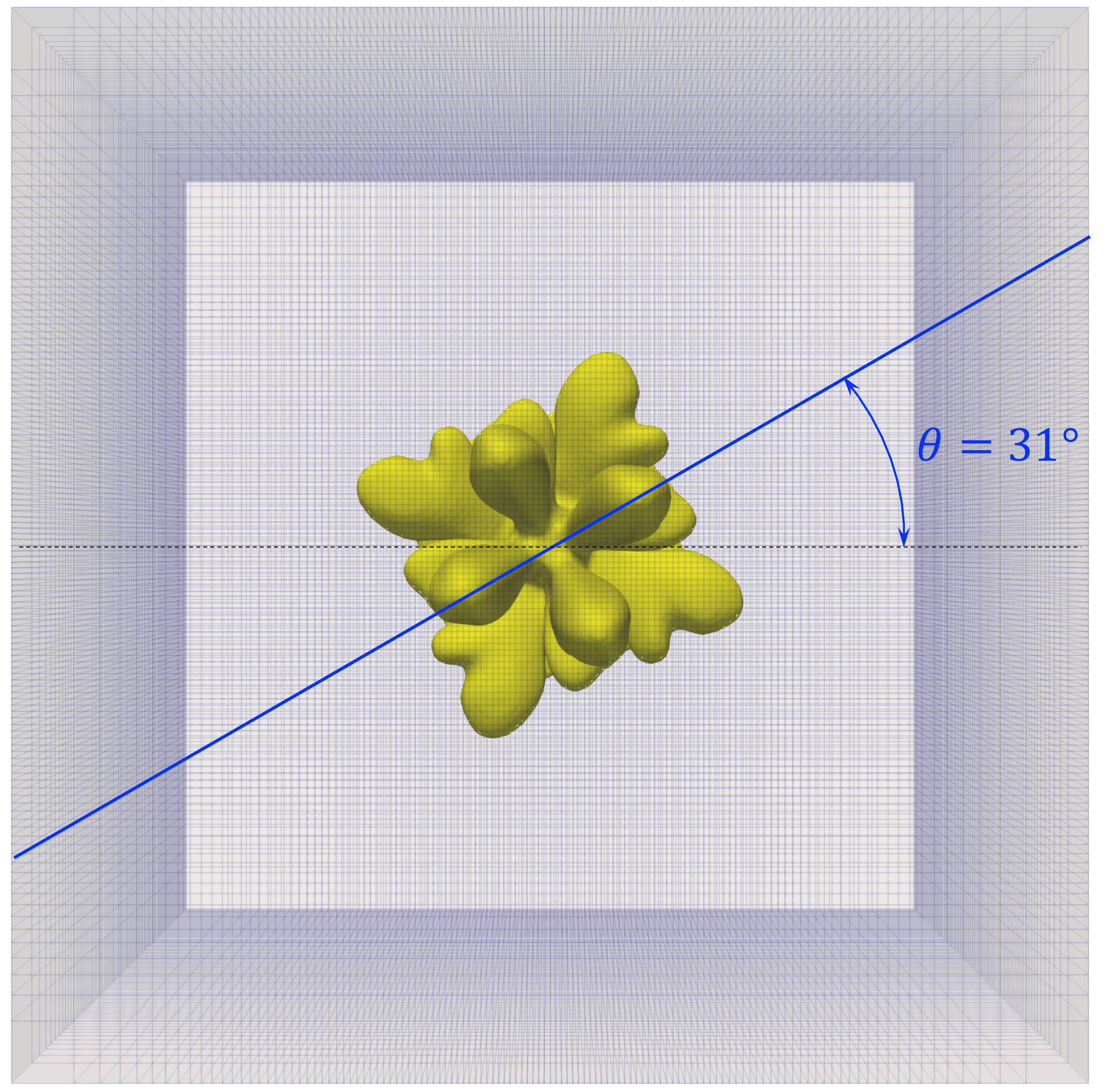}}
\caption{}
\label{fig:Aniso35_Rot3D_Top}
\end{subfigure}
\begin{subfigure}[b]{0.48\linewidth}
    \centering%
{\includegraphics[height = 6cm]{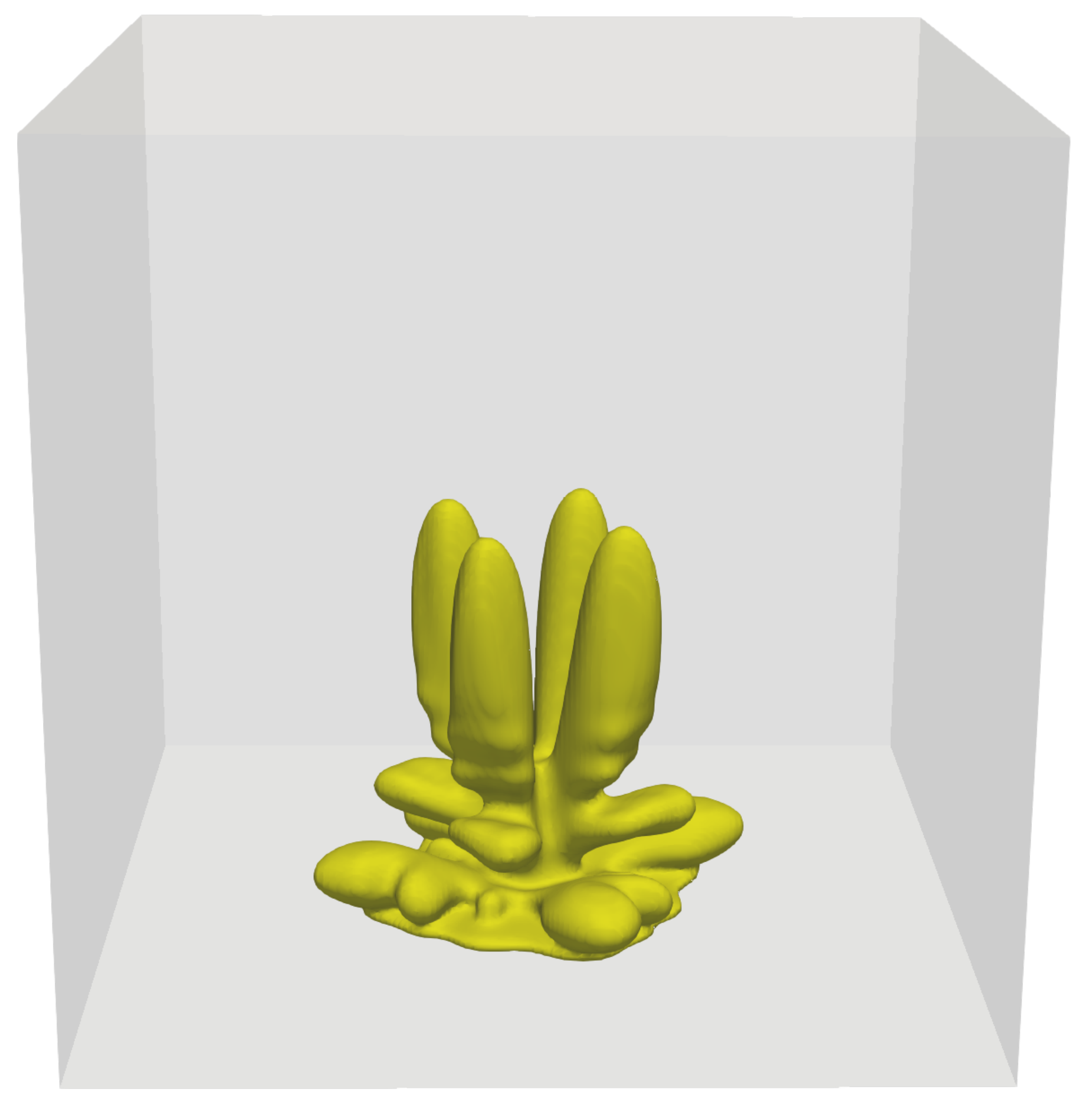}}
\caption{}
\label{fig:Ansio_Rot3D_View}
\end{subfigure}
\caption{Top \subref{fig:Aniso35_Rot3D_Top} \& perspective \subref{fig:Ansio_Rot3D_View} views of the 3D spike-like lithium dendrite simulation, with $\theta_x=35^\circ$ rotation of the surface anisotropy. Top view overlaid with mesh shows that dendrite's orientation is not aligned with the Cartesian axes. Test 11.}
\label{fig:Aniso35_Rot3D}
\end{figure}

We use~\eqref{eq:3Daniso_orientation} to compute the gradient of the phase-field variable ($\nabla\xi$) and use it in the surface anisotropy expression~\eqref{eq:SurfEnerg_VarDeriv3D}. Therefore, we can assign random values to each of the rotation angles $\left(\theta_x,\theta_y,\theta_z\right)$ to control the preferred growth direction of the lithium dendrite and side branches. We test the proposed strategy (Test 10) by applying a $\theta_x=35^\circ$ rotation ($\theta_y=\theta_z=0^\circ$) to the lithium surface anisotropy when using a Cartesian mesh. Figure~\ref{fig:Aniso35_Rot3D} shows the simulated spike-like lithium dendrite morphology after $t=0.7\left[s\right]$. The top-view analysis~\ref{fig:Aniso35_Rot3D_Top} reveals that this rotation resulted in a dendrite rotation of about $\theta=31^\circ$ under the applied anisotropy angle.

\subsection{Mesh size effect for different surface anisotropy representations}
\label{subsection:Aniso3D_MeshSenst}

Following Section~\ref{section:Senst_3D}, we study the spatial sensitivity of the modified surface representation under mesh refinement (Test 12). Given spike-like lithium dendrite symmetry (see Figure~\ref{fig:SingleSeedAniso_evolut}),  we use symmetry condition of Section~\ref{subsubsection:Symmetry_3D} to reduce the computation cost and improve the mesh resolution. Thus, we model only one-quarter of the domain, using a $200\times100\times100$ tensor-product mesh with a mesh spacing of $0.25\left[\mu m\right]$ in the region of interest (bottom half of the domain). The node mapping produces a finer mesh to properly capture the smaller phase-field interface thickness $\delta_{PF}=1\left[\mu m\right]$ (4 elements in the interface~\cite{ arguello2022phase}) and the steepest gradients of $\widetilde{\zeta}_{+}$ and $\phi$.

\begin{figure}[h!]
    \centering%
{\includegraphics[height = 10cm]{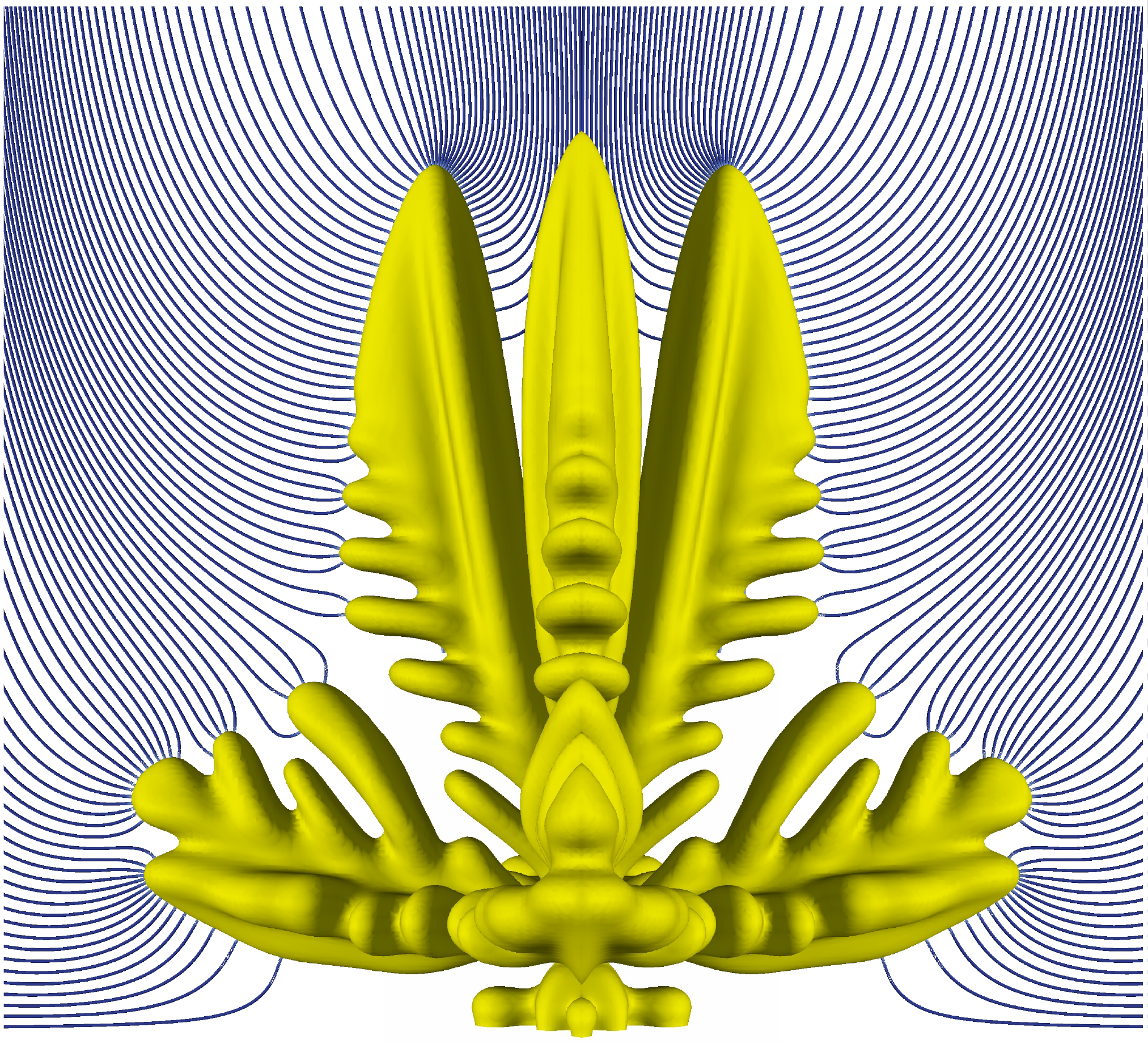}}
\caption{Overlay of electric field distribution (blue streamlines) with dendrite morphology under the modified surface anisotropy representation at time $t = 0.7\left[s\right]$. We combine in this figure 4 symmetric copies. Streamline plane set at $y=40\left[\mu m\right]$. Test 12.}
\label{fig:ElectField_Aniso3DPF1}
\end{figure}

Figure~\ref{fig:ElectField_Aniso3DPF1} shows the simulated lithium dendrite (isosurface plot of the phase-field variable $\xi=0.5$). The simulation forms a spike-like and highly branched pattern. We calculate the electric field distribution by differentiating the resolved electric potential $\vec{E}=-\nabla\phi$. Figure~\ref{fig:ElectField_Aniso3DPF1} shows how the electric field localizes in the vicinity of the dendrite tip~\cite{ ARGUELLO2022104892}. In agreement with previous numerical experiments~\cite{ ARGUELLO2022104892}, the electric field distribution leads to an enriched concentration of the lithium-ion it induces due to the strong migration from the surrounding regions~\cite{ doi:10.1063/1.4905341}. The dendrite growth does not occur perpendicular to the stack but at an angle of about $22^\circ$ between main trunks (cf. Figure~\ref{fig:Aniso35_Rot3D})). The stacks repel each other, similar to previous multiple-nuclei results~\cite{ ARGUELLO2022104892}. 

\begin{figure}[h!]
\begin{subfigure}[b]{0.45\linewidth}
    \centering%
{\includegraphics[height = 5cm]{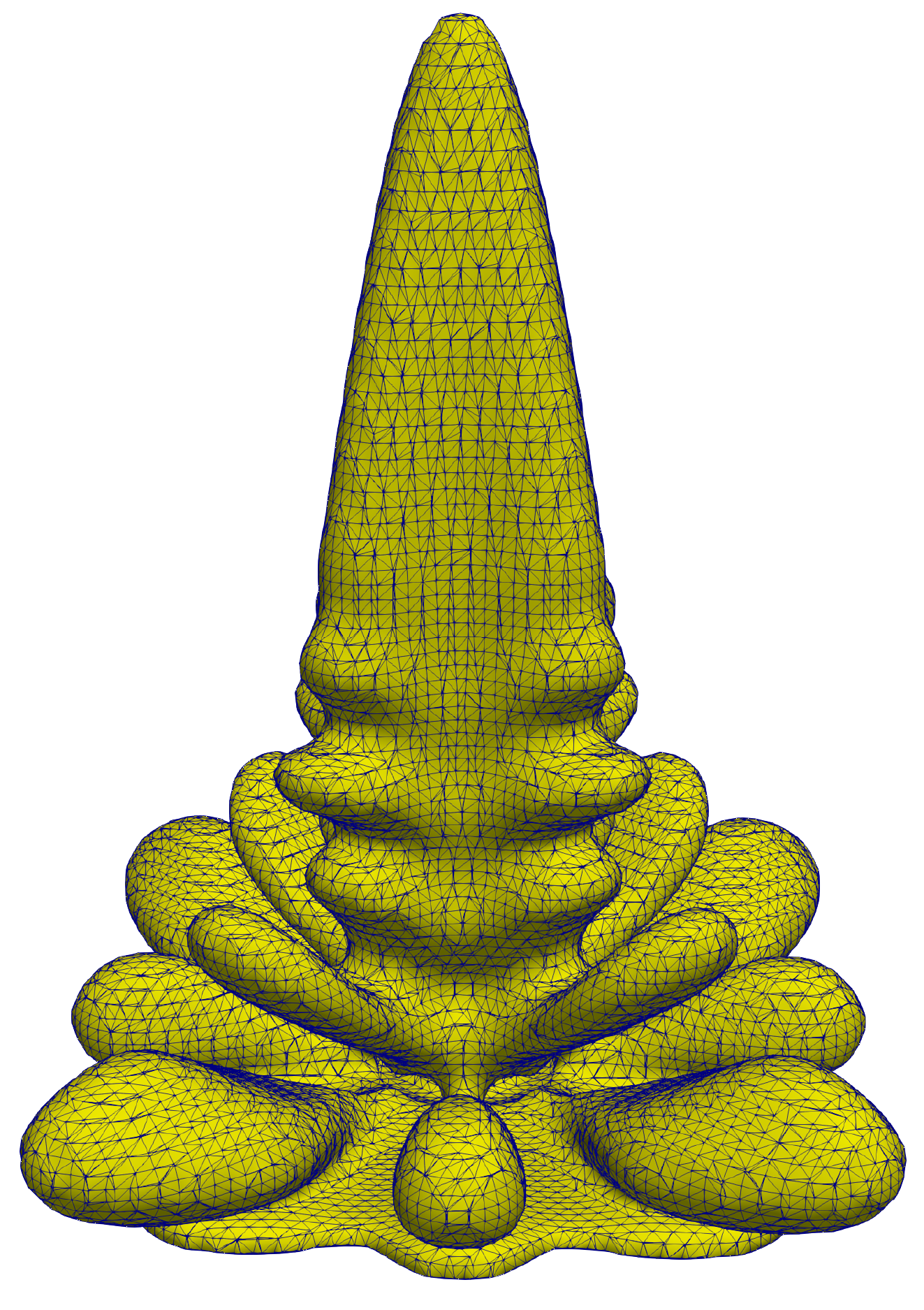}}
\caption{Test 4; $\ t=0.58\left[s\right]$}
\label{fig:NonImproved_Corse}
\end{subfigure}
\begin{subfigure}[b]{0.45\linewidth}
    \centering%
    {\includegraphics[height = 5cm]{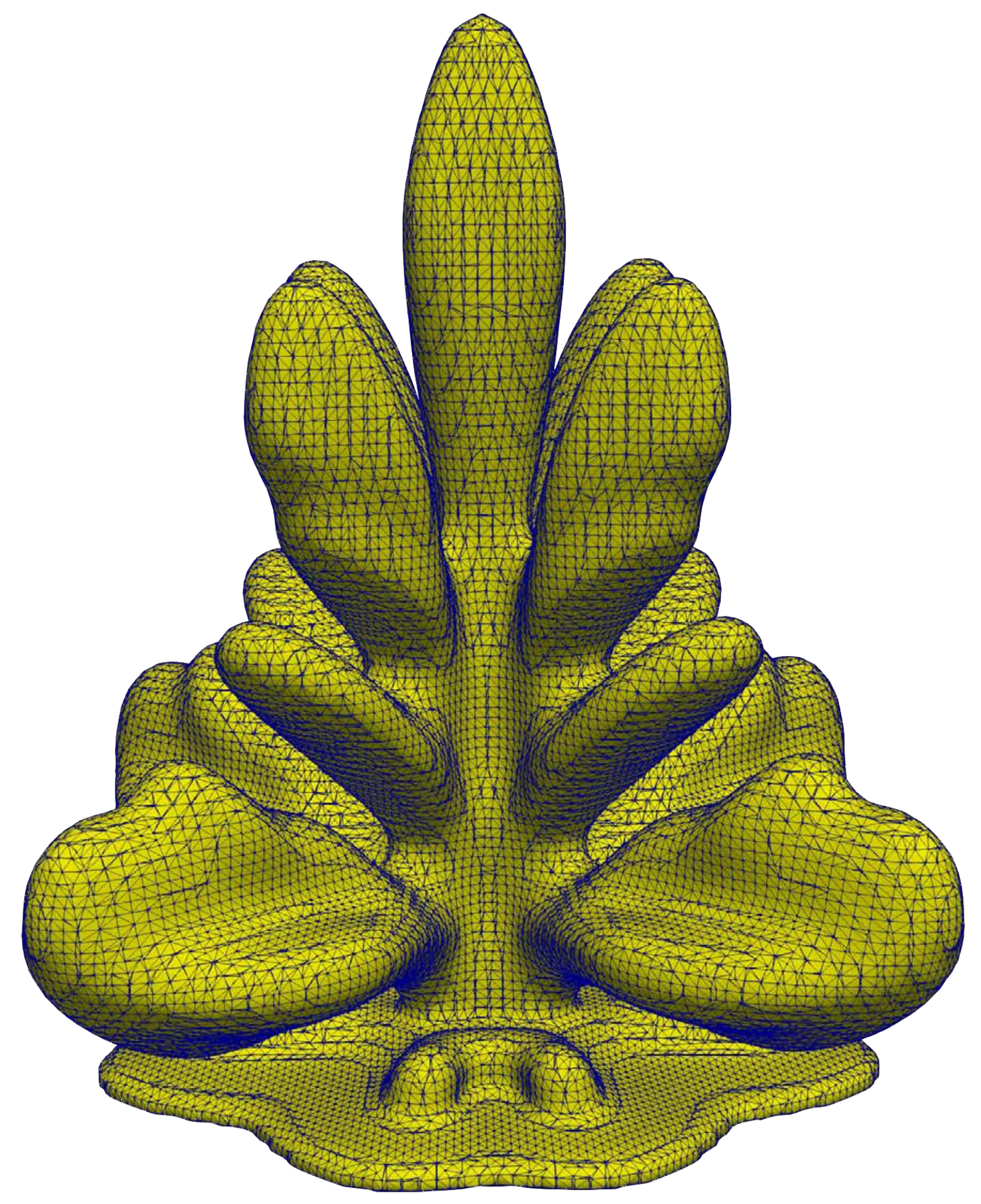}}
    \caption{Test 7; $\ t=0.80\left[s\right]$}
\label{fig:Improved_Corse}
\end{subfigure}

\vspace{0.3cm}
\begin{subfigure}[b]{0.45\linewidth}
    \centering%
    {\includegraphics[height = 5cm]{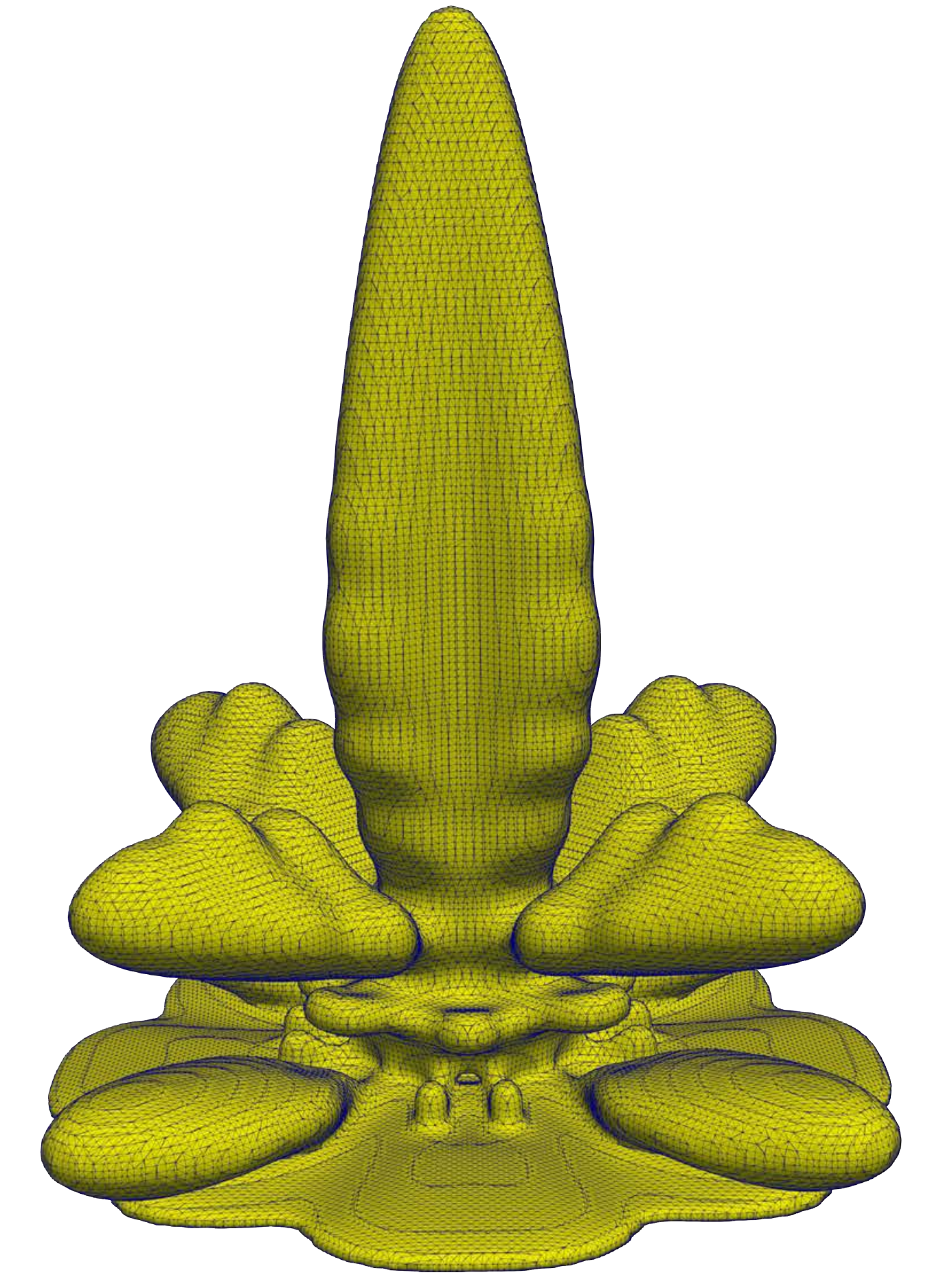}}
    \caption{Test 6; $\ t=0.52\left[s\right]$}
    \label{fig:NonImproved_Fine}
    \end{subfigure}
    \begin{subfigure}[b]{0.45\linewidth}
    \centering%
    {\includegraphics[height = 5cm]{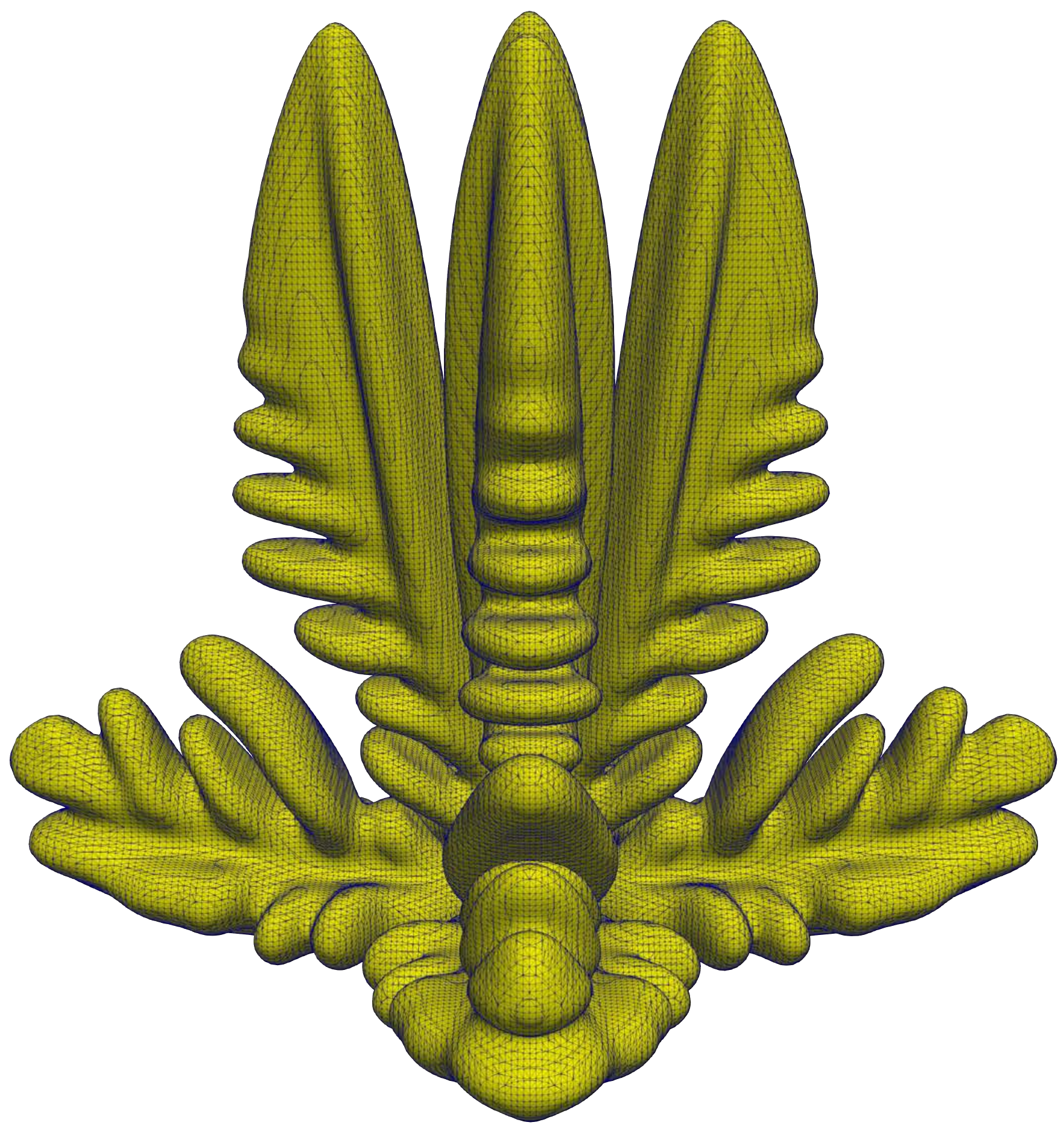}}
    \caption{Test 12; $\ t=0.71\left[s\right]$}
    \label{fig:Improved_Fine}
\end{subfigure}
\caption{Comparison of fully developed lithium dendrite morphologies under $\phi_b=-0.7\left[V\right]$ charging potential. The electrodeposited lithium is represented with a yellow isosurface plot of the phase-field variable $\xi$, overlaid with the mesh grid. Top row (\subref{fig:NonImproved_Corse} \& \subref{fig:Improved_Corse}) presents simulation results obtained under coarser mesh resolution (h=$0.5\left[\mu m\right]$ \& $\delta_{PF}=1.5\left[\mu m\right]$). Bottom row (\subref{fig:NonImproved_Fine} \& \subref{fig:Improved_Fine}) depicts results obtained under finer resolution (h=$0.25\left[\mu m\right]$ \& $\delta_{PF}=1\left[\mu m\right]$). Left column (\subref{fig:NonImproved_Corse} \& \subref{fig:NonImproved_Fine}) correspond to simulated morphologies using the non-modified surface anisotropy representation, and the right column (\subref{fig:Improved_Corse} \& \subref{fig:Improved_Fine}) allocates dendritic patterns under the modified anisotropy representation. We use dendrite's common height ($H=45\left[\mu m\right]$) as the basis of our comparison. Cube domain set as $80 \times 80 \times 80 \left[\mu m^3\right]$ in all cases.}
\label{fig:NonVsImproved_CorseVsFine}
\end{figure}

Figure~\ref{fig:NonVsImproved_CorseVsFine} compares the effect of the mesh resolution and phase-field interface thickness on the simulated morphologies, with and without the presence of the modified surface anisotropy term. For the modified surface anisotropy representation, smaller phase-field interface thickness ($\delta_{PF}$) and finer mesh resolution ($\text{h}$) lead to more branched and detailed dendritic patterns. However, in the standard case, finer mesh resolution leads to less branched micorstructures (see Figure~\ref{fig:NonImproved_Fine}).

Despite the morphological differences mentioned above, the computed lithium electrodeposition average rate in this case (10,800 $\left[\mu m^3/s\right]$) is within analogous simulation results under coarser mesh resolution (9,150 $\left[\mu m^3/s\right]$, see Figure~\ref{fig:VolvsSurf_3DSeedAniso}), as well as simulation result using the standard anisotropy representation, using coarse and fine mesh options (10,100 to 12,400$\left[\mu m^3/s\right]$, see Figure~\ref{fig:VolumeRate_SingleSeed3D}). These results show that using the modified surface anisotropy representation is robust in relation to the rate of electrodeposition (volume of lithium metal deposited over time), showing relatively low sensitivity to numerical parameters of our choice ($\delta_{PF}$ and $\mathscr{R}$).  In practice, the amount of dendritic lithium directly reduces the Coulombic efficiency of the battery~\cite{ Adams2018}. Therefore, we envisage a future application of our model in evaluating Coulombic efficiency reduction due dendite's formation in rechargeable lithium batteries.

\section{Experimental-scale 3D simulations of lithium dendrite formation}
\label{section:3D_LargeScale}

This section evaluates the performance of the modified surface anisotropy model (see Section~\ref{subsection:Aniso_Improved_GovEq}) in experimental-scale interelectrode distances. We map the nodal distribution concentrating the nodes in the region of interest, inspired by experimental and simulation results. The increased domain size affects the lithium electrodeposition behavior by increasing the interelectrode distance. We discuss the lithium dendrite propagation rates and morphologies for different charging voltages.

\subsection{Meshing strategy for experimental-scale 3D simulations}
\label{subsection:Large3D_Meshing}

\begin{figure}[h!]
    \centering%
{\includegraphics[height = 8.5cm]{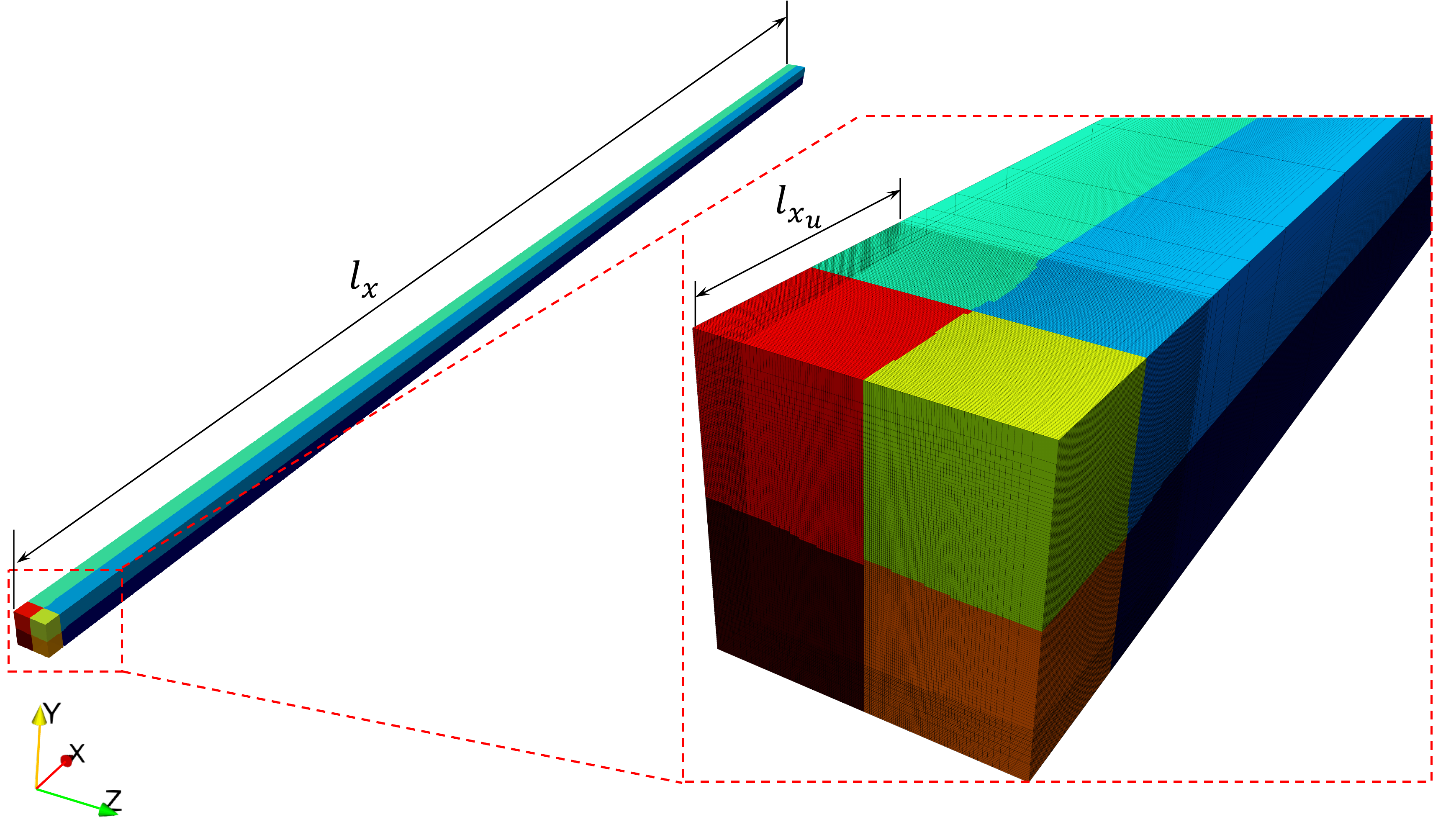}}
\caption{3D mesh partition in 8 processors, each one represented by a different color. Magnified view of the region of interest ($l_{x_{u}}\ll l_x$), showing a uniform to exponential mapping transition while moving into the bulk region of the domain.}
\label{fig:3DExp_Mesh}
\end{figure} 

The high computational cost of detailed 3D simulations of lithium electrodeposition at the whole-cell scale is a well-known challenge of electrodeposition simulations~\cite{ PhysRevE.92.011301, YURKIV2018609, ARGUELLO2022104892}. A limiting factor is the domain size, which imposes practical restrictions on the 3D simulations. Previously, we chose a domain size of  ($80 \times 80 \times 80 \left[\mu m^3\right]$) that ensures the simulation volume at an affordable computational cost. But this short domain induces (electrode separation of $l_x = 80 \left[\mu m\right]$) dendrite growth rates that are two orders of magnitude higher than those observed experimentally~\cite{ ARGUELLO2022104892}.

A detailed analysis of lithium dendrite experiments reveals that, despite the interelectrode separation distance in experimental cells, which ranges from 1 to 10$\left[mm\right]$~\cite{ NISHIKAWA201184, nishida2013optical, YUFIT2019485}, the lithium dendrites effectively occupy up to 20\% of the interelectrode space. Thus, we focus on this area of interest, the region/volume of the experimental cell where lithium dendrites develop, near the anode surface. Furthermore, previous simulation results show that the spatial distribution of the variables in the bulk region (outside the area of interest) exhibit either constant values, such as $\xi=0$ and $\widetilde{\zeta}_{+}=1$, or small electric potential gradients $\nabla\phi$. This weak variation indicates that only a few elements may adequately capture the bulk behavior. At the same time, we assign most computational resources to the area presenting the steepest gradients of $\xi$, $\widetilde{\zeta}_{+}$ and $\phi$, representing a small portion of the whole domain. 

This section applies the modified anisotropy representation in 3D simulations targeting experimental time and length scales. We describe a simple meshing strategy that exploits the aforementioned distribution by combining uniform node's mapping in the portion of the physical domain where the lithium electrodeposition process occurs (finer and regular mesh), with an exponential increment of the mesh size as we move away from the electrode into the electrolyte's bulk region. Thus, we use a 3D structured mesh with eight-node hexahedral elements. Within the bulk region, in particular in the $x$-direction $x_r=2^j \times x_u$ with $j=1,2,...,n$; where $x_u$ is the node's $x$ coordinate normalized by $l_x$, before mapping (uniform distribution), and $x_r$ is the node's mapped coordinate.  The exponential function transitions smoothly by doubling the element size when moving away from the area of interest into the bulk region. This focussed-mesh distribution in the area of interest and subsequent stretching allow us to achieve experimental interelectrode distances in only a few additional elements. Consequently, although the detailed portion of our domain ($l_{x_{u}}$) remains the same ($80 \times 80 \times 80 \left[\mu m^3\right]$), we are now able to avoid simulations with higher-than-normal dendrite's growth rates, by achieving experimental interelectrode distances ($l_x$ up-to $5000 \left[\mu m\right]$). 

Thus, we select a geometrical unit that characterizes a real cell structure~\cite{ YURKIV2018609, Trembacki_2019}. We choose a computational domain of $5000 \times 80 \times 80 \left[\mu m^3\right]$. Figure~\ref{fig:BC_Seed3D} summarizes the boundary conditions we apply. Lateral dimensions remain unchanged in this case ($l_y=l_z=80 \left[\mu m^3\right]$), which along with periodic boundary conditions applied on the lateral faces, generates a $80\left[\mu m\right] \times 80 \left[\mu m\right]$ nucleation arrangement surrounding the simulated morphology (neighbouring dendrites). The implemented approach constitutes a more realistic alternative than modeling a single isolated dendrite~\cite{ NISHIKAWA201184, nishida2013optical}. Furthermore, neighboring dendrites act as a barrier (charge repulsion effect) that limits the side development of the simulated electrodeposit beyond the domain's boundaries.

We use a $180\times 100\times 100$ tensor-product mesh, with a mesh size of $0.4\left[\mu m\right]$ in the region of interest ($l_{x_{u}}$). We partition the mesh into eight processors. Figure~\ref{fig:3DExp_Mesh} identifies each core with a different color, showing that the tensor-product mesh can efficiently allocate resources in the region of interest ($l_{x_{u}}$).  

\subsection{Experimental-scale 3D simulations.}
\label{subsection:Large3D}

This section presents 3D phase-field simulations of lithium dendrite formation to study dendritic patterns formed under $\phi_b=-0.7\left[V\right]$ (Test 13) and $\phi_b=-1.4\left[V\right]$ charging potential (Test 14), using experimental-scale interelectrode distance ($l_x = 5000 \left[\mu m\right]$). We use artificial nucleation regions, ellipsoidal protrusions (seeds) with semi-axes $4\left[\mu m\right]\times2\left[\mu m\right]\times2\left[\mu m\right]$, and centres located at $\left(y,x,z\right)=\left(0,38,38\right)$, $\left(0,42,38\right)$, $\left(0,38,42\right)$ and $\left(0,42,42\right)$~\cite{ ARGUELLO2022104892}. We modify the initial condition, by introducing a constant electric potential gradient in the liquid electrolyte region, from $\phi=\phi_b$ at the electrode-electrolyte interface, to $\phi=0$ at $x=l_x$ (cathode), which corresponds to the experimental observations by Nishida et al.~\cite{nishida2013optical}. They measured the initiation periods (time transient) for dendrite precursors to start to grow (become visible under an optical microscope) between 4 to 140 s~\cite{nishida2013optical}; shorter initiation times occur under larger applied current density. Therefore, sufficiently developed dendrite nuclei may take several seconds to appear, depending on the electrodeposition conditions. This time is sufficient for developing the electric potential gradient in the electrolyte. In addition, the initial conditions for $\xi$ and $\widetilde{\zeta}_{+}$ remain the same, with a smooth transition between the solid seed (lithium metal anode) and the surrounding liquid electrolyte region~\cite{ ARGUELLO2022104892}. 

\begin{figure}[h!]
\begin{subfigure}[b]{0.32\linewidth}
    \centering%
{\includegraphics[height = 4.5cm]{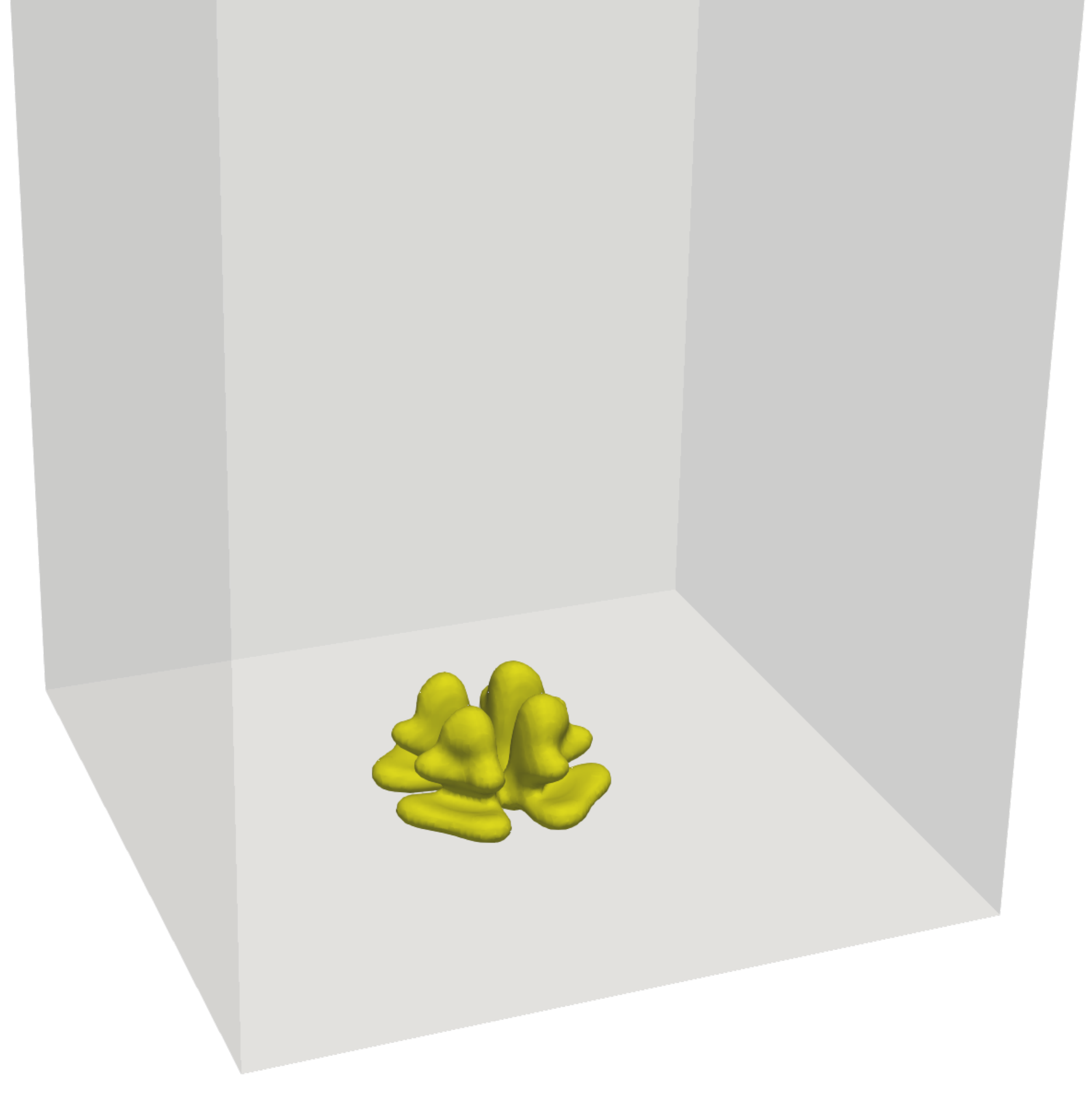}}
\caption{$t = 5\left[s\right]$.}
\end{subfigure}
\begin{subfigure}[b]{0.32\linewidth}
    \centering%
    {\includegraphics[height = 4.5cm]{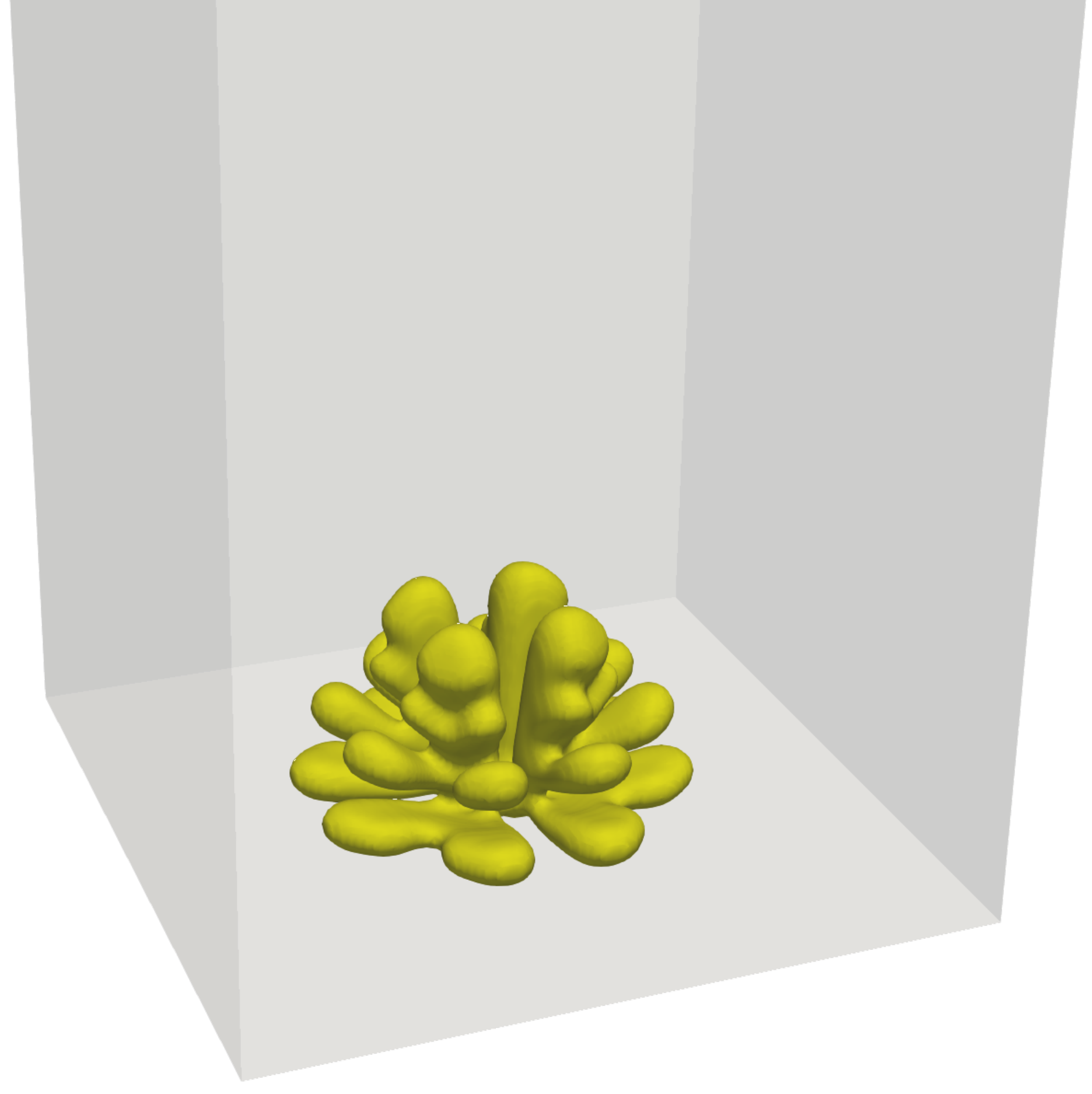}}
    \caption{$t = 30\left[s\right]$.}
\end{subfigure}
\begin{subfigure}[b]{0.32\linewidth}
    \centering%
    {\includegraphics[height = 4.5cm]{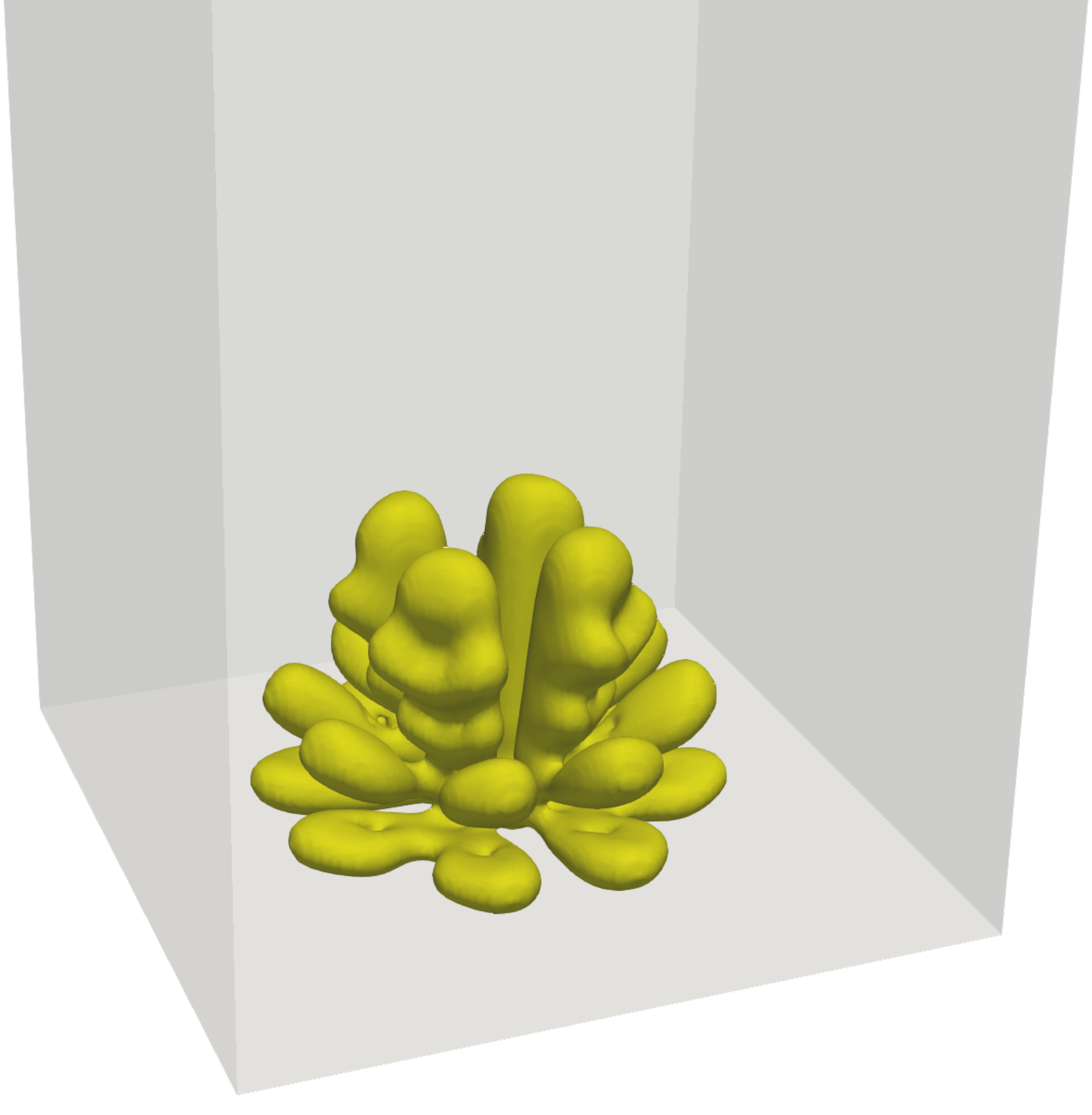}}
    \caption{$t = 50\left[s\right]$.}
\end{subfigure}
\begin{subfigure}[b]{0.32\linewidth}

\vspace{0.3cm}
    \centering%
    {\includegraphics[height = 4.5cm]{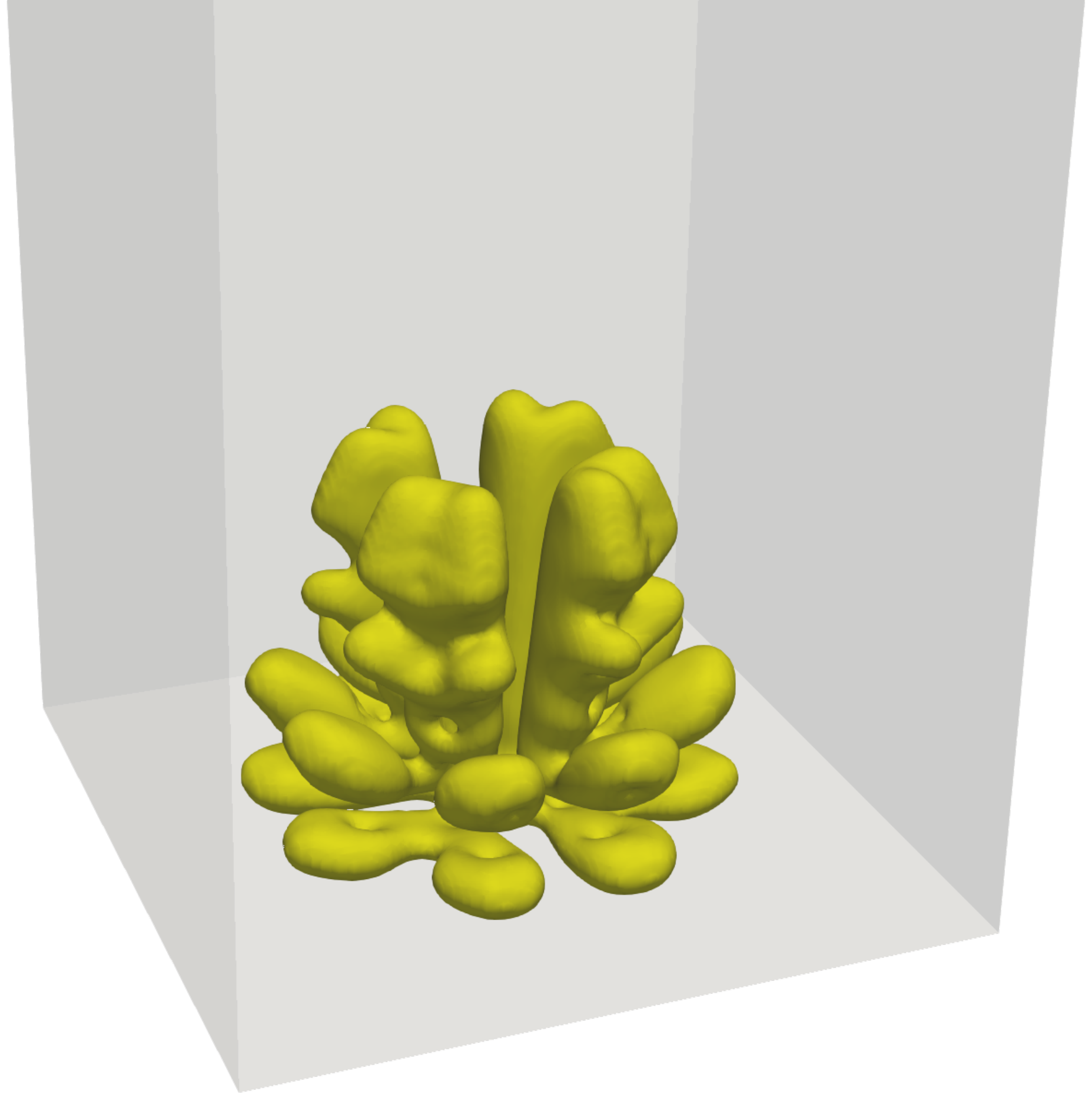}}
    \caption{$t = 90\left[s\right]$.}
\end{subfigure}
\begin{subfigure}[b]{0.32\linewidth}
    \centering%
    {\includegraphics[height = 4.5cm]{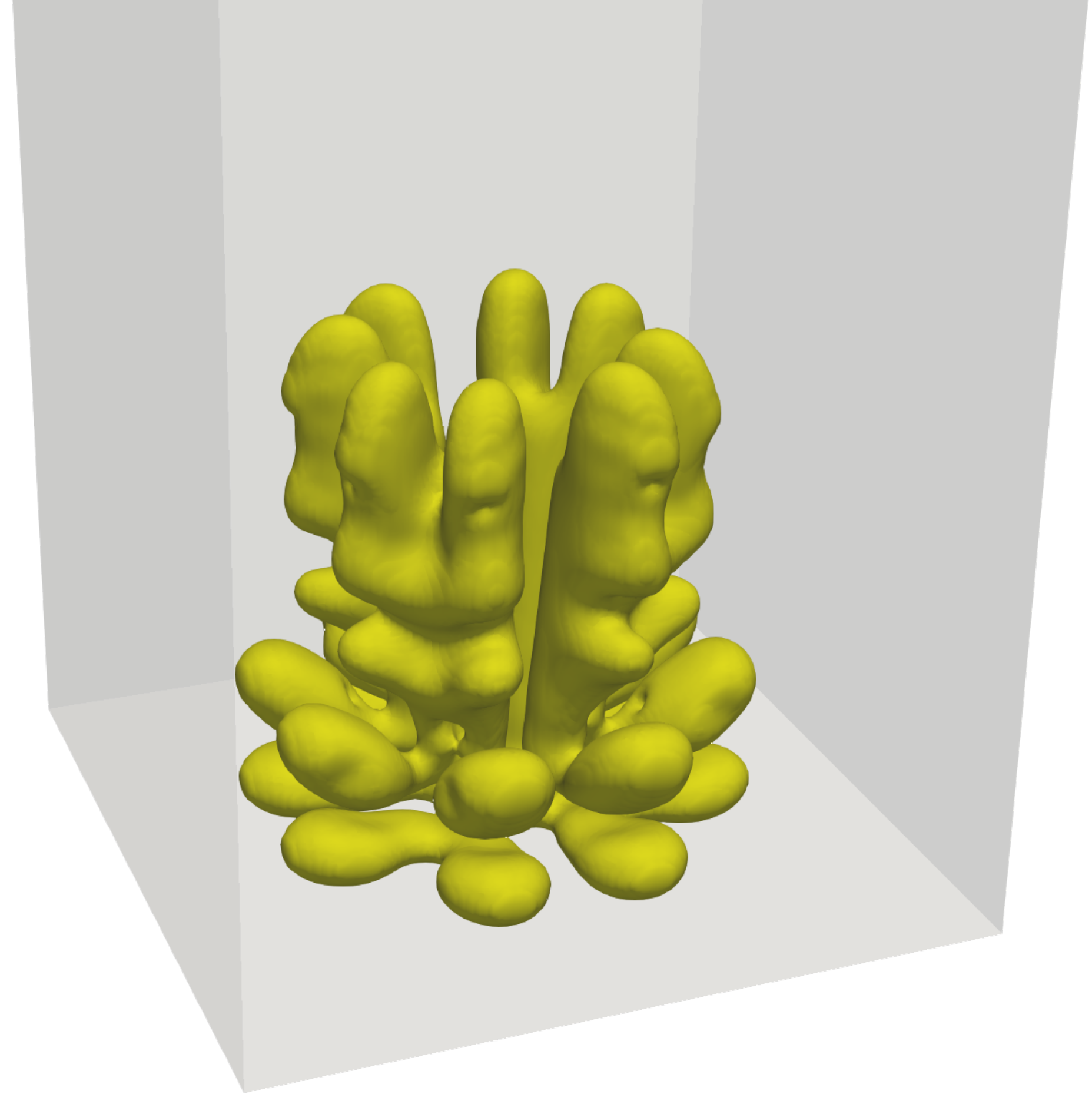}}
    \caption{$t = 150\left[s\right]$.}
\end{subfigure}
\begin{subfigure}[b]{0.32\linewidth}
    \centering%
    {\includegraphics[height = 4.5cm]{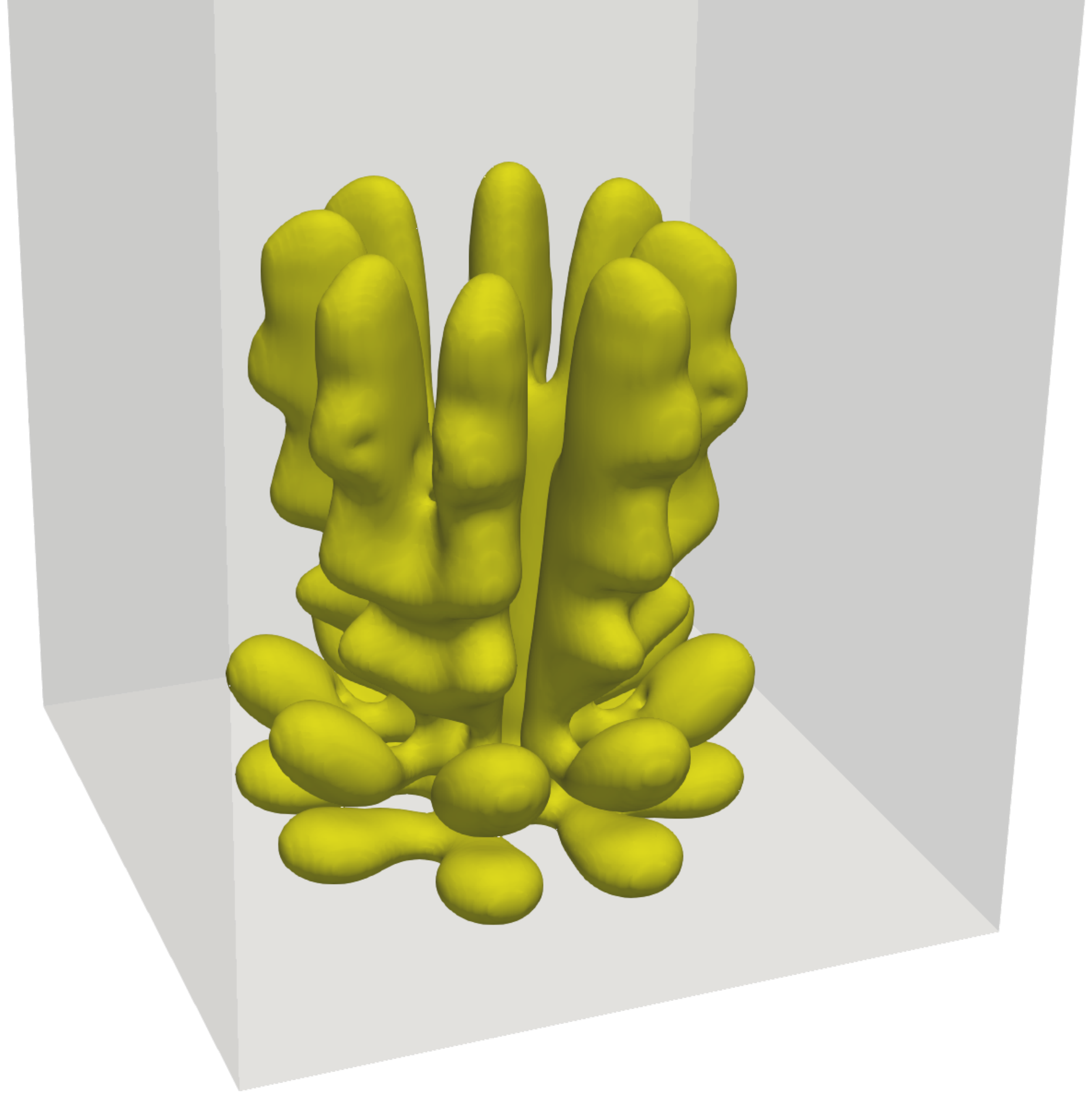}}
    \caption{$t = 210\left[s\right]$.}
\end{subfigure}
\caption{3D lithium dendrite simulation with modified anisotropy representation, under $\phi_b=-0.7\left[V\right]$ charging potential. The electrodeposited lithium is represented with a yellow isosurface plot of the phase-field variable $\xi=0.5$. Hexagonal domain set as $5000 \times 80 \times 80 \left[\mu m^3\right]$. Test 13.}
\label{fig:3DExp_0.7V_evolut}
\end{figure}

\begin{figure} [h!]
\centering%
\begin{subfigure}[b]{0.45\linewidth}
    \centering%
{\includegraphics[height = 5cm]{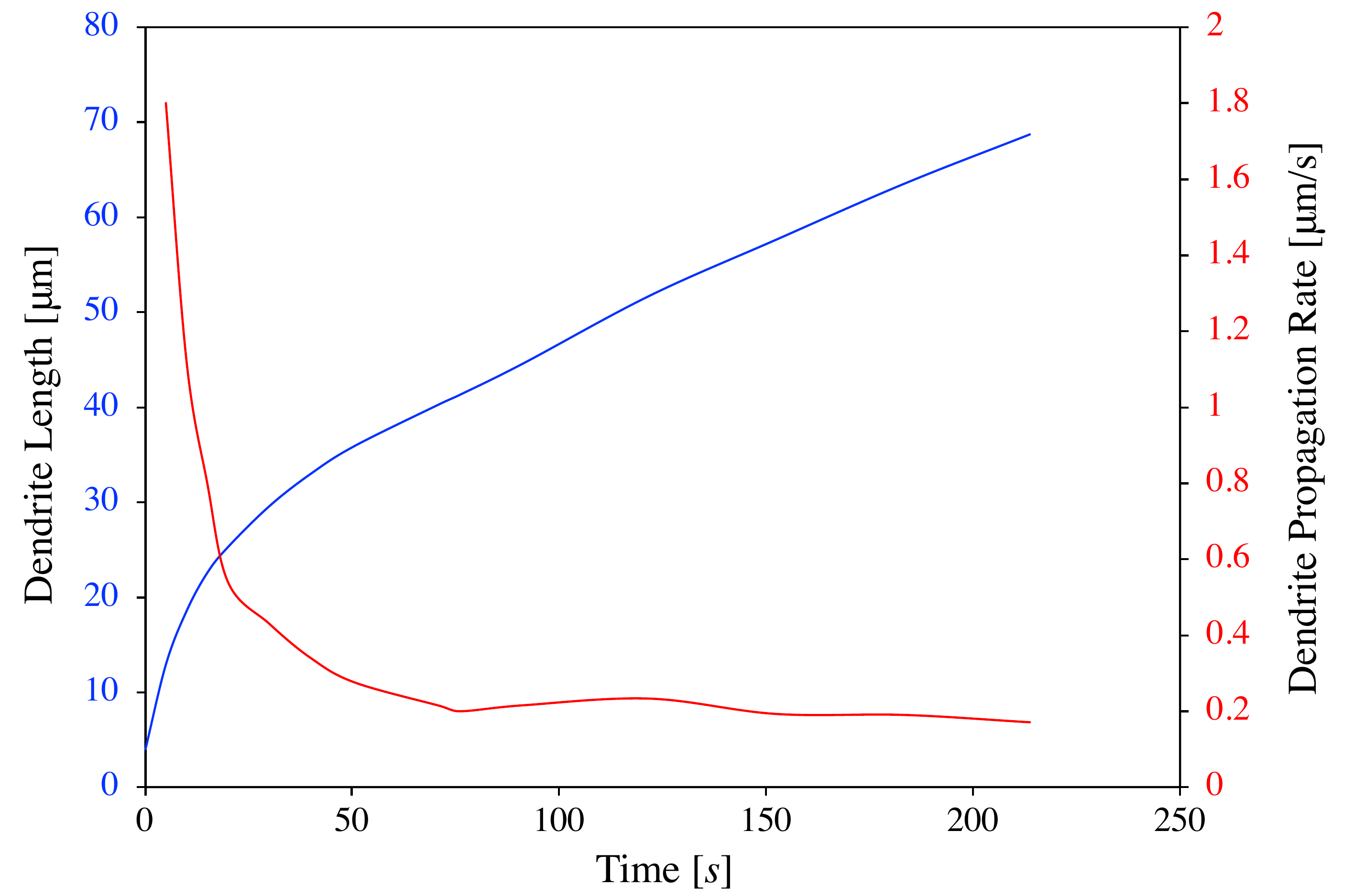}}
\caption{Test 13.}
\label{fig:Propagation_3DExp_070V}
\end{subfigure}
\begin{subfigure}[b]{0.45\linewidth}
    \centering%
{\includegraphics[height = 5cm]{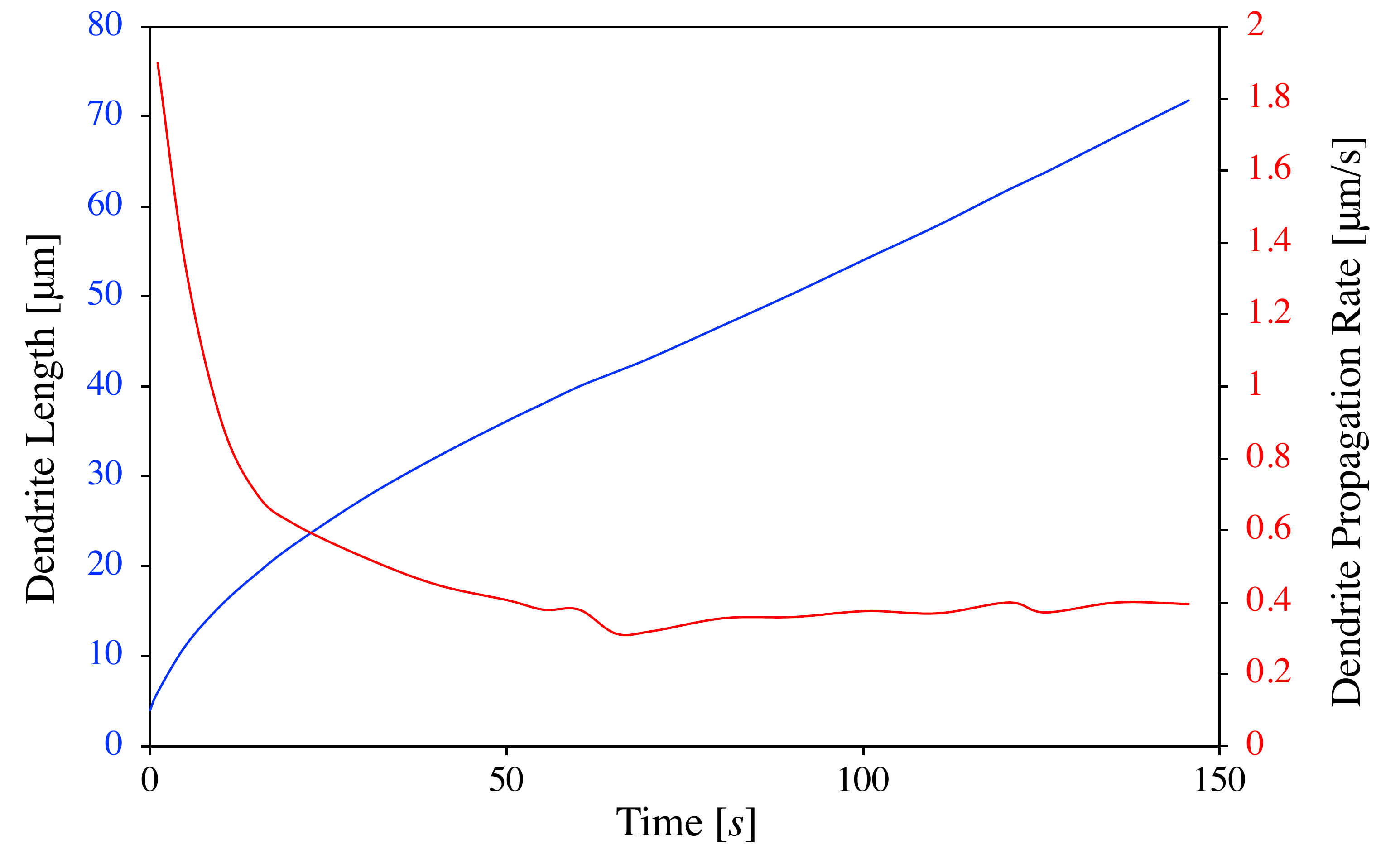}}
\caption{Test 14.}
\label{fig:Propagation_3DExp_140V}
\end{subfigure}
\caption{Simulated 3D lithium dendrite propagation plot. Dendrite length (blue) \& propagation rate (red) vs time for applied voltages: (\subref{fig:Propagation_3DExp_070V}) $\phi_b=-0.7\left[V\right]$ (Test 13), and (\subref{fig:Propagation_3DExp_140V}) $\phi_b=-1.4\left[V\right]$ (Test 14).}
\label{fig:Propagation_3DExp}
\end{figure}

Figure~\ref{fig:3DExp_0.7V_evolut} shows the morphological evolution of the simulated lithium dendrite (isosurface plot of the phase-field variable $\xi$) under $\phi_b=-0.7\left[V\right]$ charging potential (Test 13). This setup yields realistic simulation time scales due to the larger interelectrode distance we employ~\cite{ NISHIKAWA201184}. Stationary propagation rates (dendrite's tip speed) of around $0.2 \left[\mu m / s \right]$ are reached after 70 s of simulation (see Figure~\ref{fig:Propagation_3DExp_070V}). The simulated growth rates are larger than those reported by Nishikawa et al.~\cite{ NISHIKAWA201184} in experimental measurements of lithium dendrite growth in 1M LiPF$_6$ electrolyte ($0.06 \left[\mu m / s \right]$) due in part to the higher (almost double) applied current density in our model. Our results are within the range of lithium dendrite growth rates reported by Nishida et al.~\cite{ nishida2013optical} ($0.25-0.55\left[\mu m s^{-1}\right]$) using a different electrolyte type (LiTFSI). Therefore, further well-controlled experimental studies, with detailed characterization of the system parameters, combined with modeling will be necessary to improve the correlation~\cite{ AKOLKAR201484}. Unlike previous 3D simulations, forming spike-like patterns~\cite{ARGUELLO2022104892}, this case yields less branched, blunt tip, finger-like morphologies. The dendrite morphology consists of four main trunks growing from each nucleus, with pairs of orthogonal branches developing to the sides. The observed morphological difference is a consequence of the spatial distribution of the electrostatic potential in the electrolyte ($\phi$). Although the applied electric potential remains the same ($\phi_b=-0.7\left[V\right]$) the larger interelectrode distance results in a significantly different electric field distribution ($\vec{E}=-\nabla\phi$). The electric field surrounding the electrodeposit region can be 60 times smaller than in previous simulations lowering the current density (consistent with the change ratio in the interelectrode distance $5000\left[\mu m\right]/80\left[\mu m\right]=62.5$). This weaker current density results in a weaker action of the electric migration forces over the distribution of lithium ions in the electrolyte. Thus, lithium ions are less prone to accumulate around dendrite tips due to the counteracting influence of the concentration diffusion gradient, producing less branched and blunt morphologies.

The lower electric field effect is in agreement with experimental observations by Chae et al.~\cite{ chae2022modification}, where a variation of the separation between the electrodes revealed a considerable difference in the electrochemical deposition of lithium (experiments under 1$\left[mA/cm^{-2}\right]$ applied current density). Chae et al.~\cite{ chae2022modification} observed that the lithium deposition behavior and morphology changed from "hazardous" needle- and moss-like dendritic structures to "safer" morphologies (smooth and round shaped surface) as interelectrode spacing increases. The variation of lithium deposition behavior was ascribed to a difference in the Li-ion concentration distribution. Thus, when under shorter interelectrode separation ($<500\left[\mu m\right]$), lithium electrodeposition occurs closer to the high Li-ion concentration regions (formed by the release of Li-ions from the counter electrode), producing a non-uniform directional deposition of lithium. Sharp dendritic structures can grow and penetrate porous separators, which are potentially dangerous as they can create a short battery circuit~\cite{ BAI20182434}. On the other hand, larger electrode separations ($2000$ and $4000\left[\mu m\right]$) lead to a more uniform deposition, without any angular edges or sharp tips, due to lower Li-ion concentration and electric potential gradients~\cite{ chae2022modification}.  Although the current density applied in the present simulation is lower than in previous numerical examples, it remains well above the limiting current density of the system, about $i_{lim}=2\left[mA/cm^{-2}\right]$~\cite{C6EE01674J}. For over-limiting current densities applied to the cell,  the rate of lithium deposition overcomes the rate of solid-electrolyte interface formation, allowing the lithium deposit to grow almost free from the influence of the SEI~\cite{ BAI20182434}.

\begin{figure}[h!]
    \centering%
{\includegraphics[height = 7.5cm]{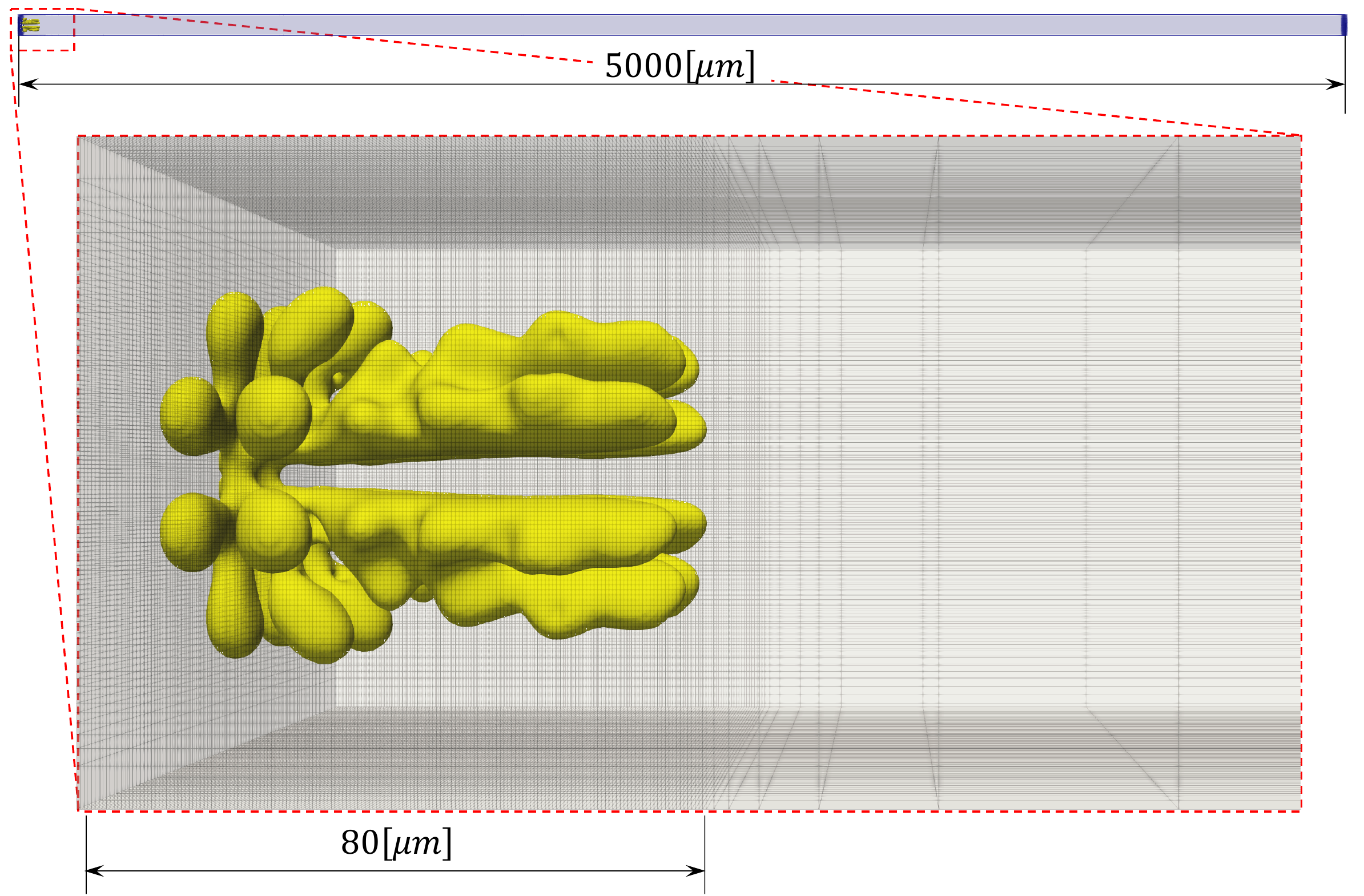}}
\caption{3D mesh overlaid with simulated lithium dendrite morphology at $t = 210\left[s\right]$ ($\phi_b=-0.7\left[V\right]$). Magnified view of the region of interest ($l_{x_{u}}\ll l_x$), showing a uniform to exponential mapping transition while moving into the bulk region.Test 13.}
\label{fig:3DExp_Mesh_070V}
\end{figure}

\begin{figure}[h!]
\begin{subfigure}[b]{0.31\linewidth}
    \centering%
{\includegraphics[height = 4.3cm]{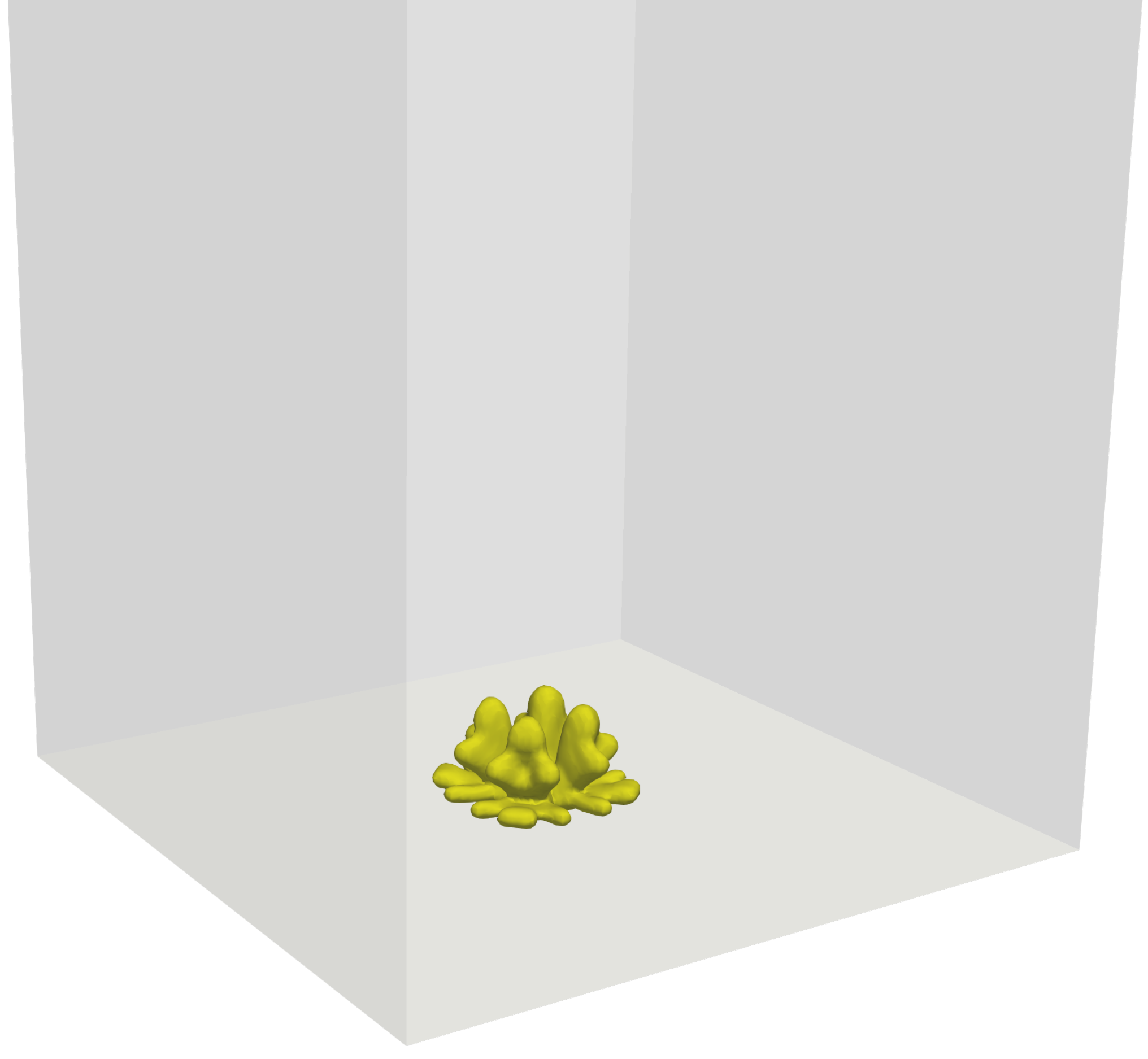}}
\caption{$t = 5\left[s\right]$.}
\end{subfigure}
\begin{subfigure}[b]{0.31\linewidth}
    \centering%
    {\includegraphics[height = 4.3cm]{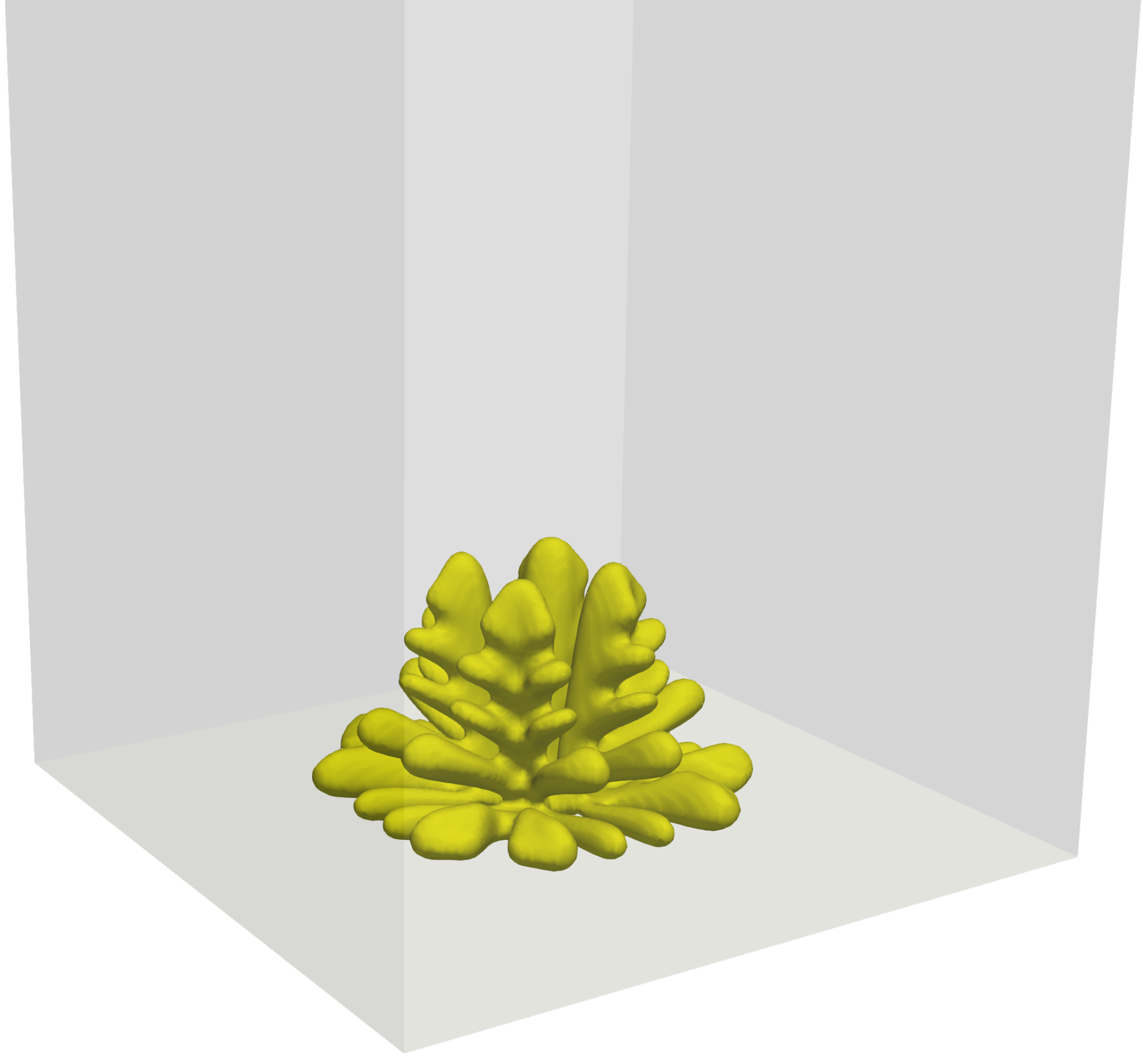}}
    \caption{$t = 30\left[s\right]$.}
\end{subfigure}
\begin{subfigure}[b]{0.31\linewidth}
    \centering%
    {\includegraphics[height = 4.3cm]{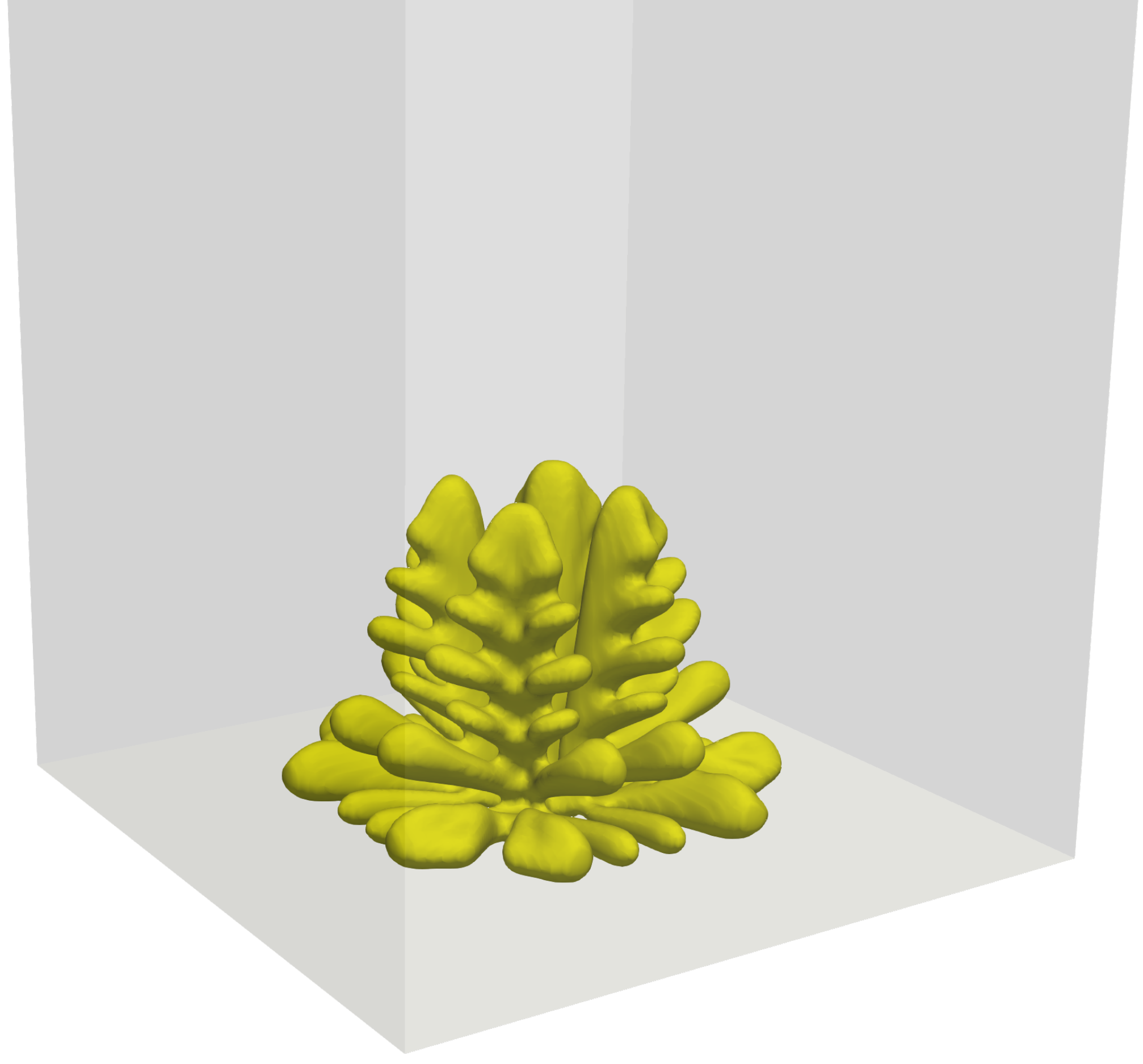}}
    \caption{$t = 50\left[s\right]$.}
\end{subfigure}

\vspace{0.3cm}
\begin{subfigure}[b]{0.31\linewidth}
    \centering%
    {\includegraphics[height = 4.3cm]{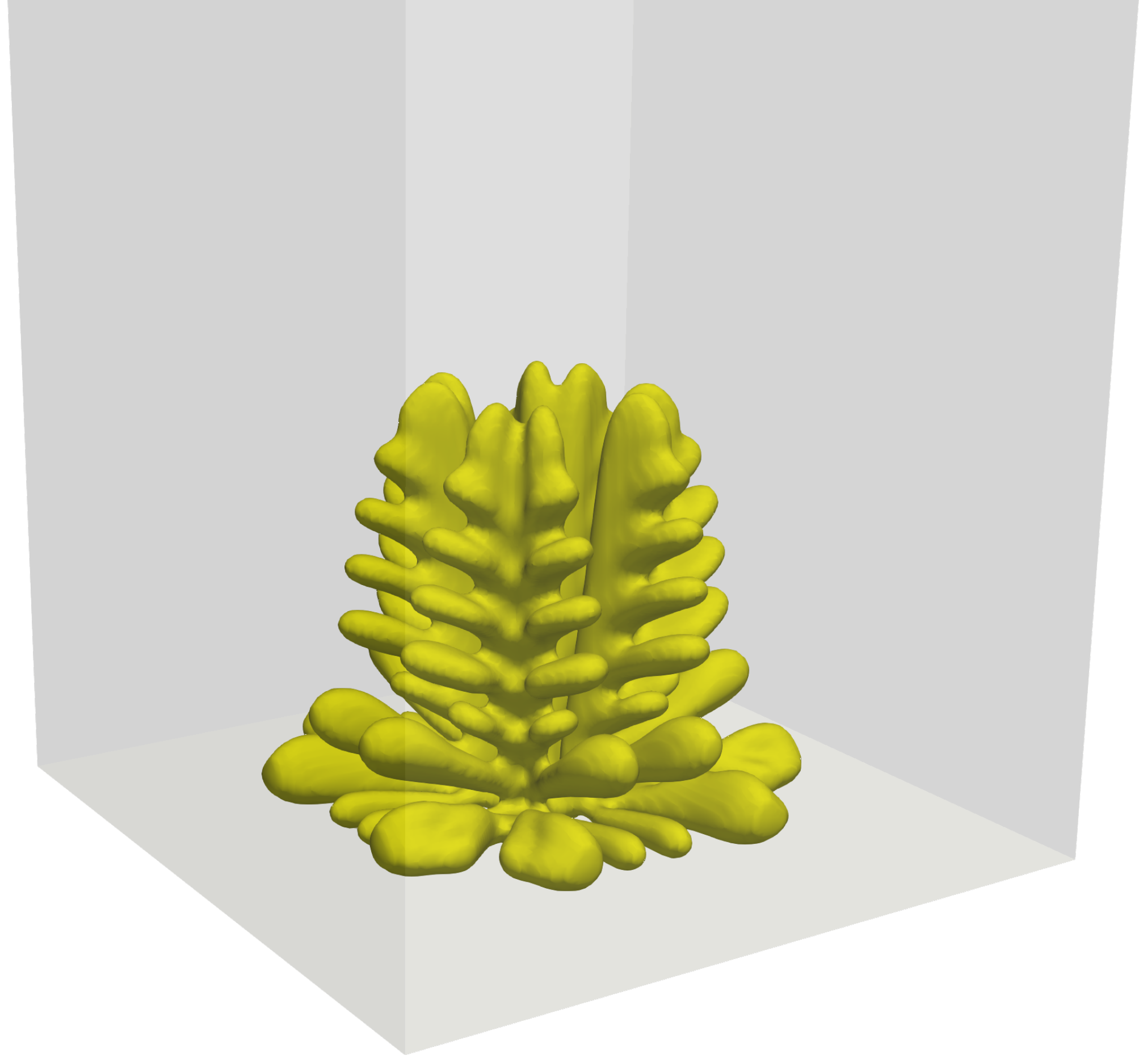}}
    \caption{$t = 80\left[s\right]$.}
\end{subfigure}
\begin{subfigure}[b]{0.31\linewidth}
    \centering%
    {\includegraphics[height = 4.3cm]{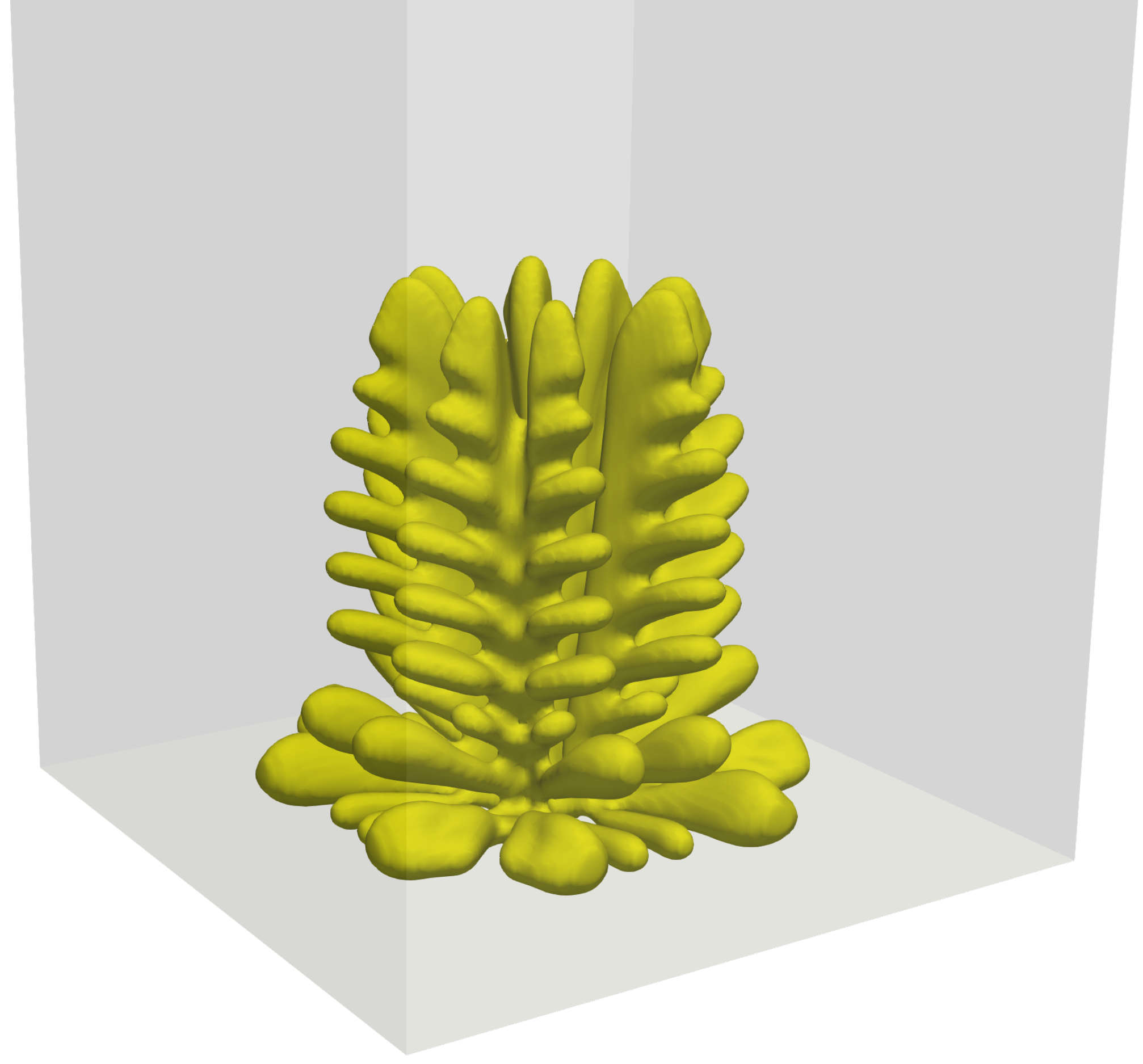}}
    \caption{$t = 110\left[s\right]$.}
\end{subfigure}
\begin{subfigure}[b]{0.31\linewidth}
    \centering%
    {\includegraphics[height = 4.3cm]{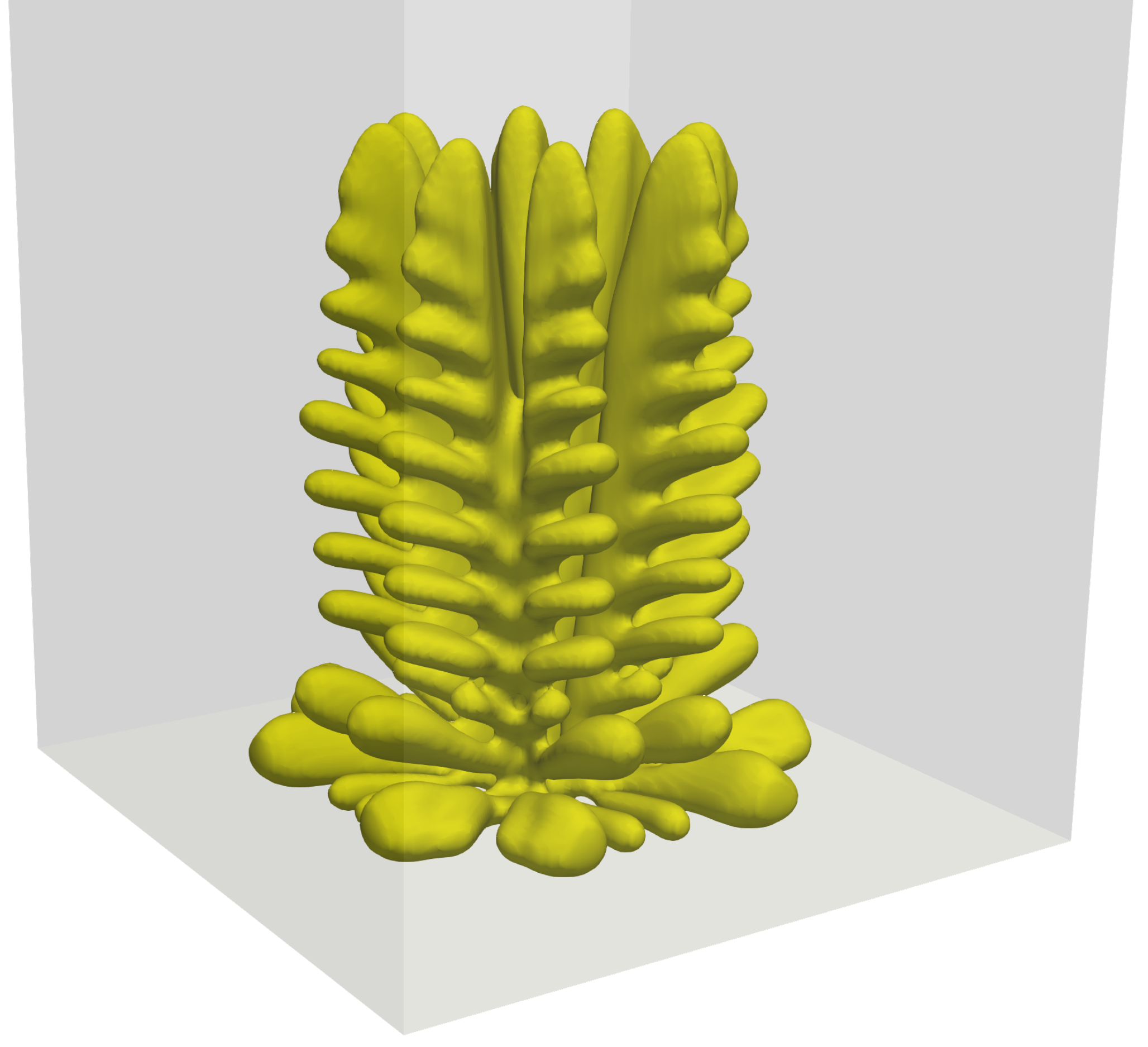}}
    \caption{$t = 150\left[s\right]$.}
\end{subfigure}
\caption{3D lithium dendrite simulation with modified anisotropy representation, under $\phi_b=-1.4\left[V\right]$ charging potential.The electrodeposited lithium is represented with a yellow isosurface plot of the phase-field variable $\xi$. Hexagonal domain set as $5000 \times 80 \times 80 \left[\mu m^3\right]$. Test 14.}
\label{fig:3DExp_1.40V_evolut}
\end{figure}

Figure~\ref{fig:3DExp_Mesh_070V} shows that the fully developed lithium dendrite morphology ($t = 210\left[s\right]$) resides within the region of interest (well-resolved portion of the domain: $\leq 80 \left[\mu m\right]$), which makes-up only 1.6\% of the whole domain. Although the system size is similar to previous 3D simulations presented in this work (5,400,000 degrees of freedom), the temporal evolution is slower, increasing the computational time to four times. 

Next, we present a 3D phase-field simulation of lithium dendrite formation under more negative applied voltage $\phi_b=-1.4\left[V\right]$ (Test 14). We use the setup of the previous experiment, with the sole difference of the applied voltage $\phi_b$. We adjust the interfacial mobility parameter $L_{\sigma}$ to the newly applied electro potential to achieve the right balance between the phase-field interface energy term and the electrochemical reaction contribution (see Table~\ref{table:steup_parameters})~\cite{ arguello2022phase}. Figure~\ref{fig:3DExp_1.40V_evolut} depicts the evolution of the lithium dendrite ($\xi$ isosurface). As in the previous experimental-scale case, we obtain realistic simulated time scale, with stationary dendrite propagation rates of about $0.4\left[\mu m / s \right]$ (see Figure~\ref{fig:Propagation_3DExp_140V}). The higher propagation rate in this case is due to the higher applied current density (from $\phi_b=-0.7$ to $-1.4\left[V\right]$ charging potential), which agrees with experimental results, where higher current densities produce faster electrodeposition and dendrite propagation rates~\cite{ Monroe_2003, NISHIKAWA201184, AKOLKAR201323}. Furthermore, we see that computed dendrite propagation rates are within the range of lithium dendrite experiments reported by Nishida et al.~\cite{ nishida2013optical}.

\begin{figure}[h!]
    \centering%
{\includegraphics[height = 7cm]{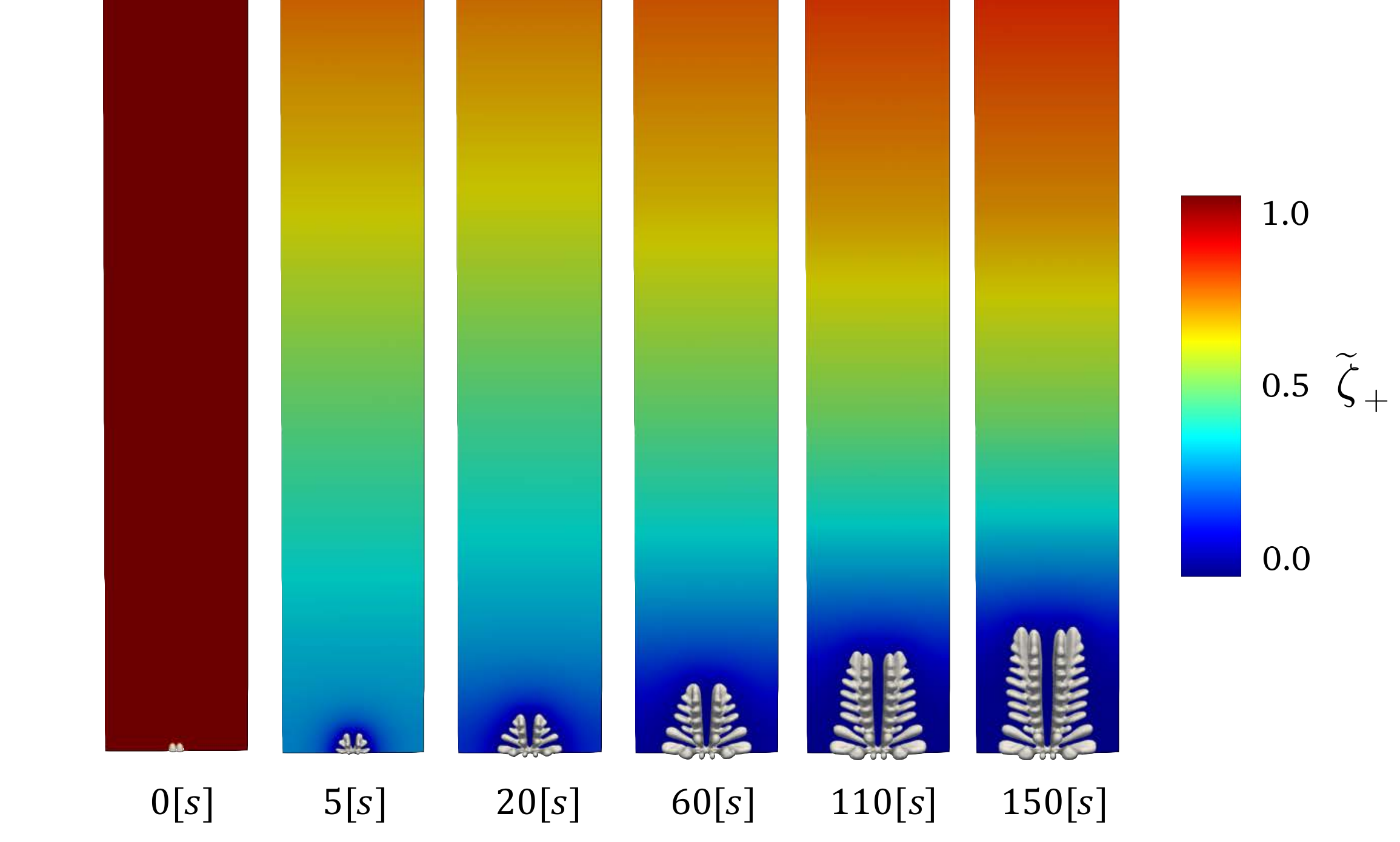}}
\caption{Evolution of the spatial distribution of lithium-ion concentration, overlaid with dendrite morphology. Contour plane set at $y=35\left[\mu m\right]$, display of first $400\left[\mu m\right]$ portion of the domain. Experimental interelectrode distance $l_x=5000\left[\mu m\right]$, and applied voltage $\phi_b=-1.4\left[V\right]$. Test 14.}
\label{fig:Li-ion_Conc_3DExp_140V}
\end{figure}

The simulation produces a spike-like, symmetric, and highly branched pattern, with morphological resemblance to previous dendritic deposits obtained under shorter interelectrode distance~\cite{ ARGUELLO2022104892}. The microstructure consists of four main trunks growing from each nucleus, with pairs of orthogonal branches developing to the sides. Figure~\ref{fig:Li-ion_Conc_3DExp_140V} depicts the evolution of the simulated spike-like dendritic morphology, together with the spatial distribution of the lithium-ion concentration $\widetilde{\zeta}_{+}$ in the electrolyte region. 

Figure~\ref{fig:Li-ion_Conc_3DExp_140V}  shows the evolution of the Li-ion concentration profile extending over $400\left[\mu m\right]$ in the stack direction ($x$); where the deposition process depletes the lithium-ion concentration close to the electrode (shown in blue). This behavior contrasts with smaller-scale simulations presented earlier in this work and in previous work~\cite{ ARGUELLO2022104892}, where Li-ion concentration enriches the dendrite tips due to large electric migration forces (see Figure~\ref{fig:3D_Aniso_CompareVsSingleSeed}). This dendrite-tip enrichment can happen in a close-to-short-circuit condition (short interelectrode distance). Nevertheless, our simulations indicate that lower electro-potential gradients, such as those obtained under experimental-scale interelectrode distances, do not seem to generate high Li-ion concentration around the dendrite tips (competition between electric migration forces and Li-ion concentration diffusion gradient). This observation is in agreement with experimental measurements of Li-ion surface concentration by Nishida et al.~\cite{ nishida2013optical}, where the concentration of Li-ion near the electrode surface was reduced from 1 M (initially) to less than 0.1 M, after a few tens of seconds of electrodeposition, depending on the experiment's conditions.

\begin{figure}[h!]
\begin{subfigure}[b]{0.45\linewidth}
    \centering%
    {\includegraphics[height = 5.4cm]{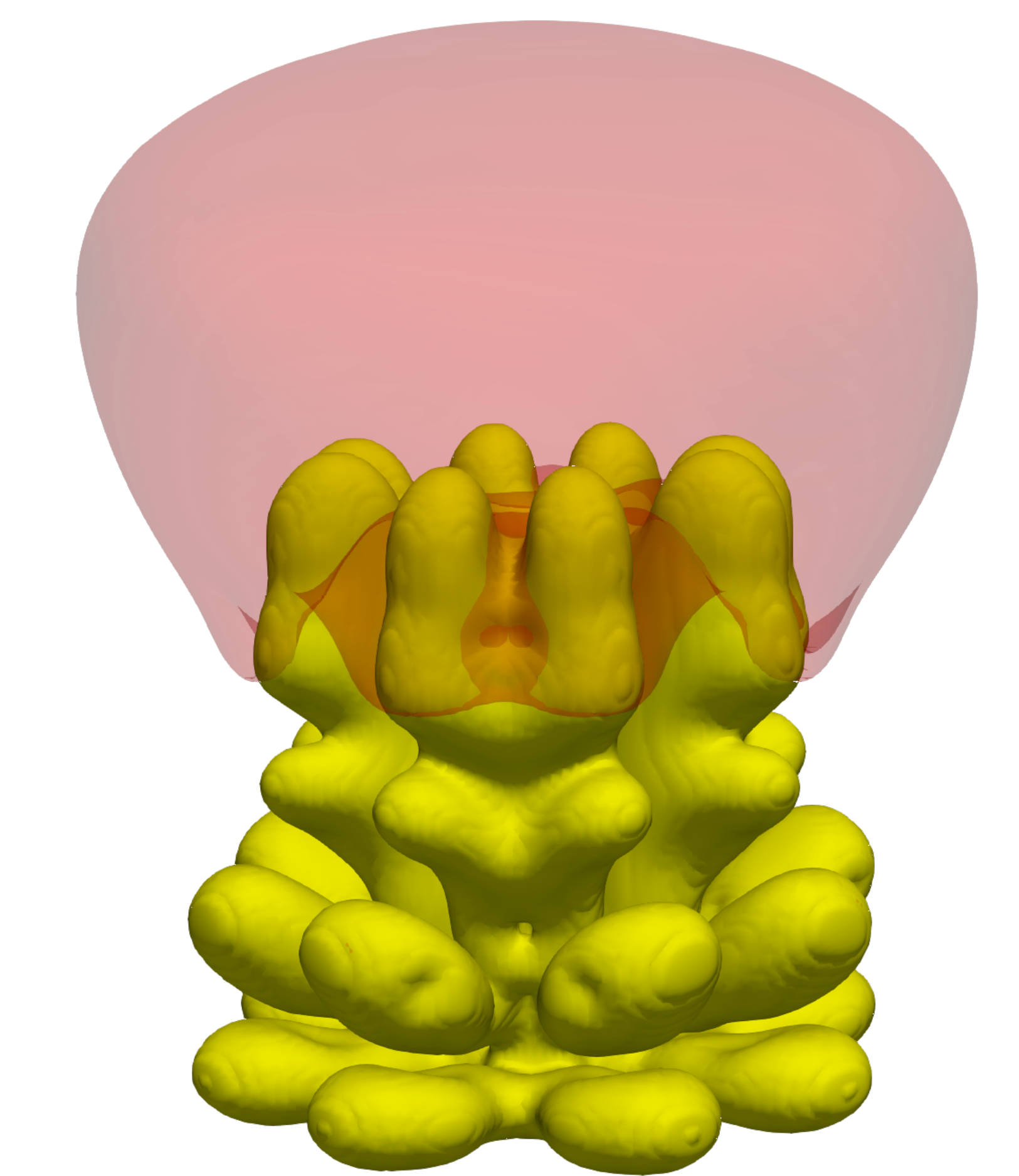}}
    \caption{$\phi_b = -0.7\left[V\right]; \ t=100\left[s\right]$. Test 13.}
    \label{fig:GradCLi_3DExp_070V}
    \end{subfigure}
    \begin{subfigure}[b]{0.45\linewidth}
    \centering%
    {\includegraphics[height = 5cm]{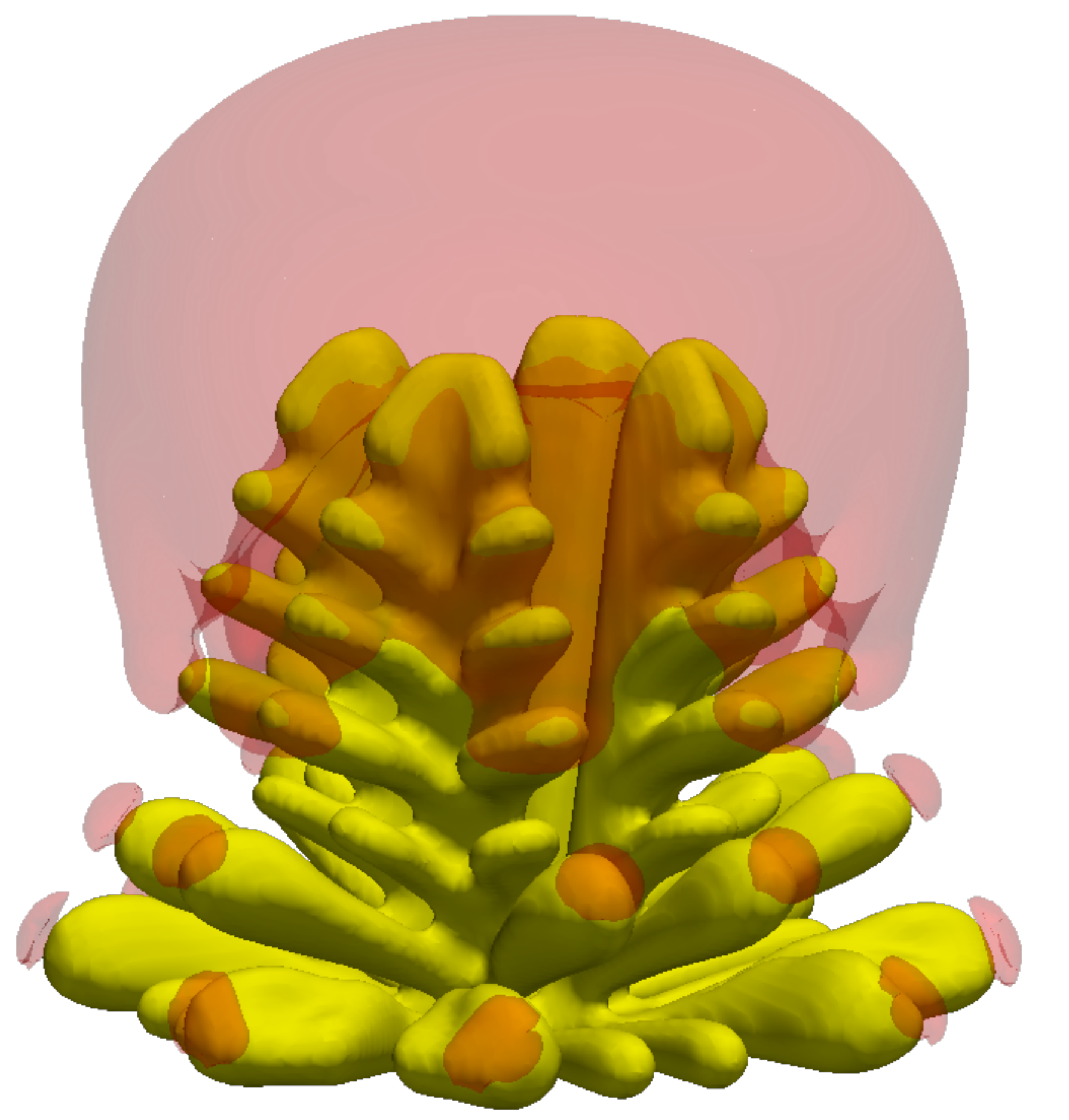}}
    \caption{$\phi_b = -1.4\left[V\right]; \ t=65\left[s\right]$. Test 14.}
    \label{fig:GradCLi_3DExp_140V}
\end{subfigure}
\caption{Comparison of lithium-ion concentration gradients for (\subref{fig:GradCLi_3DExp_070V}) $\phi_b = -0.7\left[V\right]$ (Test 13), and (\subref{fig:GradCLi_3DExp_140V}) $\phi_b = -1.4\left[V\right]$ (Test 14), applied voltage. Electrolyte regions with higher lithium-ion concentration gradient ($\|\nabla\widetilde{\zeta}_{+}\|>0.005$) represented with red volumes. Interelectrode distance $l_x=5000\left[\mu m\right]$. We use dendrite's common height ($H=45\left[\mu m\right]$) as the basis of our comparison.}
\label{fig:GradPhiVsGradCLi_3DExp_070Vs140V}
\end{figure}

We obtain apparent morphological differences from the previous dendritic lithium electrodeposition simulation under the experimental-scale domain (compare with Figure~\ref{fig:3DExp_0.7V_evolut}). Although in this case, the electric field ($\vec{E}=-\nabla\phi$) surrounding the electrodeposit region remains low relative to previous simulations with shorter interelectrode separation $\sim30$ times smaller; the larger charging voltage ($\phi_b=-1.4\left[V\right]$)  induces a spike-like and highly branched dendrite (over-limiting current density condition). This result agrees with previous two-dimensional phase-field studies investigating the effect of the applied voltage on the electrodeposit's morphological structure. Increasing the applied voltage produces faster dendrite formation with the tip splitting phenomenon~\cite{ CHEN2015376}, changing from a needle or finger-like structure to a tip splitting or spike-like pattern~\cite{ MU2019100921}. The reactive term of the phase-field equation, see~\eqref{eq:linealPF_Aniso}, is exponentially affected by the electric potential through $\eta_a=\phi\ -\text{E}^\Theta$. The applied voltage increases the degree of polarization on the electrode, affecting the deposition and accumulation of lithium on the anode surface, which leads to changes in the morphology of lithium dendrites~\cite{MU2019100921}. One verifies this by inspecting the Li-ion concentration gradient $\|\nabla\widetilde{\zeta}_{+}\|$ in the electrolyte region surrounding the dendrites morphologies. Figure~\ref{fig:GradPhiVsGradCLi_3DExp_070Vs140V} shows a comparison between the experimental-scale simulation results obtained under different charging voltages: $\phi_b = -0.7\left[V\right]$ (Figure~\ref{fig:GradCLi_3DExp_070V}), and $\phi_b = -1.4\left[V\right]$ (Figure~\ref{fig:GradCLi_3DExp_140V}). Electrolyte regions with higher lithium-ion concentration gradients ($\|\nabla\widetilde{\zeta}_{+}\|>0.005$) are represented with red volumes. Thus,  higher lithium-ion concentration gradients appear in the vicinity of the dendrites' tips and side branches in Figure~\ref{fig:GradCLi_3DExp_140V} leading to a spike-like, highly branched dendritic lithium (resembling the previously observed electric-migration versus Li-ion diffusion gradient competition happening here at a smaller scale). In contrast, Figure~\ref{fig:GradCLi_3DExp_070V}, under lower applied voltage, only presents higher lithium-ion concentration gradients in the vicinity of upper tips of the dendrite triggering vertical and less branched growth. Therefore, the spike-like lithium morphologies forming under over-limiting current density (fast battery charge)~\cite{ jana2019electrochemomechanics} can occur either using a large electric field ($\vec{E}=-\nabla\phi$) surrounding the electrodeposit region (close-to-short-circuit condition) or under a large applied voltage $\phi_b$ (fast battery charge). This strong forcing produces strong electric migration forces, causing lithium cations to move from less concentrated surrounding regions (i.e., lithium-ion depletion of valley regions) and accumulate around dendrite tips, triggering tip-growing and highly branched dendritic lithium.

\begin{figure}[h!]
    \centering%
{\includegraphics[height = 9cm]{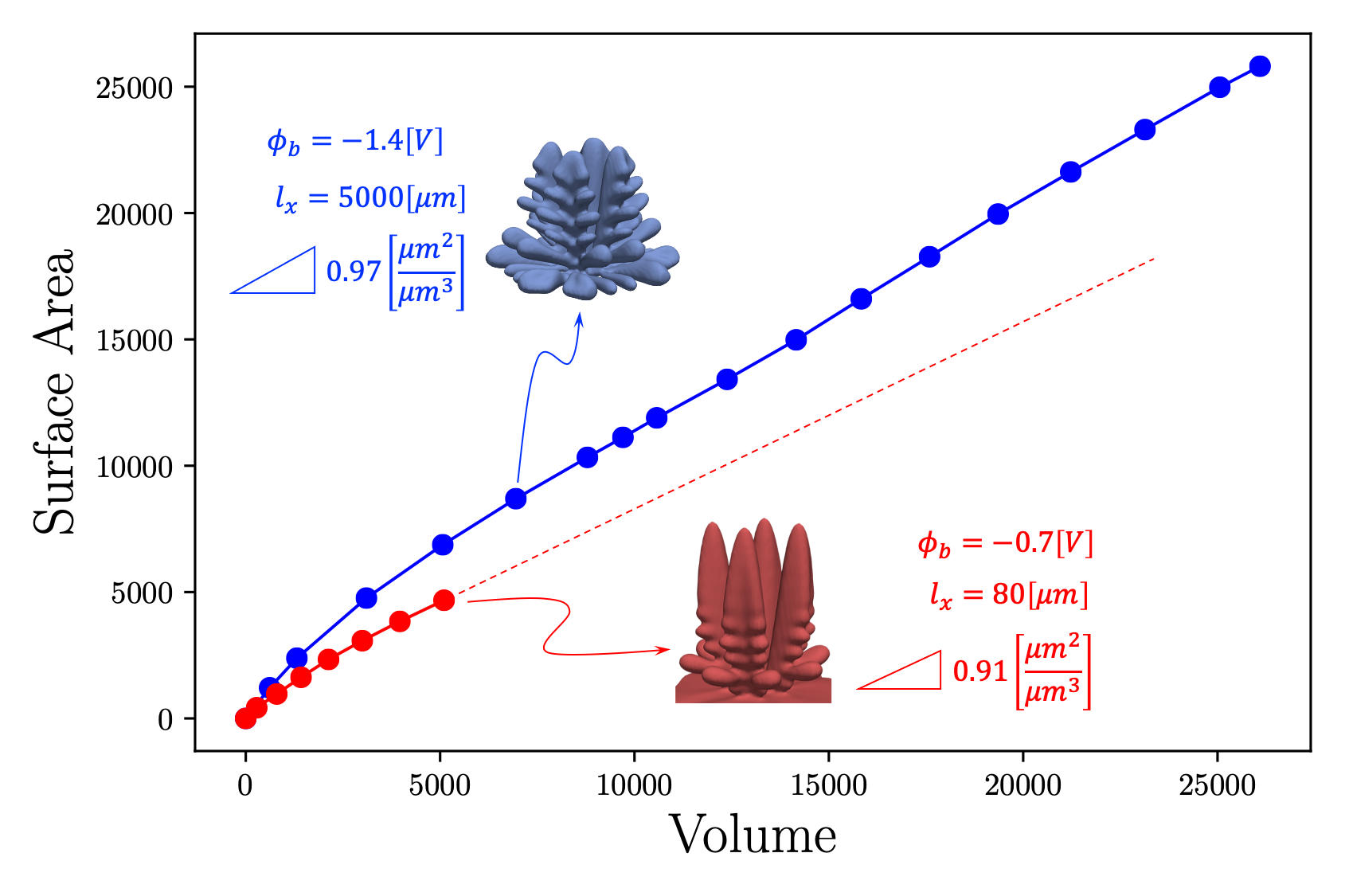}}
\caption{Morphological comparison between 3D simulations of spike-like multi-nuclei dendrite growth, smaller-scale with non-modified anisotropy representation (red)~\cite{ARGUELLO2022104892} (reproduced with Journal's permission), and experimental-scale with modified anisotropy representation (blue - Test 14), in terms of the evolution of volume vs surface area ratio.}
\label{fig:SurfVsVol_3DExp_Spike}
\end{figure}

Following~\cite{ YUFIT2019485}, we characterize the morphology by tracking the dendrites' volume-specific area ($\mu m^2 / \mu m^3$). Figure~\ref{fig:SurfVsVol_3DExp_Spike} compares the growth of the deposited volume versus the surface area for the 3D spike-like lithium pattern we simulate (short interelectrode separation vs experimental-scale results).

\begin{figure}[h!]
    \centering%
{\includegraphics[height = 9cm]{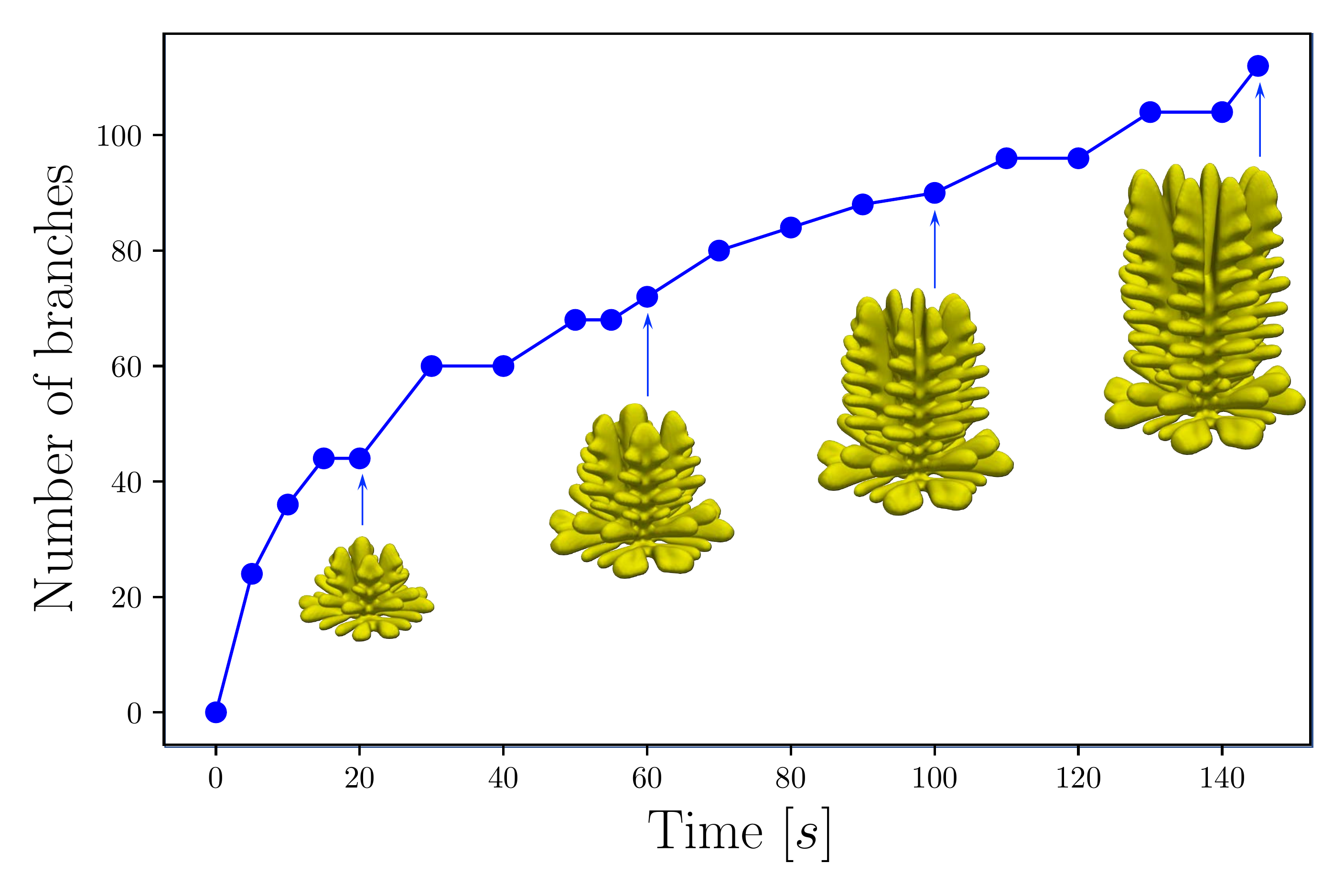}}
\caption{Morphological analysis of 3D spike-like dendrite growth simulation in terms of number of side branches developed over time. Experimental interelectrode distance $l_x=5000\left[\mu m\right]$, and applied voltage $\phi_b=-1.4\left[V\right]$.Test 14.}
\label{fig:SideBranches_3DExp_Spike}
\end{figure}

Despite differences in the time and length scales between these simulations, we obtain similar volume-specific area average ratios; 0.91 and 0.97 $\left[\mu m^2 / \mu m^3\right]$, for smaller-scale and experimental-scale simulations, respectively. The higher $\text{surface area/volume}$ ratio indicates a more branched shape in the experimental-scale simulation. Both cases are within the volume-specific results reported for experimental formation of dendrites in zinc batteries (0.86 and 1.04 $\left[\mu m^2 / \mu m^3\right]$)~\cite{ YUFIT2019485} (the literature lacks experimental data for quantitative characterization of the spike-like lithium morphologies). 

\begin{remark}
The similar area/volume average ratios between the dendritic microstructures formed using the experimental-scale simulation domain and the deposition patterns obtained under the short interelectrode distance setup (close-to-short-circuit condition) opens the possibility of using small-scale (lower-cost) 3D simulations. For example, the earlier ones in this work may be a useful testing tool to assess and adjust different 3D strategies before moving into more expensive, well-resolved larger-scale 3D simulations.
\end{remark}

Figure~\ref{fig:SideBranches_3DExp_Spike} tracks the number of side branches formed over time. The simulation produces stationary ratios of about 0.5 branch per second $\left[1/s\right]$. We compute the number of branches by visual inspection of the simulated morphologies, where we consider new protuberances as incipient branches. Given the lack of experimental data in the literature for quantitative characterization of the spike-like lithium morphologies, we rely on experimental results for zinc dendrites. Yufit et al.~\cite{ YUFIT2019485} report values between 0.19 and 0.92 branches per second $\left[1/ s\right]$ for experimental formation of "spruce tree"-like dendrites in zinc batteries under $\phi_b=-1.6\left[V\right]$ applied voltage, and $3000 \left[\mu m\right]$ interelectrode separation. Thus, we observe consistency of the simulated branching dynamic with experimental data.

\begin{figure} [h!]
    \centering%
{\includegraphics[height = 7cm]{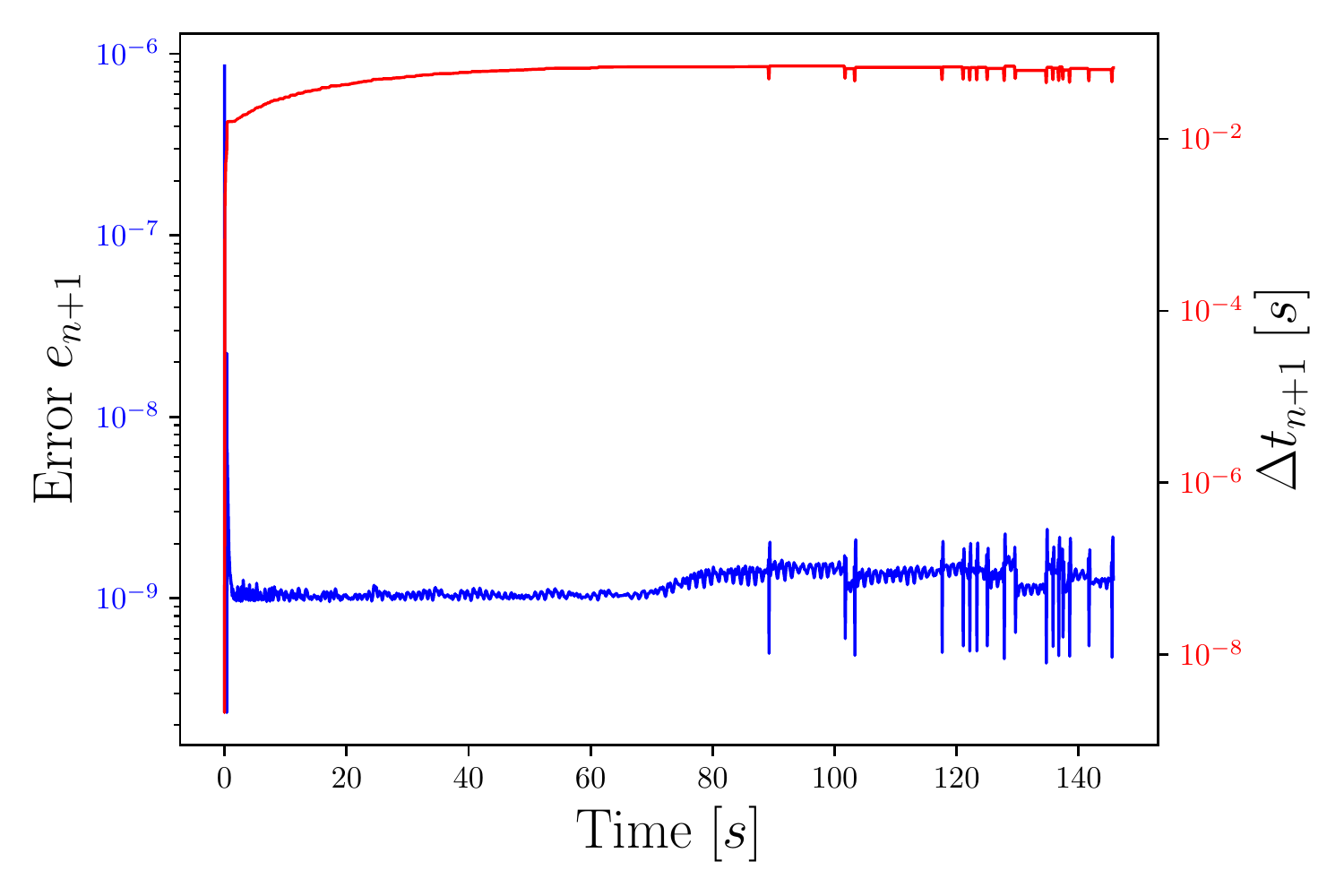}}
\caption{Time adaptivity plot for 3D lithium dendrite growth simulation under $\phi_b=-1.4\left[V\right]$ charging potential, and experimental-scale interelectrode distance ($5000 \left[\mu m\right]$). Test 14.}
\label{fig:DeltaTvsT_MultipleSeed3D_Exp_140V}
\end{figure}

Figure~\ref{fig:DeltaTvsT_MultipleSeed3D_Exp_140V} shows the behaviour of the time-adaptive scheme, throughout the $150\left[s\right]$ of the simulation. Starting with a small time-step of $\Delta t_0 = 10^{-8}\left[s\right]$ to initially achieve convergence, followed by an increase in size, until reaching a stationary value of about $\Delta t_{n+1}=0.05\left[s\right]$ (almost two orders of magnitude larger than previous simulations under smaller interelectrode distance~\cite{ARGUELLO2022104892}). The weighted truncation error $e_{n+1}$ (blue) stays close to the minimum tolerance limit ($10^{-9}$) during the whole simulation. The estimated error does not grow exponentially as in previous cases~\cite{ARGUELLO2022104892} since the lithium dendrite remains far away from the positive electrode (propagation rate does not accelerate).

\begin{figure} [h!]
    \centering%
{\includegraphics[height = 8.5cm]{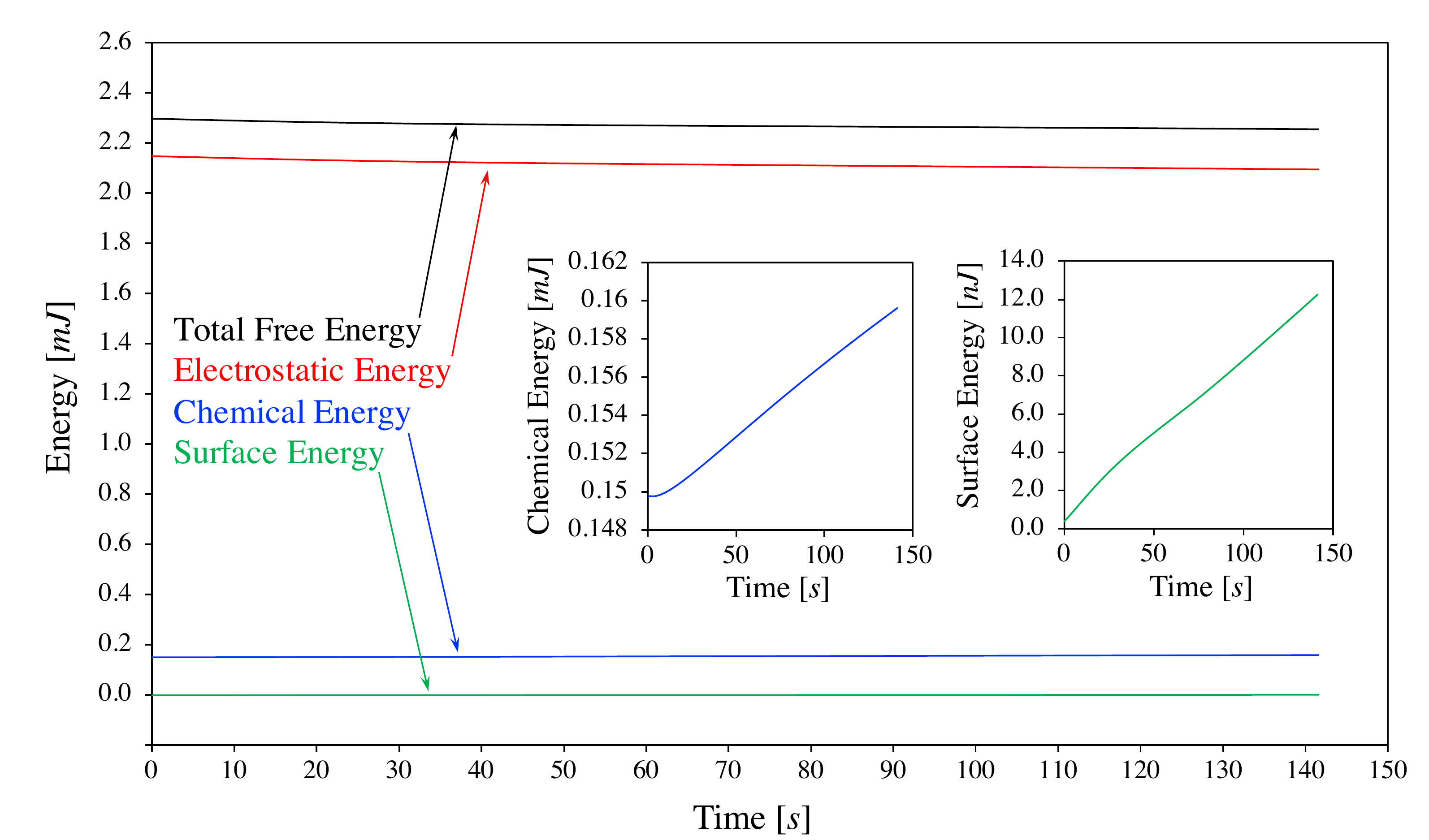}}
\caption{Energy time series for 3D spike-like dendrite growth simulation with modified anisotropy representation, under $\phi_b=-1.4\left[V\right]$ applied voltage, and interelectrode distance $l_x=5000\left[\mu m\right]$. The insets plot the increasing chemical and surface energy in smaller scale for better appreciation. Test 14.}
\label{fig:EnergyvsT_3D_Exp_140V}
\end{figure}

Standard discrete approximations do not inherit the a priori nonlinear stability relationship satisfied by phase-field models, expressed as a time-decreasing free-energy functional (see, e.g.,~\cite{ GOMEZ20115310, Sarmiento:2017, Vignal:2017} for discussions on energy stable time-marching methods). Therefore, we study the energetic evolution of our system. Figure~\ref{fig:EnergyvsT_3D_Exp_140V} shows the evolution of the Gibbs free energy of the system $\Psi$, using our adaptive time integration scheme for the experimental-scale phase-field simulation. We plot the total energy curve (black), as well as three additional energy curves that correspond to each one of its terms, namely, the Helmholtz (chemical) free energy $\int_{V}\text{f}_{\text{ch}}dV$ (blue), surface energy $\int_{V}\text{f}_{\text{grad}}dV$ (green), and electrostatic energy $\int_{V}\text{f}_{\text{elec}}dV$ (red), as the figure indicates.   Figure~\ref{fig:EnergyvsT_3D_Exp_140V} shows that the total systems' discrete free energy does not increase with time. Thus, we obtain discrete energy stable results in experimental-scale simulations using our second-order backward-difference (BDF2) time-adaptive marching scheme~\cite{ ARGUELLO2022104892}, although the method is not provably stable energetically. Moreover, while the system's surface and chemical energies grow as the area of the lithium deposit increase, the electrostatic energy decreases in time. This behavior, previously observed in smaller scale 3D simulations, is consistent with the electrodeposition process, where the system stores the applied electrostatic energy as electrochemical energy as the battery charges. The inset in Figure~\ref{fig:EnergyvsT_3D_Exp_140V} shows that the surface energy of the fully developed pattern is almost four times larger than the surface energy computed in~\cite{ ARGUELLO2022104892} for the smaller-scale simulation. The proportionately four-times larger surface area in the experimental-scale case (see Figure~\ref{fig:SurfVsVol_3DExp_Spike}) explains this scaling.

\section{Conclusions}
\label{section:concl}

We use phase-field modeling to investigate the electrodeposition process that forms dendrites within lithium-metal batteries (LMB). These simulations of lithium dendrite formation explain 3D highly branched "spike-like" dendritic morphologies formed under high current density (fast battery charge). These dendritic patterns grow fast across the electrolyte region and penetrate through porous separators, becoming hazardous for battery operation~\cite{ BAI20182434}. We analyze the dendrite formation in domains with various sizes using both, short ($80\left[\mu m\right]$) and experimental-scale ($5000\left[\mu m\right]$) interelectrode separation. The use of symmetry boundary conditions is adequate to exploit the symmetric nature of the spike-like lithium morphologies~\citep{ TATSUMA20011201}, reducing the computational cost down to 25\% of the original requirement.

Through a resolution sensitivity analysis, we asses the mesh-induced effect on the simulated 3D dendrite morphology, propagation rates (dendrite's height vs time), electrodeposition rates (dendrite's volume vs time), and energy levels. The simulated electrodeposition rate (volume of lithium metal deposited over time) is the least sensitive to the numerical parameters of our choice ($\delta_{PF}$ and $\mathscr{R}$), while dendrite's propagation rate shows the strongest sensitivity. These results have practical significance since the amount of dendritic lithium produced during charge is directly linked to the reduction in Coulombic efficiency of the battery~\citep{ Adams2018}, and the growth rate is related to battery short-circuit predictions~\cite{ROSSO20065334}. Therefore, future work may evaluate the Coulombic efficiency reduction due to dendrite formation in rechargeable lithium batteries.

We mimic lithium's cubic crystal structure and surface anisotropy by using a 3D four-fold anisotropy model of~\cite{ GEORGE2002264} to simulate crystal growth. We implement a modified 3D representation of the surface anisotropy. This modified model improves the simulation results being less sensitive to the mesh orientation. Furthermore, we introduce a surface anisotropy-based strategy that deals with randomness and uncertainty when determining the preferred growth direction of the dendrite crystal in the battery. Well-resolved simulations showed that the modified model preserves the robustness in the rate of lithium electrodeposition. We test the modified 3D surface anisotropy representation at experimental-scale interelectrode distances (higher computational cost). We simulate two charging voltages ($\phi_b=-0.7\left[V\right]$ and $\phi_b=-1.4\left[V\right]$), revealing details about the mechanism behind spike-like dendrite growth at experimental scale. Furthermore, we verify measured morphological parameters, such as simulated dendrite propagation rates, volume-specific area, and side branching rates, within the reported ranges for experimental electrodeposition of spike- or tree-like metal dendrites.

Unlike simulations using shorter interelectrode separation, we observe no enrichment of Li-ion concentration surrounding the dendrite morphology at experimental scale ($\widetilde{\zeta}_{+}<1$). However, electric migration forces continue to cause lithium cations to move from less concentrated surrounding regions and accumulate around dendrite tips (identified as higher lithium-ion concentration gradients $\|\nabla\widetilde{\zeta}_{+}\|>0.005$), triggering spike-growing and highly branched dendritic lithium in the case of $\phi_b=-1.4\left[V\right]$ charging potential. In contrast, under 50\% lower applied voltage ($\phi_b=-0.7\left[V\right]$), high lithium-ion concentration gradients are only present in the vicinity of the upper tips of the dendrite, triggering vertical and less branched growth, with smoother and rounder surface shapes~\cite{ chae2022modification}.

Thus, our analysis at the experimental scale confirms what was previously observed under smaller-scale simulations: dendrite formation is connected to the competition between the lithium cation diffusion and electric migration forces, generating an uneven distribution of Li$^+$ on the electrode surface~\cite{ ARGUELLO2022104892}. This fact gives insight into inhibition strategies focusing on enhancing the diffusion of lithium ions to achieve a more uniform concentration field on the anode surface, leading to lower dendrite formation propensity \cite{ SUNDSTROM1995599, Li2017, Zheng2017, Qian2015, Suo2013, KIM2018517, Cheng2016, Yang2005, Aoxuan2019, TAN201667, Crowther_2008, Wlasenko2010, Li2018, Iverson2008}.

Given our understanding of the process, in future work we may add other physical aspects to the simulation; our 3D phase-field model coupled with additional fields will allow us to gain insight into other aspects of dendrite formation and assess some of the proposed strategies for dendrite suppression. Thus, strategies from 2D phase-field models available in the literature could be followed; for example, the current model does not consider heat transfer to simulate the thermal effect during the lithium dendrite growth process. Thermal-induced ion-diffusion may allow us to study dendrite suppression under high operating temperatures~\citep{ YAN2018193, qiao2022quantitative}. Also, the contribution from transport (forced advection) will allow us to study the effect of electrolyte hydrodynamics on the dendrite morphology in flow batteries~\citep{ wang2019phase, parekh2020controlling}, and electrochemical-mechanical phase-field models to study the role of stress in lithium dendrites~\citep{ jana2019electrochemomechanics, YURKIV2018609}. We will also develop provably unconditionally stable second-order time accurate methods that may deliver larger time-step sizes for phase-field models~\citep{ GOMEZ20115310, Sarmiento:2017, Wu:2014, hawkins2012numerical, Vignal:2017}, adaptive mesh refinement strategies~\citep{ sakane2022parallel}, and improvement of the parallel computation efficiency~\citep{ sakane2022parallel, mu2020simulation}. 

\section{Acknowledgments}
\label{section:acknow}

This work was supported by the sponsorship of a Curtin International Postgraduate Research Scholarship (CIPRS) and the Aberdeen-Curtin Alliance PhD Scholarship. This publication was also made possible in part by the Professorial Chair in Computational Geoscience at Curtin University. This project has received funding from the European Union's Horizon 2020 research and innovation programme under the Marie Sklodowska-Curie grant agreement No 777778 (MATHROCKS). The Curtin Corrosion Centre and the Curtin Institute for Computation kindly provide ongoing support. 

\bibliographystyle{unsrt}
\bibliography{references}



\end{document}